\documentclass[aps,twocolumn,superscriptaddress,preprintnumbers,floats]{revtex4}
\usepackage[colorlinks, citecolor=blue,anchorcolor=red,menucolor=red, linkcolor=red,filecolor=red,urlcolor=blue,frenchlinks=red]{hyperref}

\usepackage{amsfonts}
\usepackage{amsmath}
\usepackage{amssymb}
\usepackage{CJKutf8}
\usepackage{color}
\usepackage{comment}
\usepackage{epsfig}
\usepackage{epstopdf}
\usepackage{float}
\usepackage{graphicx}
\usepackage{tikz}
\usepackage{booktabs}
\usepackage{indentfirst}
\usepackage{longtable,lscape}
\usepackage{mathrsfs}
\usepackage{mathtools}
\usepackage{morefloats}
\usepackage{pifont}
\usepackage{txfonts}
\usepackage{multirow}
\usepackage{makecell}
\usepackage{slashed}
\usepackage{braket}
\usepackage{bm}
\usepackage{microtype}

\begin{document}

\title{Charmed baryon semileptonic decays in a relativistic three-quark model}

\author{Ru-Hui Ni}
\affiliation{School of Physical Sciences, University of Chinese Academy of Sciences, Beijing 100049, China}
\author{Zhen-Yang Wang}
\affiliation{Physics Department, Ningbo University, Zhejiang 315211, China}
\author{Jia-Jun Wu}
\email[Corresponding author: ]{wujiajun@ucas.ac.cn}
\affiliation{School of Physical Sciences, University of Chinese Academy of Sciences, Beijing 100049, China}
\affiliation{Southern Center for Nuclear Science Theory (SCNT), Institute of Modern Physics, Chinese Academy of Sciences, Huizhou 516000, China}
\author{Bing-Song Zou}
\email[Corresponding author: ]{zoubs@mail.tsinghua.edu.cn}
\affiliation{Department of Physics and Center for High Energy Physics, Tsinghua University, Beijing 100084, China}

\begin{abstract}
In this work, we study the spin-$1/2\to1/2$ semileptonic decays of singly charmed baryons ($\Lambda_c^+$, $\Xi_c^{0,+}$, and $\Omega_c^0$) into the light baryon octet within a relativistic three-quark model.
The constituent quark masses and spatial wave functions are determined by the baryon mass spectrum. For the physical $\Xi_c$ states, the light flavor $\mathrm{SU}(3)$ breaking ($m_s > m_{u,d}$) naturally induces a coherent mixing between the flavor antitriplet and flavor sextet configurations, which is completely fixed by the mass eigenstates. Consequently, no adjustable parameters are introduced in calculating the weak transition amplitudes.
Using these wave functions, we calculate the $c\to s,d$ helicity amplitudes with the Bakamjian--Thomas boost, including the spatial Jacobian and the Wigner rotations of the constituent spins.
The branching fractions, $q^2$ distributions, longitudinal polarizations, and form factors are then obtained from these amplitudes.
For the $\Lambda_c^+\to\Lambda\ell^+\nu_\ell$ and $\Lambda_c^+\to n\ell^+\nu_\ell$ modes, our branching fractions and form factors are in good agreement with the experimental data and Lattice QCD results.
For $\Xi_c^0\to\Xi^-e^+\nu_e$, we obtain branch ratio $\mathcal B_{\ell} \simeq 4.0~\%$, which is consistent with recent Lattice QCD calculations but lies well above the current experimental average.
Clarifying the origin of this discrepancy calls for further dedicated efforts from both experimental and theoretical sides.
For the $\Omega_c^0\to\Xi^-\ell^+\nu_\ell$ decay, the spin-$1$ $ss$ spectator yields a positive longitudinal polarization of the final $\Xi^-$, in contrast to the negative polarizations in the predominantly antitriplet decay modes.
Our predictions for the $q^2$ spectra, angular asymmetries, and final baryon polarizations may provide useful references for future measurements of singly charmed baryon semileptonic decays.
\end{abstract}


\maketitle

\section{Introduction}
\label{sec:intro}

Semileptonic decays of singly charmed baryons probe the hadronic matrix elements of the $c\to s,d$ weak current.
With the leptonic current fixed by the Standard Model, the decay amplitude depends on how this active quark transition is combined with the spectator structure inside the baryon.
The charm quark occupies a unique intermediate mass scale between light flavor quarks and the heavy quark limit: it is significantly heavier than the $u$, $d$, and $s$ quarks, but not heavy enough to satisfy the heavy quark expansion criterion.
This intermediate nature means neither of the two foundational symmetries widely used in hadron decay studies (i.e., heavy quark symmetry and light flavor $\mathrm{SU}(3)$ symmetry) can be applied rigorously to singly charmed baryon semileptonic decays.
The remaining channel-to-channel variation in decay amplitudes originates from three distinct sources: the dynamics of the active charm quark transition, the spin configuration of the spectator light quark, and explicit $\mathrm{SU}(3)$ breaking effects embedded in the baryon wave functions.

The $\Lambda_c^+$ semileptonic modes provide the scalar spectator reference.
For the Cabibbo-favored process $\Lambda_c^+\to\Lambda\ell^+\nu_\ell$, the form factor ratio, absolute branching fractions, differential distributions, lepton flavor universality ratio, and angular asymmetries have been measured~\cite{CLEO:2004txf,BESIII:2015ysy,BESIII:2016ffj,BESIII:2022ysa,BESIII:2023jxv}.
For the Cabibbo-suppressed $\Lambda_c^+\to n e^+\nu_e$ channel, the BESIII Collaboration~\cite{BESIII:2024mgg} has measured the branching fraction.
Lattice QCD (LQCD) has also determined the $\Lambda_c\to\Lambda$ and $\Lambda_c\to n$ transition form factors and the corresponding decay rates~\cite{Meinel:2016dqj,Meinel:2017ggx,Bahtiyar:2021voz}.
It is nice to find that almost all calculations are consistent with the experimental measurement. 
However, the situation for $\Xi_c$ semileptonic decay is totally different.
In the neutral electron mode, the Review of Particle Physics (RPP) average of the branching fraction $\mathcal{B}_{e}[\Xi_c^0\to\Xi^-]$, based on the measurement by the Belle Collaboration and the ratio to $\Xi_c^0\to\Xi^-\pi^+$ reported by the ALICE Collaboration~\cite{Belle:2021crz,ALICE:2021bli,ParticleDataGroup:2024cfk}, lies well below the values obtained in flavor-$\mathrm{SU}(3)$ analyses~\cite{He:2021qnc}, phenomenological studies~\cite{Geng:2019bfz,Faustov:2019ddj,Geng:2020gjh,Aliev:2021wat,Duan:2022yia,Geng:2022yxb,Aliev:2025zbk,Geng:2026bnk}, and LQCD calculations~\cite{Zhang:2021oja,Farrell:2025gis}.
This discrepancy between theoretical predictions and experimental data raises key questions concerning the role of the strange spectator and $\Xi_c$--$\Xi_c'$ mixing.
Furthermore, the $\Omega_c^0\to\Xi^-\ell^+\nu_\ell$ transition then changes the spectator spin itself.
Here the scalar light pair is replaced by a spin-$1$ $ss$ pair, and the spin recoupling changes the projection of the active quark transition onto the recoil baryon helicity amplitudes.

For these channel comparisons, symmetry arguments provide only the first layer of information.
Heavy quark symmetry yields model-independent leading-order relations for heavy hadrons in the infinite-mass limit~\cite{Mannel:1990vg,Korner:1991ph,Korner:1994nh}, but it receives non-negligible $1/m_c$ corrections due to the finite charm quark mass~\cite{Korner:1994nh,Cheng:1995fe}. 
On the other hand, light flavor $\mathrm{SU}(3)$ symmetry, which is successfully proved in hyperon decays~\cite{Cabibbo:1963yz, Cabibbo:2003cu}, relate the antitriplet transition amplitudes and their symmetry-breaking patterns~\cite{Savage:1989qr,Lu:2016ogy,Geng:2017mxn,Geng:2019bfz,He:2021qnc}.
They can connect one channel with another, but they do not give the absolute size or the full $q^2$ behavior of the form factors.

There are various theoretical methods to calculate the $q^2$ dependence of the hadronic form factors. 
LQCD provides direct form factor determinations for the $\Lambda_c\to\Lambda, n$ and $\Xi_c\to\Xi$ transitions~\cite{Meinel:2016dqj, Bahtiyar:2021voz, Meinel:2017ggx, Zhang:2021oja, Farrell:2025gis}, but such calculations are not yet available for the full set of Cabibbo-suppressed channels, nor for the $\Omega_c^0\to\Xi^-$ mode.
Furthermore, LQCD yields form factor information for subleading Lorentz structures that are currently inaccessible to experimental measurements. 
While lattice calculations cannot by themselves reveal the detailed structure of baryon wave functions, they provide valuable data complementary to experimental measurements.

The remaining channels have also been studied within QCD sum rules (QCDSR) and light-cone sum rules (LCSR)~\cite{Liu:2009sn,Khodjamirian:2011jp,Li:2016qai,Zhang:2023nxl,Azizi:2011mw,Liu:2010bh,Aliev:2021wat,Duan:2020xcc,Duan:2022yia,Aliev:2025zbk,Aliev:2025cko}. 
Quark models offer another widely used framework: in nonrelativistic, relativistic, covariant, and light-front formulations, the $c\to s,d$ transition matrix element is either derived from model baryon wave functions or represented by covariant three-quark currents~\cite{Perez-Marcial:1989sch,Hussain:1990ai,Ivanov:1996fj,Pervin:2005ve,Pervin:2006ie,Ebert:2006rp,Gutsche:2014zna,Gutsche:2015rrt,Faustov:2016yza,Hussain:2017lir,Faustov:2019ddj,Zhao:2018zcb,Geng:2020fng,Geng:2020gjh,Li:2021qod,Geng:2022fsr}. 
The $q^2$ dependence is then fitted or reconstructed via monopole, dipole, double-pole, or related parametrizations, depending on whether the calculation is performed at zero recoil, in the spacelike region, or throughout the physical phase space.

With the exception of LQCD, however, these methods typically constrain the relevant amplitude parameters by fitting to $\Lambda_c\to\Lambda$ data, and subsequently extend the results to $\Xi_c$ and $\Omega_c$ decays by invoking flavor $\mathrm{SU}(3)$ symmetry. 
For instance, in sum rule and quark model calculations, the continuum thresholds, baryonic current couplings, and wave function size parameters simultaneously govern both the decay rates and the extrapolated $q^2$ behavior of the form factors. 
Consequently, successfully reproducing the $\Lambda_c$ channels does not by itself disentangle how much of a prediction arises from the recoil treatment, how much from the overlap between initial and final baryon wave functions, and how much from the decay rate information inherited from the $\Lambda_c$ input.

This ambiguity has concrete phenomenological consequences. 
For the $\Xi_c\to\Xi$ decays, simple $\mathrm{SU}(3)$ scaling from $\Lambda_c$ yields a branching fraction well above the experimental value. 
Because the $\Xi_c$ system involves both a strange spectator quark and potential $\Xi_c$--$\Xi_c'$ configuration mixing, it remains unclear whether this difference is related to configuration mixing, modified wave function overlap and recoil kinematics, or the normalization used in the experimental extraction.
The $\Omega_c^0\to\Xi^-$ mode presents a conceptually distinct puzzle. 
Here the scalar light diquark is replaced by a spin-$1$ $ss$ spectator, so the transition acquires spin recoupling contributions that are entirely absent from the scalar spectator $\Lambda_c$ modes.
Taken together, these two issues, the $\Xi_c$ rate discrepancy on the one hand, and the spin recoupling effects in $\Omega_c$ decays, call for a calculation in which the baryon wave functions and the relativistic recoil kinematics are treated consistently within a single framework.

For this purpose, we extend the relativistic three-quark model (R3QM) formalism, previously used for hyperon semileptonic decays~\cite{Ni:2026arb}, to singly charmed baryons.
The constituent quark masses and the rest frame baryon wave functions are fixed by the baryon mass spectrum.
After the spectrum is fixed, no additional free parameters are introduced in calculating the weak transition amplitudes.
The differences among the channels are then traced to the baryon wave functions and the relativistic recoil kinematics, rather than to an empirical $\Lambda_c$ rate input.

We calculate the $c\to s,d$ transition helicity amplitudes by applying the Bakamjian--Thomas boost to the rest frame wave functions, including the spatial Jacobian and the Wigner rotations of the constituent spins.
The branching fractions, $q^2$ distributions, and longitudinal polarizations are obtained from these amplitudes.
The invariant form factors are then extracted from the helicity amplitudes by analytical inversion, without imposing a monopole, dipole, or double-pole $q^2$ form.

With the helicity amplitudes and form factors obtained in this way, the channel comparison can be organized in three steps.
The $\Lambda_c$ modes first test the scalar spectator part of the calculation.
Next, the $\Xi_c\to\Xi$ modes introduce a strange spectator and $\Xi_c$--$\Xi_c'$ configuration mixing. 
They provide an independent test of how the strange spectator, configuration mixing, and recoil kinematics affect the decay rates without carrying over empirical $\Lambda_c$ branching fractions.
The $\Omega_c^0\to\Xi^-$ mode then changes the spectator spin itself.
In this channel, the scalar pair is replaced by a spin-$1$ $ss$ spectator, and the helicity amplitudes and final baryon polarization directly reflect the corresponding spin recoupling.
Comparisons with the available experimental data and LQCD results then test the scalar spectator channels and provide the reference point for the spin-$1$ spectator case.
For channels not yet measured, our predicted rates, $q^2$ distributions, and polarization observables can provide useful references for future measurements by the Belle II Collaboration, BESIII Collaboration, and LHCb Collaboration.

The remainder of this paper is organized as follows. 
Sec.~\ref{sec:semileptonic_framework} outlines the R3QM theoretical framework for the semileptonic transitions. 
Sec.~\ref{sec:results} presents our numerical predictions and provides systematic comparisons against experimental data, LQCD results, and other phenomenological predictions.
A summary is given in Sec.~\ref{sec:summary}.

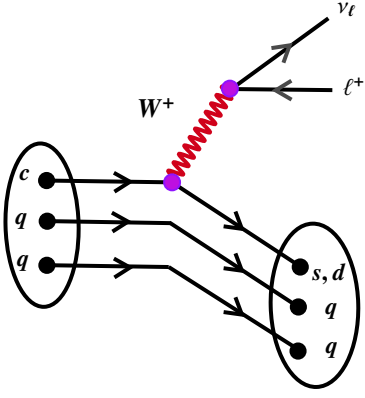
\begin{figure}[t]
\centering
\tikzset{every picture/.style={line width=0.75pt}} 

\begin{tikzpicture}[x=0.75pt,y=0.75pt,yscale=-1,xscale=1]
	\draw [color={rgb, 255:red, 0; green, 0; blue, 0 }  ,draw opacity=1 ][fill={rgb, 255:red, 0; green, 0; blue, 0 }  ,fill opacity=1 ][line width=1.5]    (156.72,164.35) -- (220.72,206.35) ;
	\draw  [color={rgb, 255:red, 208; green, 2; blue, 27 }  ,draw opacity=1 ][line width=1.5]  (155.86,163.68) .. controls (157.64,164.33) and (159.35,164.96) .. (159.59,164.58) .. controls (159.83,164.19) and (158.55,162.91) .. (157.2,161.57) .. controls (155.85,160.23) and (154.57,158.95) .. (154.81,158.56) .. controls (155.05,158.18) and (156.75,158.81) .. (158.54,159.46) .. controls (160.33,160.12) and (162.03,160.74) .. (162.28,160.36) .. controls (162.52,159.98) and (161.24,158.7) .. (159.89,157.35) .. controls (158.54,156.01) and (157.25,154.73) .. (157.49,154.35) .. controls (157.74,153.97) and (159.44,154.59) .. (161.23,155.24) .. controls (163.02,155.9) and (164.72,156.52) .. (164.96,156.14) .. controls (165.21,155.76) and (163.92,154.48) .. (162.57,153.14) .. controls (161.22,151.79) and (159.94,150.51) .. (160.18,150.13) .. controls (160.42,149.75) and (162.13,150.37) .. (163.92,151.03) .. controls (165.7,151.68) and (167.41,152.31) .. (167.65,151.92) .. controls (167.89,151.54) and (166.61,150.26) .. (165.26,148.92) .. controls (163.91,147.58) and (162.62,146.29) .. (162.87,145.91) .. controls (163.11,145.53) and (164.81,146.15) .. (166.6,146.81) .. controls (168.39,147.47) and (170.09,148.09) .. (170.34,147.71) .. controls (170.58,147.33) and (169.3,146.05) .. (167.95,144.7) .. controls (166.6,143.36) and (165.31,142.08) .. (165.55,141.7) .. controls (165.8,141.31) and (167.5,141.94) .. (169.29,142.59) .. controls (171.08,143.25) and (172.78,143.87) .. (173.02,143.49) .. controls (173.27,143.11) and (171.98,141.83) .. (170.63,140.48) .. controls (169.28,139.14) and (168,137.86) .. (168.24,137.48) .. controls (168.48,137.1) and (170.19,137.72) .. (171.97,138.38) .. controls (173.76,139.03) and (175.47,139.65) .. (175.71,139.27) .. controls (175.95,138.89) and (174.67,137.61) .. (173.32,136.27) .. controls (171.97,134.92) and (170.68,133.64) .. (170.93,133.26) .. controls (171.17,132.88) and (172.87,133.5) .. (174.66,134.16) .. controls (176.45,134.82) and (178.15,135.44) .. (178.4,135.06) .. controls (178.64,134.67) and (177.35,133.39) .. (176,132.05) .. controls (174.65,130.71) and (173.37,129.43) .. (173.61,129.05) .. controls (173.86,128.66) and (175.56,129.29) .. (177.35,129.94) .. controls (179.14,130.6) and (180.84,131.22) .. (181.08,130.84) .. controls (181.33,130.46) and (180.04,129.18) .. (178.69,127.83) .. controls (177.34,126.49) and (176.06,125.21) .. (176.3,124.83) .. controls (176.54,124.45) and (178.25,125.07) .. (180.03,125.73) .. controls (181.82,126.38) and (183.53,127) .. (183.77,126.62) .. controls (184.01,126.24) and (182.73,124.96) .. (181.38,123.62) .. controls (180.03,122.27) and (178.74,120.99) .. (178.99,120.61) .. controls (179.23,120.23) and (180.93,120.85) .. (182.72,121.51) .. controls (184.51,122.16) and (186.21,122.79) .. (186.46,122.41) .. controls (186.7,122.02) and (185.41,120.74) .. (184.06,119.4) .. controls (182.71,118.06) and (181.43,116.78) .. (181.67,116.39) .. controls (181.92,116.01) and (183.62,116.64) .. (185.41,117.29) .. controls (186.19,117.58) and (186.95,117.86) .. (187.58,118.05) ;
	\draw [color={rgb, 255:red, 0; green, 0; blue, 0 }  ,draw opacity=1 ][line width=1.5]    (97.7,205.94) -- (155.03,206.8) ;
	\draw  [line width=1.5]  (90.45,144.86) .. controls (100.57,144.6) and (109.25,162.72) .. (109.82,185.34) .. controls (110.39,207.95) and (102.65,226.5) .. (92.52,226.76) .. controls (82.4,227.02) and (73.73,208.89) .. (73.16,186.28) .. controls (72.59,163.67) and (80.33,145.12) .. (90.45,144.86) -- cycle ;
	\draw  [color={rgb, 255:red, 0; green, 0; blue, 0 }  ,draw opacity=1 ][fill={rgb, 255:red, 0; green, 0; blue, 0 }  ,fill opacity=1 ] (90.11,205.94) .. controls (90.11,203.83) and (91.81,202.13) .. (93.91,202.13) .. controls (96.01,202.13) and (97.7,203.83) .. (97.7,205.94) .. controls (97.7,208.04) and (96.01,209.75) .. (93.91,209.75) .. controls (91.81,209.75) and (90.11,208.04) .. (90.11,205.94) -- cycle ;
	\draw  [color={rgb, 255:red, 0; green, 0; blue, 0 }  ,draw opacity=1 ][fill={rgb, 255:red, 0; green, 0; blue, 0 }  ,fill opacity=1 ] (89.97,163.44) .. controls (89.97,161.34) and (91.67,159.63) .. (93.76,159.63) .. controls (95.86,159.63) and (97.56,161.34) .. (97.56,163.44) .. controls (97.56,165.55) and (95.86,167.25) .. (93.76,167.25) .. controls (91.67,167.25) and (89.97,165.55) .. (89.97,163.44) -- cycle ;
	\draw  [color={rgb, 255:red, 0; green, 0; blue, 0 }  ,draw opacity=1 ][fill={rgb, 255:red, 0; green, 0; blue, 0 }  ,fill opacity=1 ] (90.28,183.84) .. controls (90.28,181.73) and (91.98,180.02) .. (94.07,180.02) .. controls (96.17,180.02) and (97.87,181.73) .. (97.87,183.84) .. controls (97.87,185.94) and (96.17,187.65) .. (94.07,187.65) .. controls (91.98,187.65) and (90.28,185.94) .. (90.28,183.84) -- cycle ;
	\draw [color={rgb, 255:red, 0; green, 0; blue, 0 }  ,draw opacity=1 ][fill={rgb, 255:red, 0; green, 0; blue, 0 }  ,fill opacity=1 ][line width=1.5]    (97.87,183.84) -- (156.03,184.8) ;
	\draw [color={rgb, 255:red, 0; green, 0; blue, 0 }  ,draw opacity=1 ][fill={rgb, 255:red, 0; green, 0; blue, 0 }  ,fill opacity=1 ][line width=1.5]    (93.76,163.44) -- (156.72,164.35) ;
	\draw  [color={rgb, 255:red, 0; green, 0; blue, 0 }  ,draw opacity=1 ][line width=1.5]  (127.53,159.55) -- (135.62,164.05) -- (127.8,169.03) ;
	\draw [line width=1.5]    (235.12,82.1) -- (186.45,117.56) ;
	\draw [color={rgb, 255:red, 0; green, 0; blue, 0 }  ,draw opacity=1 ][line width=1.5]    (186.45,117.56) -- (237.12,118.1) ;
	\draw  [color={rgb, 255:red, 144; green, 19; blue, 254 }  ,draw opacity=1 ][fill={rgb, 255:red, 189; green, 16; blue, 224 }  ,fill opacity=1 ] (152.72,164.35) .. controls (152.72,162.14) and (154.51,160.35) .. (156.72,160.35) .. controls (158.93,160.35) and (160.72,162.14) .. (160.72,164.35) .. controls (160.72,166.55) and (158.93,168.35) .. (156.72,168.35) .. controls (154.51,168.35) and (152.72,166.55) .. (152.72,164.35) -- cycle ;
	\draw [color={rgb, 255:red, 0; green, 0; blue, 0 }  ,draw opacity=1 ][line width=1.5]    (155.35,184.46) -- (220.07,226.84) ;
	\draw [color={rgb, 255:red, 0; green, 0; blue, 0 }  ,draw opacity=1 ][line width=1.5]    (155.03,206.8) -- (221.03,248.8) ;
	\draw  [color={rgb, 255:red, 0; green, 0; blue, 0 }  ,draw opacity=1 ][fill={rgb, 255:red, 0; green, 0; blue, 0 }  ,fill opacity=1 ] (216.11,248.94) .. controls (216.11,246.83) and (217.81,245.13) .. (219.91,245.13) .. controls (222.01,245.13) and (223.7,246.83) .. (223.7,248.94) .. controls (223.7,251.04) and (222.01,252.75) .. (219.91,252.75) .. controls (217.81,252.75) and (216.11,251.04) .. (216.11,248.94) -- cycle ;
	\draw  [color={rgb, 255:red, 0; green, 0; blue, 0 }  ,draw opacity=1 ][fill={rgb, 255:red, 0; green, 0; blue, 0 }  ,fill opacity=1 ] (217.23,206.8) .. controls (217.23,204.69) and (218.93,202.98) .. (221.03,202.98) .. controls (223.12,202.98) and (224.82,204.69) .. (224.82,206.8) .. controls (224.82,208.9) and (223.12,210.61) .. (221.03,210.61) .. controls (218.93,210.61) and (217.23,208.9) .. (217.23,206.8) -- cycle ;
	\draw  [color={rgb, 255:red, 0; green, 0; blue, 0 }  ,draw opacity=1 ][fill={rgb, 255:red, 0; green, 0; blue, 0 }  ,fill opacity=1 ] (216.28,226.84) .. controls (216.28,224.73) and (217.98,223.02) .. (220.07,223.02) .. controls (222.17,223.02) and (223.87,224.73) .. (223.87,226.84) .. controls (223.87,228.94) and (222.17,230.65) .. (220.07,230.65) .. controls (217.98,230.65) and (216.28,228.94) .. (216.28,226.84) -- cycle ;
	\draw  [line width=1.5]  (227.11,185.06) .. controls (239.24,184.75) and (249.54,202.83) .. (250.11,225.44) .. controls (250.68,248.06) and (241.32,266.64) .. (229.19,266.95) .. controls (217.06,267.26) and (206.76,249.18) .. (206.19,226.57) .. controls (205.62,203.95) and (214.99,185.37) .. (227.11,185.06) -- cycle ;
	\draw  [color={rgb, 255:red, 0; green, 0; blue, 0 }  ,draw opacity=1 ][line width=1.5]  (127.53,202.55) -- (135.62,207.05) -- (127.8,212.03) ;
	\draw  [color={rgb, 255:red, 0; green, 0; blue, 0 }  ,draw opacity=1 ][line width=1.5]  (126.53,179.55) -- (134.62,184.05) -- (126.8,189.03) ;
	\draw  [color={rgb, 255:red, 144; green, 19; blue, 254 }  ,draw opacity=1 ][fill={rgb, 255:red, 189; green, 16; blue, 224 }  ,fill opacity=1 ] (181.72,117.35) .. controls (181.72,115.14) and (183.51,113.35) .. (185.72,113.35) .. controls (187.93,113.35) and (189.72,115.14) .. (189.72,117.35) .. controls (189.72,119.55) and (187.93,121.35) .. (185.72,121.35) .. controls (183.51,121.35) and (181.72,119.55) .. (181.72,117.35) -- cycle ;
	\draw  [color={rgb, 255:red, 74; green, 74; blue, 74 }  ,draw opacity=1 ][line width=1.5]  (206.56,97.03) -- (215.71,95.65) -- (212.58,104.36) ;
	\draw  [color={rgb, 255:red, 74; green, 74; blue, 74 }  ,draw opacity=1 ][line width=1.5]  (216.67,122.87) -- (208.66,118.22) -- (216.57,113.38) ;
	\draw  [color={rgb, 255:red, 0; green, 0; blue, 0 }  ,draw opacity=1 ][line width=1.5]  (188.5,221.98) -- (191.75,230.66) -- (182.57,229.39) ;
	\draw  [color={rgb, 255:red, 0; green, 0; blue, 0 }  ,draw opacity=1 ][line width=1.5]  (188.5,178.98) -- (191.75,187.66) -- (182.57,186.39) ;
	\draw  [color={rgb, 255:red, 0; green, 0; blue, 0 }  ,draw opacity=1 ][line width=1.5]  (188.5,199.98) -- (191.75,208.66) -- (182.57,207.39) ;
	
	\draw (138,120.4) node [anchor=north west][inner sep=0.75pt]  [font=\normalsize]  {$\boldsymbol{W^{+}}$};
	\draw (78.16,156.37) node [anchor=north west][inner sep=0.75pt]  [color={rgb, 255:red, 0; green, 0; blue, 0 }  ,opacity=1 ]  {$\boldsymbol{c}$};
	\draw (76.93,199.21) node [anchor=north west][inner sep=0.75pt]  [color={rgb, 255:red, 0; green, 0; blue, 0 }  ,opacity=1 ]  {$\boldsymbol{q}$};
	\draw (76.09,178.09) node [anchor=north west][inner sep=0.75pt]  [color={rgb, 255:red, 0; green, 0; blue, 0 }  ,opacity=1 ]  {$\boldsymbol{q}$};
	\draw (238,72.4) node [anchor=north west][inner sep=0.75pt]    {$\nu \boldsymbol{_{\ell }}$};
	\draw (241,110.4) node [anchor=north west][inner sep=0.75pt]    {$\ell \boldsymbol{^{+}}$};
	\draw (225.16,203.37) node [anchor=north west][inner sep=0.75pt]  [color={rgb, 255:red, 0; green, 0; blue, 0 }  ,opacity=1 ]  {$\boldsymbol{s,d}$};
	\draw (231.71,243.1) node [anchor=north west][inner sep=0.75pt]  [color={rgb, 255:red, 0; green, 0; blue, 0 }  ,opacity=1 ]  {$\boldsymbol{q}$};
	\draw (231.65,223.31) node [anchor=north west][inner sep=0.75pt]  [color={rgb, 255:red, 0; green, 0; blue, 0 }  ,opacity=1 ]  {$\boldsymbol{q}$};
\end{tikzpicture}
\caption{Quark level diagram for the charm baryon semileptonic transition $B_i\to B_f\ell^+\nu_\ell$, where the two light quarks act as spectators.}
\label{fig:semileptonic_decay_diagram}
\end{figure}

\section{Theoretical Framework}
\label{sec:semileptonic_framework}

\subsection{The derivation of amplitude from three-quark model}
The semileptonic decay $B_i(P_i,\lambda_i)\to B_f(P_f,\lambda_f)\ell^+\nu_\ell$ is induced by the quark transition $c\to q_f$, as illustrated in Fig.~\ref{fig:semileptonic_decay_diagram}.
The transition amplitude is given by
\begin{equation}
\mathcal M(s_\ell,s_\nu,\lambda_f,\lambda_i)=
\frac{G_F}{\sqrt{2}}V_{cq_f}^{*}
L_+^\mu(s_\ell,s_\nu)\,
g_{\mu\nu}\,
H^\nu(\lambda_f,\lambda_i;P_i,P_f),
\label{eq:Amplitude}
\end{equation}
Here, $G_F$ is the Fermi constant, and $V_{cq_f}^{*}$ is the Cabibbo--Kobayashi--Maskawa (CKM) matrix element.  
The symbol $q_f$ labels the final quark flavor, with $q_f=s,d$.  The labels $\lambda_i$ and $\lambda_f$ denote the initial and final baryon helicities, while $s_\ell$ and $s_\nu$ denote the lepton and neutrino spin labels.
The leptonic current takes the standard form
\begin{equation}
L_+^\mu(s_\ell,s_\nu)
=
\bar u_\nu(p_\nu,s_\nu)\gamma^\mu(1-\gamma^5)v_\ell(p_\ell,s_\ell),
\label{eq:leptonic_current_charm}
\end{equation}
where $p_\ell$ and $p_\nu$ are the four momenta of the lepton and neutrino. 
The baryon matrix element is
\begin{align}
H^\nu(\lambda_f,\lambda_i;P_i,P_f)
&=
\sum_{j=1}^{3}
\langle B_f(P_f,\lambda_f)|J_j^\nu(0)|B_i(P_i,\lambda_i)\rangle .
\label{eq:H_def_charm}
\end{align}
Here, $J_j^\nu$ is the single quark $c\to q_f$ current acting on quark line $j$. 
In the impulse approximation, the other two quark lines in each term of the sum are spectators. 
The hadronic current matrix element is calculated in the rest frame of the initial baryon. 
In this frame, the baryon four momenta are $P_i=(E_i,\mathbf{0})=(M_i,\mathbf{0})$ and $P_f=(E_f,\bm P_f)$ with $E_f=\sqrt{M_f^2+\bm P_f^2}$, and the momentum transfer is $q^\nu=P_i^\nu-P_f^\nu$.

In the present calculation, the rest frame wave functions of the ground state baryons ($J^P=1/2^+$) are expanded in the simple harmonic oscillator basis with $N_{\mathrm{osc}}\leq2$ and orbital angular momentum $L=0$, with the configuration mixing coefficients determined by the baryon mass spectrum~\cite{Ni:2026arb}.
In terms of these basis states, the physical light octet final state and a singly charmed initial state with definite light pair flavor symmetry $\tau\in\{A,S\}$ are expanded as
\begin{align}
\left|B_f^{\mathbf 8},\frac12^+\right\rangle
&=
\sum_{\kappa_f=1}^{3}
c_{\kappa_f}^{B_f,1/2}
\left|B_{f,\kappa_f}^{\mathbf 8},\frac12^+\right\rangle,
\nonumber\\
\left|B_i^\tau,\frac12^+\right\rangle
&=
\sum_{\kappa_i=1}^{4}
c_{\kappa_i}^{B_i,1/2}
\left|B_{i,\kappa_i}^{\tau},\frac12^+\right\rangle.
\label{eq:charm_configuration_state_expansions}
\end{align}
Here, $\kappa_f=1,2,3$ and $\kappa_i=1,2,3,4$ label the configuration basis states for the light octet and singly charmed baryons, respectively.
For states with definite flavor symmetry, such as $\Lambda_c^+$ ($\tau=A$) and $\Omega_c^0$ ($\tau=S$), the mixing coefficients satisfy the normalization conditions $\sum_{\kappa_i=1}^{4}|c_{\kappa_i}^{B_i,1/2}|^2=1$ and $\sum_{\kappa_f=1}^{3}|c_{\kappa_f}^{B_f,1/2}|^2=1$.
For the $\Xi_c$--$\Xi'_c$ system, the mass difference between the strange and light quarks ($m_s > m_{u,d}$) breaks the light flavor symmetry, which induces a mixing between the flavor-$A$ and flavor-$S$ configurations. 
The physical states are represented by the coherent superposition
$
\left|\Xi_c^{(\prime)},\frac12^+\right\rangle
=
\sum_{\tau=A,S}\sum_{\kappa_i=1}^{4}
c_{\kappa_i,\tau}^{\Xi_c^{(\prime)}}
\left|B_{i,\kappa_i}^{\tau},\frac12^+\right\rangle,
$
with the normalization condition $\sum_{\tau=A,S}\sum_{\kappa_i=1}^{4}|c_{\kappa_i,\tau}^{\Xi_c^{(\prime)}}|^2=1$.

To account for the permutation symmetry of the three constituent quarks, each singly charmed basis state is constructed from three ordered components.
Taking the component with the charm quark in slot 3 as the representative state $|B_{i,\kappa_i}^{\tau,[3]}\rangle$, the fully symmetrized basis state is written as
\begin{equation}
\left|B_{i,\kappa_i}^{\tau}\right\rangle
=
\frac{1}{\sqrt3}
\sum_{a=1}^{3}
\left|B_{i,\kappa_i}^{\tau,[a]}\right\rangle.
\label{eq:single_charm_slot_state}
\end{equation}
Here, the component $|B_{i,\kappa_i}^{\tau,[a]}\rangle$ with the charm quark in slot $a$ is obtained via the permutation operator $\widehat P_{a3}$,
$
\left|B_{i,\kappa_i}^{\tau,[a]}\right\rangle \equiv \widehat P_{a3} \left|B_{i,\kappa_i}^{\tau,[3]}\right\rangle,
$
which interchanges quarks in slots $a$ and 3 in both spatial and spin-flavor spaces.
Since components with the charm quark in different slots are mutually orthogonal in flavor space, the prefactor $1/\sqrt3$ ensures unit normalization.

Substituting the state expansions in Eq.~\eqref{eq:charm_configuration_state_expansions} into the transition matrix element, the physical hadronic decay amplitudes for initial states with definite flavor symmetry ($\Lambda_c^+$, $\Omega_c^0$) and for the $\Xi_c$--$\Xi'_c$ system are given by
\begin{align}
&H^\nu(\lambda_f,\lambda_i;P_i,P_f)
\nonumber \\
&=
\sum_{\kappa_f=1}^{3}\sum_{\kappa_i=1}^{4}
\left(c_{\kappa_f}^{B_f,1/2}\right)^*
c_{\kappa_i}^{B_i,1/2}
\sum_{j=1}^{3}
H_{\kappa_f\kappa_i}^{\nu;\tau,j}
(\lambda_f,\lambda_i;P_i,P_f),
\label{eq:charm_configuration_mixed_hadronic_sum}
\\
&H_{\Xi_c^{(\prime)}\to B_f}^{\nu}
(\lambda_f,\lambda_i;P_i,P_f)
\nonumber \\
&=
\sum_{\tau=A,S}\sum_{\kappa_f=1}^{3}\sum_{\kappa_i=1}^{4}
\left(c_{\kappa_f}^{B_f,1/2}\right)^*
c_{\kappa_i,\tau}^{\Xi_c^{(\prime)}}
\sum_{j=1}^{3}
H_{\kappa_f\kappa_i}^{\nu;\tau,j}
(\lambda_f,\lambda_i;P_i,P_f),
\label{eq:charm_xi_configuration_mixed_hadronic_sum}
\end{align}
where $H_{\kappa_f\kappa_i}^{\nu;\tau,j}$ denotes the single quark transition amplitude on line $j$.
For a given quark line $j$, the flavor changing weak current $J_j^\nu(0)$ acts solely on the charm quark.
Substituting the three slot state expansion in Eq.~\eqref{eq:single_charm_slot_state}, only the component with the charm quark in slot $j$ ($a=j$) gives a nonzero contribution, which simplifies the single line matrix element to
\begin{equation}
H_{\kappa_f\kappa_i}^{\nu;\tau,j}
(\lambda_f,\lambda_i;P_i,P_f)
=
\frac{1}{\sqrt3}
\left\langle
B_{f,\kappa_f}^{\mathbf 8}(P_f,\lambda_f)
\right|
J_j^\nu(0)
\left|
B_{i,\kappa_i}^{\tau,[j]}(P_i,\lambda_i)
\right\rangle.
\label{eq:charm_current_line_slot_reduction}
\end{equation}

To incorporate the relativistic recoil effects via the Bakamjian--Thomas (BT) boost~\cite{Ni:2026arb}, the transition amplitude on line $j$ is calculated in the initial baryon rest frame as
\begin{widetext}
\begin{align}
H_{\kappa_f\kappa_i}^{\nu;\tau,j}
(\lambda_f,\lambda_i;P_i,P_f)
={}&
\frac{1}{\sqrt3}
\sqrt{2E_i\,2E_f}
\int\frac{d^3\bm k_\rho}{(2\pi)^3}
\frac{d^3\bm k_\lambda}{(2\pi)^3}
\frac{\sqrt{\mathcal J_f(\{\bm p'_n\}_{n=1}^{3},\bm P_f)}}{\sqrt{4e_j e'_j}}
\left(\prod_{r\ne j}\mathcal S_r\right)
\sum_{\alpha,\beta}
\left[
\mathcal G_{\kappa_f\kappa_i}^{\tau}(j)
\right]_{\alpha\beta}
[\mathcal O^\nu_{W,j}]^{\alpha\beta}.
\label{eq:sc_full_amplitude_charm}
\end{align}
\end{widetext}
Here, $\sqrt{2E_i\,2E_f}$ is the covariant normalization of the external baryon states, and $(\bm k_\rho,\bm k_\lambda)$ are the internal Jacobi momenta in the initial baryon rest frame.
For the recoiling final baryon, the constituent quark momenta $\{\bm p'_n\}_{n=1}^{3}$ are transformed via the BT boost to its rest frame to define the final Jacobi momenta $(\bm k'_\rho,\bm k'_\lambda)$ entering the final wave function, with $\mathcal J_f(\{\bm p'_n\}_{n=1}^{3},\bm P_f)$ being the associated spatial Jacobian.
The active quark energy denominator $\sqrt{4e_j e'_j}$, with $e_j=\sqrt{m_j^2+\bm p_j^2}$ and $e'_j=\sqrt{{m'_j}^2+\bm p'_j{}^2}$, arises from the covariant normalization of the constituent quark field operators in the one body weak current.
For the two spectator lines ($r\ne j$), their spin overlaps under the relativistic boost are characterized by $\mathcal S_r$.

On the active quark line $j$, the Wigner rotated one body weak current operator $[\mathcal O^\nu_{W,j}]^{\alpha\beta}$ is defined as
\begin{align}
&[\mathcal O^\nu_{W,j}]^{\alpha\beta}
\nonumber\\
&=
\left[
D^\dagger(\bm k'_j,\bm P_f;m'_j,M_f)
\mathcal O^\nu(\bm p'_j,\bm p_j)
D(\bm k_j,\bm P_i;m_j,M_i)
\right]^{\alpha\beta},
\label{eq:Gamma_eff_full}
\end{align}
with $\alpha$ and $\beta$ being the final and initial active quark spin projections.
Here, $D(\bm k,\bm P;m,M)$ is the spin-$1/2$ Wigner rotation matrix associated with the BT boost~\cite{Bakamjian:1953kh,Faustov:1972rp} for a constituent quark of mass $m$ in a baryon of mass $M$.
Specifically, the matrix $D$ transforms the initial active quark spin state from the rest frame to the calculation frame (where $\bm P_i=\mathbf{0}$ and $D=\mathbf{1}$), while $D^\dagger$ rotates the final active quark spin state from the calculation frame to the final rest frame.
The bare one body $V-A$ weak transition current is given by
\begin{equation}
[\mathcal O^\nu(\bm p'_j,\bm p_j)]^{\alpha\beta}
=
\bar u_{q_f}(p'_j,\alpha)
\gamma^\nu(1-\gamma^5)
u_c(p_j,\beta),
\label{eq:quark_va_current_charm}
\end{equation}
where $u_c(p_j,\beta)$ and $u_{q_f}(p'_j,\alpha)$ are the constituent quark Dirac spinors with calculation frame momenta $p_j$ and $p'_j$, respectively.

For each spectator line $r\ne j$, the impulse approximation imposes that the spectator quark momentum is conserved in the calculation frame, $\bm p'_r=\bm p_r$. 
Due to the relative recoil between the initial and final baryons, the corresponding rest frame momenta $\bm k_r$ and $\bm k'_r$ undergo distinct Lorentz boosts. 
The spectator spin overlap is then given by
$
\mathcal S_r = \frac{1}{2} \mathrm{Tr}\!\left[D^\dagger(\bm k'_r,\bm P_f;m_r,M_f)D(\bm k_r,\bm P_i;m_r,M_i)\right],
$
where $m_r$ is the spectator quark mass, and the factor $1/2$ normalizes the trace over the two dimensional spectator spin space. 
Further details regarding the BT momentum mappings $\bm k^{(\prime)}_n\to\bm p^{(\prime)}_n$, the Jacobian $\mathcal J_f$, and the Wigner rotation matrices $D$ can be found in our previous work~\cite{Ni:2026arb}.

The transition kernel $[\mathcal G_{\kappa_f\kappa_i}^{\tau}(j)]_{\alpha\beta}$ contains the spatial, flavor, and spin overlaps between the initial component $|B_{i,\kappa_i}^{\tau,[j]}\rangle$ and the final octet state $|B_{f,\kappa_f}^{\mathbf 8}\rangle$, with its explicit forms for the $3\times4$ configuration space collected in Appendix~\ref{app:charm_transition_matrix_elements}.
In the static limit ($|\bm P_f|\to 0$), where the spatial overlap is unity and the Wigner rotations reduce to $D=\mathbf 1$, this kernel gives the static $\mathrm{SU}(6)$ spin flavor factors listed in Table~\ref{tab:slot_singlecharm_sf} of Appendix~\ref{app:static_form_factors}.
The physical hadronic matrix element $H^\nu$ is then obtained by summing the single line amplitudes in Eq.~\eqref{eq:sc_full_amplitude_charm} over the three quark lines and configuration states according to Eq.~\eqref{eq:charm_configuration_mixed_hadronic_sum} for $\Lambda_c^+$ and $\Omega_c^0$, and Eq.~\eqref{eq:charm_xi_configuration_mixed_hadronic_sum} for the $\Xi_c$--$\Xi'_c$ system.

\subsection{The relationship between form factors and amplitudes}

Due to the $V-A$ structure of the weak current, the physical hadronic transition matrix element $H^\nu(\lambda_f,\lambda_i;P_i,P_f)$ is decomposed into vector and axial-vector parts,
\begin{equation}
H^\nu(\lambda_f,\lambda_i;P_i,P_f)
=
H_V^\nu(\lambda_f,\lambda_i;P_i,P_f)
-H_A^\nu(\lambda_f,\lambda_i;P_i,P_f),
\end{equation}
where the separate matrix elements take the form
\begin{equation}
H_{V/A}^\nu(\lambda_f,\lambda_i;P_i,P_f)
\equiv
\sum_{j=1}^{3}
\left\langle B_f(P_f,\lambda_f)\left|
J_{V/A,j}^\nu(0)
\right|B_i(P_i,\lambda_i)\right\rangle,
\label{eq:HVHA_def_charm}
\end{equation}
with $J_{V,j}^\nu$ and $J_{A,j}^\nu$ denoting the vector and axial-vector one-body currents acting on the active quark line $j$.
By Lorentz covariance, the $H_V^\nu$ and $H_A^\nu$ matrix elements for the spin-$1/2\to1/2$ transitions are parametrized by the six invariant form factors $f_i(q^2)$ and $g_i(q^2)$~\cite{Weinberg:1958ut,Garcia:1985xz},
\begin{widetext}
\begin{align}
H_V^\nu
&=
\bar u_f(P_f,\lambda_f)
\left[
f_1(q^2)\gamma^\nu
-i\frac{f_2(q^2)}{M}\sigma^{\nu\kappa}q_\kappa
+\frac{f_3(q^2)}{M}q^\nu
\right]
u_i(P_i,\lambda_i),
\label{eq:ff_vector_current}
\\
H_A^\nu
&=
\bar u_f(P_f,\lambda_f)
\left[
g_1(q^2)\gamma^\nu\gamma^5
-i\frac{g_2(q^2)}{M}\sigma^{\nu\kappa}q_\kappa\gamma^5
+\frac{g_3(q^2)}{M}q^\nu\gamma^5
\right]
u_i(P_i,\lambda_i),
\label{eq:ff_axial_current}
\end{align}
\end{widetext}
where $M\equiv M_i$ and $\sigma^{\nu\kappa}\equiv i[\gamma^\nu,\gamma^\kappa]/2$.
To extract these invariant form factors from the R3QM, we match the covariant parameterizations in Eqs.~\eqref{eq:ff_vector_current} and \eqref{eq:ff_axial_current} to the Cartesian components obtained after the line and configuration sums associated with Eq.~\eqref{eq:sc_full_amplitude_charm}.
According to the analytical inversion scheme derived in Ref.~\cite{Ni:2026arb}, the six form factors can be projected out by selecting the temporal and longitudinal components $H_{V/A}^{0,3}(\tfrac12,\tfrac12)$ alongside the transverse component $H_{V/A}^{1}(-\tfrac12,\tfrac12)$. 
This matching yields the explicit relations
\begin{widetext}
\begin{align}
f_2(q^2)
&=
-\frac{M}{2M_i\mathcal N}
\left[
H_V^0\!\left(\tfrac12,\tfrac12\right)
+\frac{M_i-E_f}{|\bm P_f|}H_V^3\!\left(\tfrac12,\tfrac12\right)
+\left(\frac{1}{R}+\frac{M_i-E_f}{|\bm P_f|}\right)
H_V^1\!\left(-\tfrac12,\tfrac12\right)
\right],
\label{eq:ff_extract_f2_charm}
\\
f_1(q^2)
&=
-\frac{M_i+M_f}{M}f_2(q^2)
-\frac{H_V^1\!\left(-\tfrac12,\tfrac12\right)}{\mathcal N R},
\label{eq:ff_extract_f1_charm}
\\
f_3(q^2)
&=
\frac{M}{M_i-E_f}
\left[
\frac{H_V^0\!\left(\tfrac12,\tfrac12\right)}{\mathcal N}
-f_1(q^2)
+\frac{E_f-M_f}{M}f_2(q^2)
\right].
\label{eq:ff_extract_f3_charm}
\end{align}
\begin{align}
g_2(q^2)
&=
\frac{M}{2M_i\mathcal N}
\left[
\frac{1}{R}H_A^0\!\left(\tfrac12,\tfrac12\right)
+\frac{M_i-E_f}{R|\bm P_f|}H_A^3\!\left(\tfrac12,\tfrac12\right)
-\left(1+\frac{M_i-E_f}{R|\bm P_f|}\right)
H_A^1\!\left(-\tfrac12,\tfrac12\right)
\right],
\label{eq:ff_extract_g2_charm}
\\
g_1(q^2)
&=
\frac{H_A^1\!\left(-\tfrac12,\tfrac12\right)}{\mathcal N}
+\frac{M_i-M_f}{M}g_2(q^2),
\label{eq:ff_extract_g1_charm}
\\
g_3^{\mathrm{np}}(q^2)
&=
\frac{M}{R|\bm P_f|}
\left[
\frac{H_A^3\!\left(\tfrac12,\tfrac12\right)}{\mathcal N}
-g_1(q^2)
+\frac{M_i-E_f}{M}g_2(q^2)
\right].
\label{eq:ff_extract_g3_charm}
\end{align}
\end{widetext}
Here, the kinematic variables are $E_f=(M_i^2+M_f^2-q^2)/(2M_i)$ and $|\bm P_f|=\lambda^{1/2}(M_i^2,M_f^2,q^2)/(2M_i)$, supplemented by the dimensionless ratio $R=|\bm P_f|/(E_f+M_f)$ and the normalization factor $\mathcal N=\sqrt{2M_i(E_f+M_f)}$; the K\"all\'en function is $\lambda(a,b,c)=a^2+b^2+c^2-2ab-2ac-2bc$.
Eqs.~\eqref{eq:ff_extract_f2_charm}--\eqref{eq:ff_extract_g1_charm} explicitly give $f_1$, $f_2$, $f_3$, $g_1$, and $g_2$.
In contrast, the constituent overlap in Eq.~\eqref{eq:ff_extract_g3_charm} yields only the non-pole background $g_3^{\mathrm{np}}(q^2)$. 
The physical induced pseudoscalar form factor $g_3(q^2)$ is dominated by the pseudoscalar pole and thus approximated as~\cite{Guadagnoli:2006gj,Sasaki:2008ha}
\begin{equation}
g_3(q^2)\simeq -\frac{M(M_i+M_f)}{m_{\mathcal{P}}^2-q^2}g_1(q^2),
\label{eq:g3_reconstructed}
\end{equation}
with the pole mass $m_{\mathcal{P}}=m_{D_s}$ ($m_D$) for the $c\to s$ ($c\to d$) transition.

Recent LQCD studies of charmed baryon semileptonic decays report the vector helicity form factors $f_0$, $f_\perp$, and $f_+$, as well as the axial-vector helicity form factors $g_0$, $g_\perp$, $g_+$~\cite{Meinel:2016dqj,Meinel:2017ggx,Zhang:2021oja,Farrell:2025gis}.
To put our R3QM form factors in the same basis, the on-shell vector and axial-vector matrix elements are expressed with the covariant helicity parametrization of Refs.~\cite{Feldmann:2011xf,Detmold:2015aaa}.
This parametrization separates the timelike, transverse, and longitudinal virtual-$W^\ast$ structures and takes the form
\begin{widetext}
\begin{align}
H_V^\mu
&=
\bar u_f(P_f,\lambda_f)
\Bigg[
f_0(q^2)(M_i-M_f)\frac{q^\mu}{q^2}
+f_\perp(q^2)\left(
\gamma^\mu-\frac{2M_f}{s_+}P_i^\mu-\frac{2M_i}{s_+}P_f^\mu
\right)
\nonumber\\
&\hspace{2.0cm}
+f_+(q^2)\frac{M_i+M_f}{s_+}
\left(
P_i^\mu+P_f^\mu-(M_i^2-M_f^2)\frac{q^\mu}{q^2}
\right)
\Bigg]
u_i(P_i,\lambda_i),
\nonumber\\
H_A^\mu
&=
-\bar u_f(P_f,\lambda_f)\gamma^5
\Bigg[
g_0(q^2)(M_i+M_f)\frac{q^\mu}{q^2}
+g_\perp(q^2)\left(
\gamma^\mu+\frac{2M_f}{s_-}P_i^\mu-\frac{2M_i}{s_-}P_f^\mu
\right)
\nonumber\\
&\hspace{2.0cm}
+g_+(q^2)\frac{M_i-M_f}{s_-}
\left(
P_i^\mu+P_f^\mu-(M_i^2-M_f^2)\frac{q^\mu}{q^2}
\right)
\Bigg]
u_i(P_i,\lambda_i),
\label{eq:helicity_current_charm}
\end{align}
\end{widetext}
where $s_\pm\equiv (M_i\pm M_f)^2-q^2$. 
The helicity form factors are not new dynamical inputs in the present calculation. 
These helicity form factors are linear combinations of the $f_i(q^2)$ and $g_i(q^2)$ form factors in Eqs.~\eqref{eq:ff_vector_current} and \eqref{eq:ff_axial_current}. 
In the $f_i$--$g_i$ basis, $f_1$ and $g_1$ multiply the leading Dirac and axial-vector structures, $f_2$ and $g_2$ multiply the tensor structures, and $f_3$ and $g_3$ multiply the longitudinal $q^\mu$ structures. 
In the helicity basis, the same six coefficients are regrouped according to the virtual-$W^\ast$ polarization: $f_0$ and $g_0$ for the timelike part, $f_\perp$ and $g_\perp$ for the transverse part, and $f_+$ and $g_+$ for the longitudinal part. 
Matching the two covariant decompositions gives~\cite{Detmold:2015aaa,Detmold:2016pkz}
\begin{widetext}
\begin{align}
f_0(q^2)
&= f_1(q^2)+\frac{q^2}{M_i(M_i-M_f)}f_3(q^2),
&
f_\perp(q^2)
&= f_1(q^2)+\frac{M_i+M_f}{M_i}f_2(q^2),
&
f_+(q^2)
&= f_1(q^2)+\frac{q^2}{M_i(M_i+M_f)}f_2(q^2),
\\
g_0(q^2)
&= g_1(q^2)-\frac{q^2}{M_i(M_i+M_f)}g_3(q^2),
&
g_\perp(q^2)
&= g_1(q^2)-\frac{M_i-M_f}{M_i}g_2(q^2),
&
g_+(q^2)
&= g_1(q^2)-\frac{q^2}{M_i(M_i-M_f)}g_2(q^2).
\label{eq:helicity_invariant_map_charm}
\end{align}
\end{widetext}

These relations connect the invariant form factor basis in Eqs.~\eqref{eq:ff_vector_current} and \eqref{eq:ff_axial_current} to the helicity basis. 
In the invariant basis, $f_1$ and $g_1$ describe the leading vector and axial-vector weak vertices and appear in all helicity responses. 
The tensor form factors carry the current components associated with the quark spin flip.
The weak magnetism term $f_2$ contributes to the transverse vector response $f_\perp$ without a $q^2$ factor, while its contribution to the longitudinal response $f_+$ is proportional to $q^2$. 
The induced tensor term $g_2$ plays the same role in the axial-vector responses $g_\perp$ and $g_+$. 
The divergence form factors $f_3$ and $g_3$ are separated into the timelike responses $f_0$ and $g_0$. 
In particular, the pseudoscalar pole reconstructed in $g_3$ is carried by $g_0$, so the large-$q^2$ behavior of $g_0$ should be read together with the pole prescription above.
At $q^2=0$, the terms weighted by $q^2$ vanish. 
Then $f_0$ and $f_+$ reduce to $f_1$, and $g_0$ and $g_+$ reduce to $g_1$, while $f_\perp$ and $g_\perp$ still retain the tensor contributions.

\subsection{The experimental observation variables}

Substituting the hadronic matrix element $H^\nu$ into the decay amplitude in Eq.~\eqref{eq:Amplitude}, the differential decay width after averaging over the initial baryon spin and summing over the final polarizations is given by
\begin{align}
d\Gamma
&=
\frac{(2\pi)^4}{2M_i}\frac{1}{2J_i+1}
\sum_{\lambda_i,\lambda_f,s_\ell,s_\nu}
|\mathcal M|^2
\delta^{(4)}(P_i-P_f-p_\ell-p_\nu) \nonumber \\
&\quad \times \frac{d^3\bm P_f}{(2\pi)^3 2E_f}
\frac{d^3\bm p_\ell}{(2\pi)^3 2E_\ell}
\frac{d^3\bm p_\nu}{(2\pi)^3 2E_\nu}.
\label{eq:differential_width}
\end{align}
Integrating Eq.~\eqref{eq:differential_width} over the three-body phase space gives the partial width $\Gamma_\ell$.
The branching fraction is obtained by multiplying this width by the lifetime of the initial baryon, $\mathcal B_\ell=\Gamma_\ell\tau_{B_i}$, with $\tau_{B_i}$ taken from the RPP~\cite{ParticleDataGroup:2024cfk}.

The lepton forward-backward asymmetry is defined from the double differential rate in $q^2$ and $\cos\theta_\ell$.
The polar angle $\theta_\ell$ is defined in the $\ell^+\nu_\ell$ rest frame, with the reference axis along the direction of $\bm q$ fixed in the initial baryon rest frame.
The angle is taken between this axis and the $\ell^+$ momentum.
In the initial baryon rest frame $\bm q=-\bm P_f$, so the positive lepton axis is opposite to the final baryon momentum.
The lepton forward-backward asymmetry is~\cite{Kadeer:2005aq,BESIII:2023jxv}
\begin{equation}
A_{FB}^{\ell}(q^2)
=
\frac{
\displaystyle \left(\int_0^1-\int_{-1}^0\right)
d\cos\theta_\ell\,
\dfrac{d^2\Gamma}{dq^2\,d\cos\theta_\ell}
}{
\displaystyle \left(\int_0^1+\int_{-1}^0\right)
d\cos\theta_\ell\,
\dfrac{d^2\Gamma}{dq^2\,d\cos\theta_\ell}
}.
\label{eq:afb_charm}
\end{equation}
By subtracting the backward hemisphere from the forward hemisphere, the numerator in Eq.~\eqref{eq:afb_charm} extracts the angular terms antisymmetric under $\cos\theta_\ell\to-\cos\theta_\ell$.
In the helicity representation, this antisymmetric component is driven strictly by the parity-odd interference between the vector and axial-vector amplitudes.
The denominator then normalizes this extracted interference by the $q^2$ differential rate.

The longitudinal polarization of the final baryon is obtained from its helicity rates.
After the lepton angle is integrated over the full range, the decay rate is separated into the two final baryon helicities.
For the spin-$1/2$ final baryons considered here, $\uparrow$ and $\downarrow$ denote $\lambda_f=+1/2$ and $-1/2$, respectively.
The two helicity rates are
\begin{equation}
\frac{d\Gamma_{\uparrow,\downarrow}}{dq^2}
=
\int_{-1}^{1}
	\frac{d^2\Gamma_{\uparrow,\downarrow}}{dq^2\,d\cos\theta_\ell}\,
	d\cos\theta_\ell .
\end{equation}
The longitudinal polarization of the final baryon is defined as~\cite{Li:2021qod,Zhang:2023nxl}
\begin{equation}
\alpha_{B_f}(q^2)=
\frac{d\Gamma_\uparrow/dq^2-d\Gamma_\downarrow/dq^2}
{d\Gamma_\uparrow/dq^2+d\Gamma_\downarrow/dq^2}.
\label{eq:alpha_baryon_charm}
\end{equation}
The difference between the two helicity rates isolates the interference between the vector and axial-vector amplitudes.
Unlike the total decay rate, $\alpha_{B_f}$ is sensitive to the relative sign of these amplitudes.

\begin{figure*}[t]
	\centering
	\includegraphics[width=16.0cm]{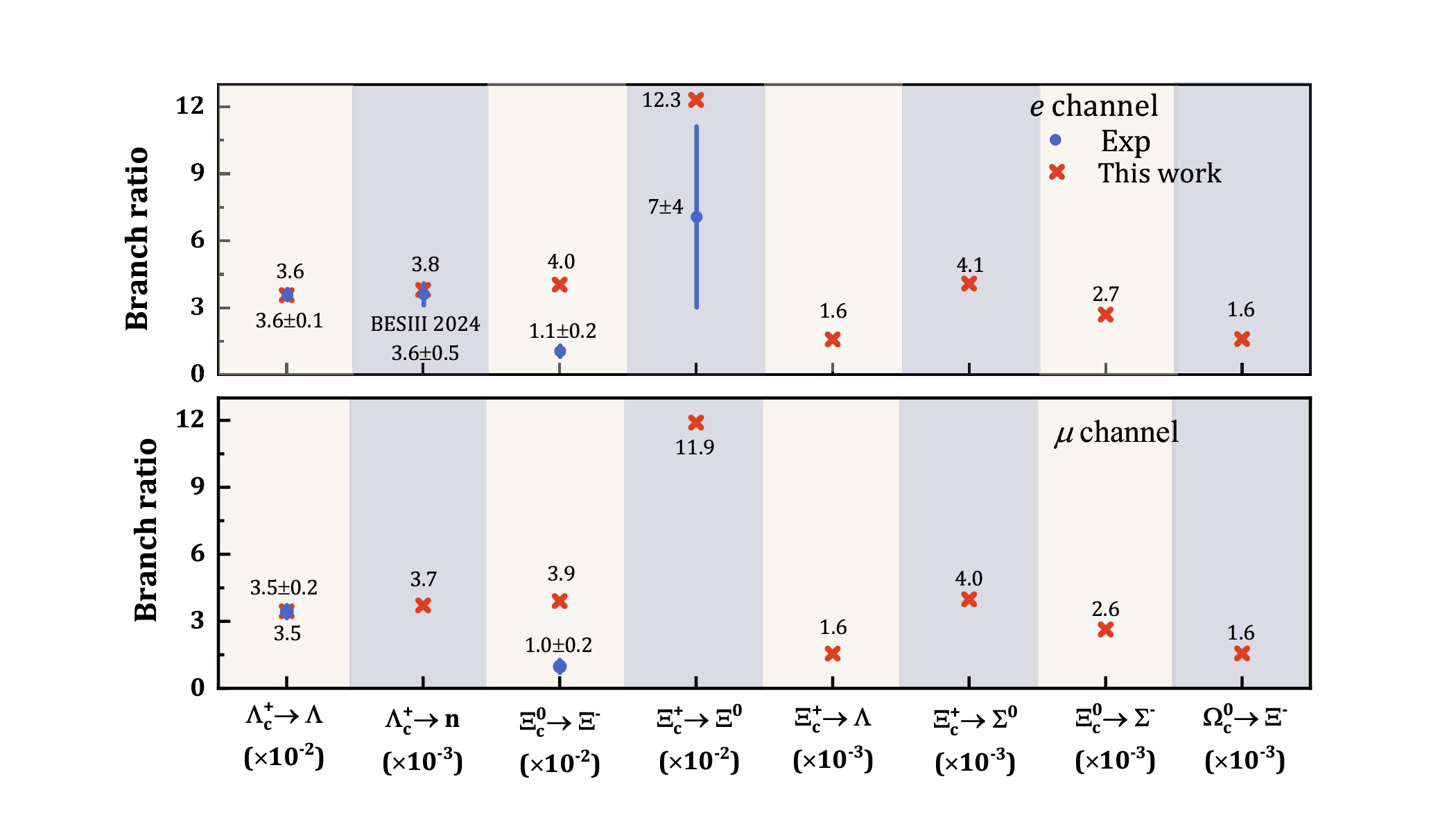}
	\vspace{-0.6 cm}
	\caption{
		Branching fraction overview for the singly charmed baryon semileptonic channels in the electron and muon modes.
		The red crosses give our R3QM results, and the blue points show the available experimental values~\cite{BESIII:2024mgg,ParticleDataGroup:2024cfk}.
	}\label{fig:Charm_Baryon_Ratio}
\end{figure*}

\subsection{Parameters and conventions}

For the numerical calculation, we take the physical baryon masses and lifetimes, lepton masses, pseudoscalar pole masses, $G_F$, and CKM matrix elements from the RPP~\cite{ParticleDataGroup:2024cfk}. 
The constituent quark masses and rest frame spatial wave functions are taken from our previous R3QM study of hyperon semileptonic decays~\cite{Ni:2026arb}. 
The quark masses are $m_{u/d}=368.5~\mathrm{MeV}$, $m_s=460.7~\mathrm{MeV}$, and $m_c=1550.8~\mathrm{MeV}$. 
No parameter in these wave functions or the single quark current is fitted to the present decay data.

In the following, the $q^2=0$ rows compare our $f_i(0)$ and $g_i(0)$ form factors with LQCD and other model predictions in the same convention as our hyperon semileptonic study~\cite{Ni:2026arb}. 
Before a literature value is listed in the tables, the baryon phase, the momentum transfer direction, the tensor sign, and the mass denominator of the subleading form factors are matched to Eqs.~\eqref{eq:ff_vector_current} and \eqref{eq:ff_axial_current}. 
In particular, $f_2$, $f_3$, $g_2$, and $g_3$ are rescaled when a reference uses $(M_i+M_f)$ instead of $M_i$ in the denominator, and $f_2$ and $g_2$ are also converted when the tensor term has the opposite sign. 
The literature values of $g_3$ are only converted between conventions. 
The pseudoscalar pole prescription of Eq.~\eqref{eq:g3_reconstructed} is applied only to our R3QM result. 
Further details regarding the matching of these form factor conventions can be found in our previous work~\cite{Ni:2026arb}.

For the angular observables, the sign of $A_{FB}^{\ell}$ is fixed by Eq.~\eqref{eq:afb_charm}. 
By definition, the polar axis for the lepton angle $\theta_\ell$ is chosen along the direction of $\bm q$ in the initial baryon rest frame, which is antiparallel to the final baryon momentum. 
Consequently, an asymmetry $A_{FB}^{\ell}$ reported with the final baryon direction as the polar axis must be multiplied by $-1$ to match our convention. 
In contrast, the longitudinal polarization $\alpha_{B_f}$ of the final baryon is an intrinsic helicity observable; its sign is determined strictly by the final baryon helicity rates and remains invariant under the choice of the lepton reference axis.

The lepton flavor ratio is $R_{\mu e}=\Gamma_\mu/\Gamma_e$.
When the tables list an overall value of $A_{FB}^{\ell}$ or $\alpha_{B_f}$, the function of $q^2$ is averaged over the physical range with weight $d\Gamma_\ell/dq^2$.

\section{Results and Discussions}
\label{sec:results}

\begin{figure*}[t]
	\centering
	\includegraphics[width=15.0cm]{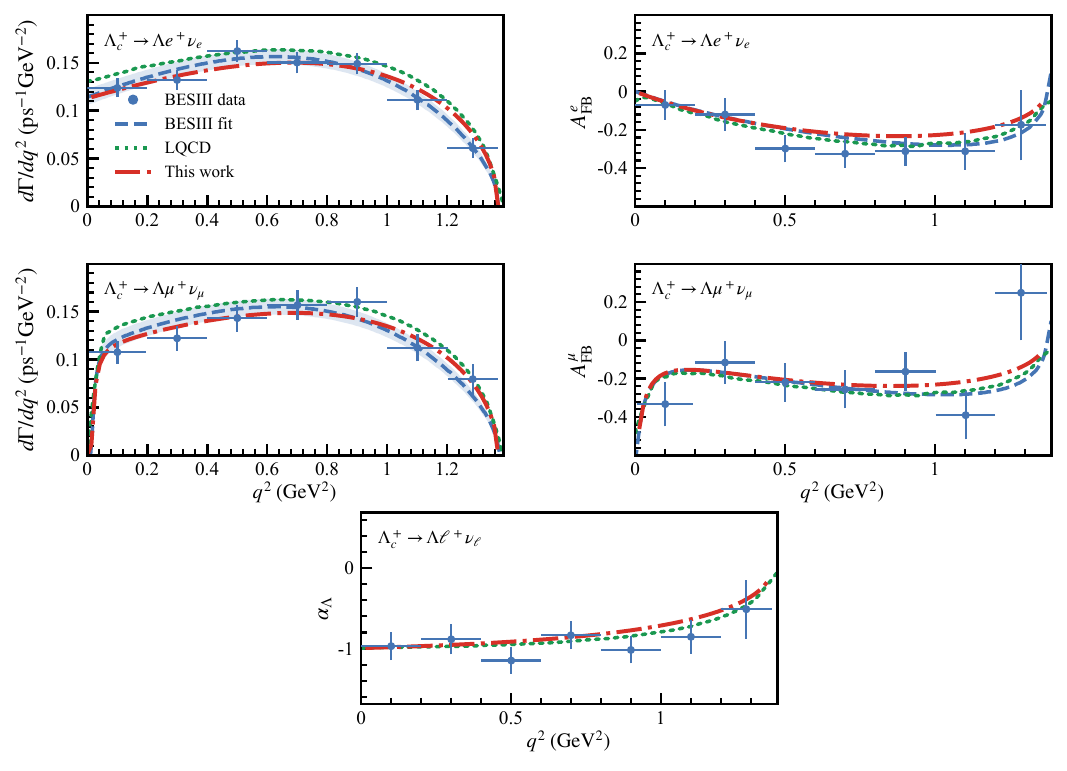}
	\vspace{-0.2 cm}
	\caption{
		Differential decay rates and angular observables for $\Lambda_c^+\to\Lambda\ell^+\nu_\ell$.
		The lower central panel gives the $q^2$ dependence of the longitudinal polarization $\alpha_{\Lambda}$ of the final $\Lambda$.
		The $\alpha_{\Lambda}$ curve is shown as a rate-weighted average over the kinematically allowed lepton modes.
		The red dash-dotted and green dotted curves represent the results from our R3QM and the LQCD calculations~\cite{Meinel:2016dqj}, which are compared with the BESIII Collaboration measurements~\cite{BESIII:2023jxv} (blue points) and their corresponding fits (blue dashed curves with error bands).
	}\label{Lambda_c_decay}
\end{figure*}

\begin{figure*}[t]
	\centering
	\includegraphics[width=15.0cm]{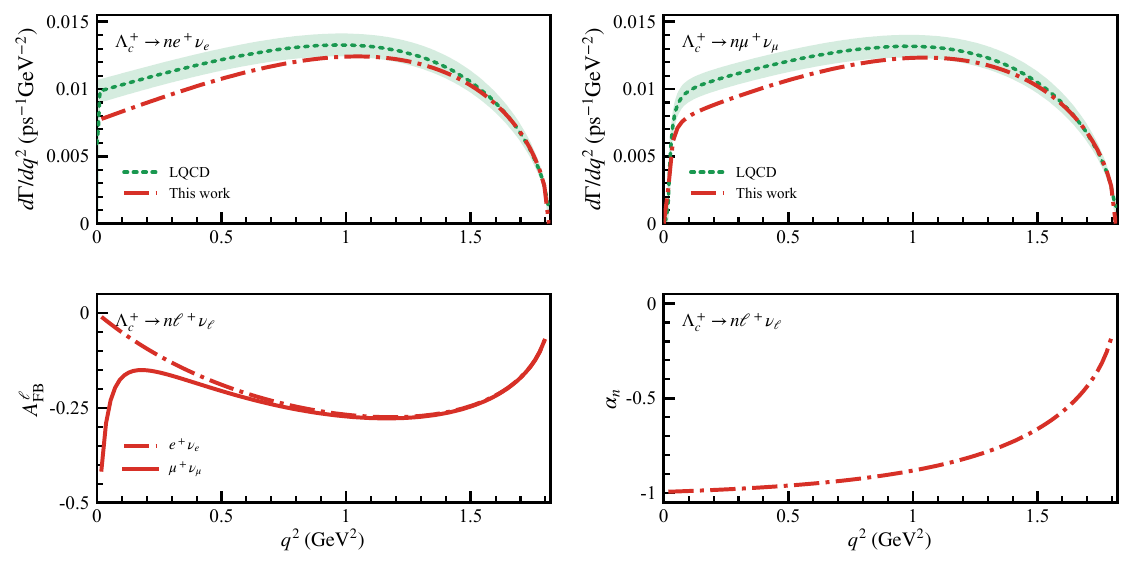}
	\vspace{-0.6 cm}
	\caption{
		Differential decay rates and angular observables for the semileptonic transition $\Lambda_c^+\to n\ell^+\nu_\ell$ as functions of $q^2$.
		Green dotted curves with bands are the LQCD results of Ref.~\cite{Meinel:2017ggx} in the two upper panels.
		These LQCD results are given for $(d\Gamma/dq^2)/|V_{cd}|^2$ and are multiplied here by $|V_{cd}|^2$ with $|V_{cd}|=0.22503$~\cite{ParticleDataGroup:2024cfk}.
		Red dash-dotted curves are our R3QM results, except that the muon curve in the $A_{\rm FB}^{\ell}$ panel is shown as a red solid line.
		The $\alpha_n$ curve is the electron and muon average with $d\Gamma/dq^2$ as the averaging factor.
	}\label{fig:Lambdac_n_observables}
\end{figure*}

\begin{figure*}[t]
	\centering
	\includegraphics[width=15.0cm]{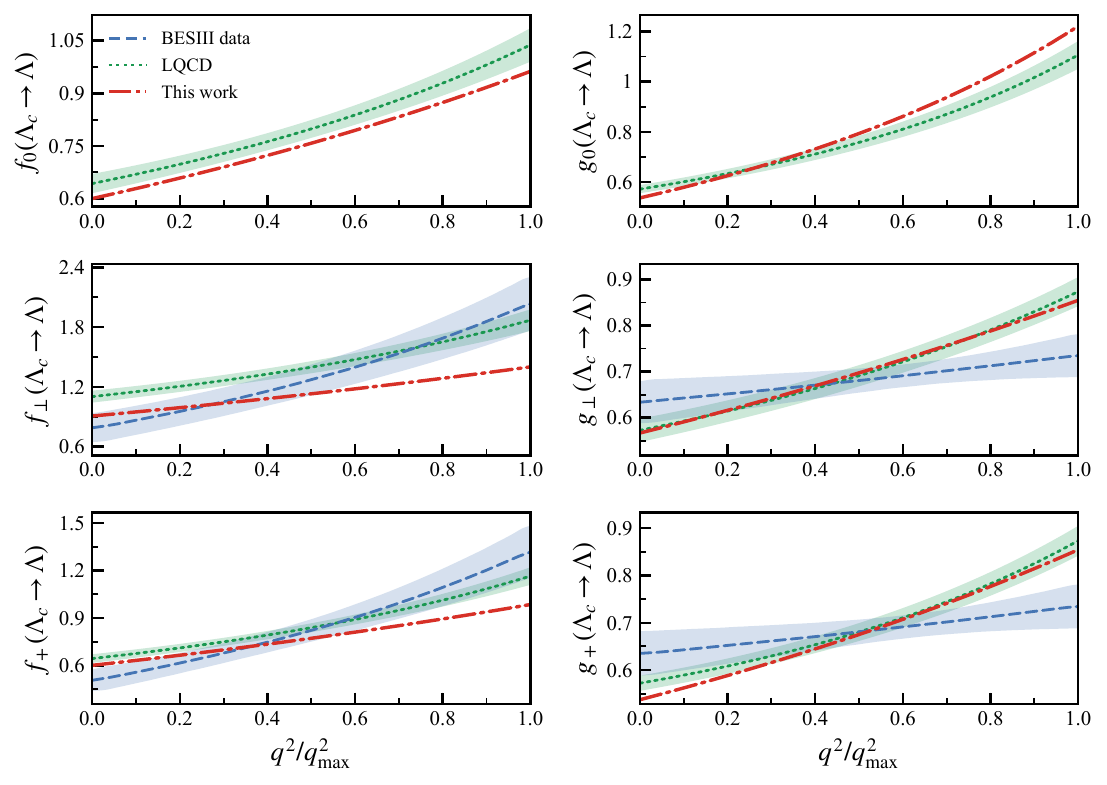}
	\vspace{-0.2 cm}
	\caption{
		Helicity basis form factors for $\Lambda_c^+\to\Lambda\ell^+\nu_\ell$ as functions of $q^2/q^2_{\rm max}$.
		Blue dashed curves with bands are the form factors extracted by the BESIII Collaboration~\cite{BESIII:2023jxv}, green dotted curves with bands are the LQCD result of Ref.~\cite{Meinel:2016dqj}, and red dash-dotted curves are our R3QM results.
	}\label{fig:Lambdac_Lambda_form_factor}
\end{figure*}

\begin{figure*}[t]
	\centering
	\includegraphics[width=15.0cm]{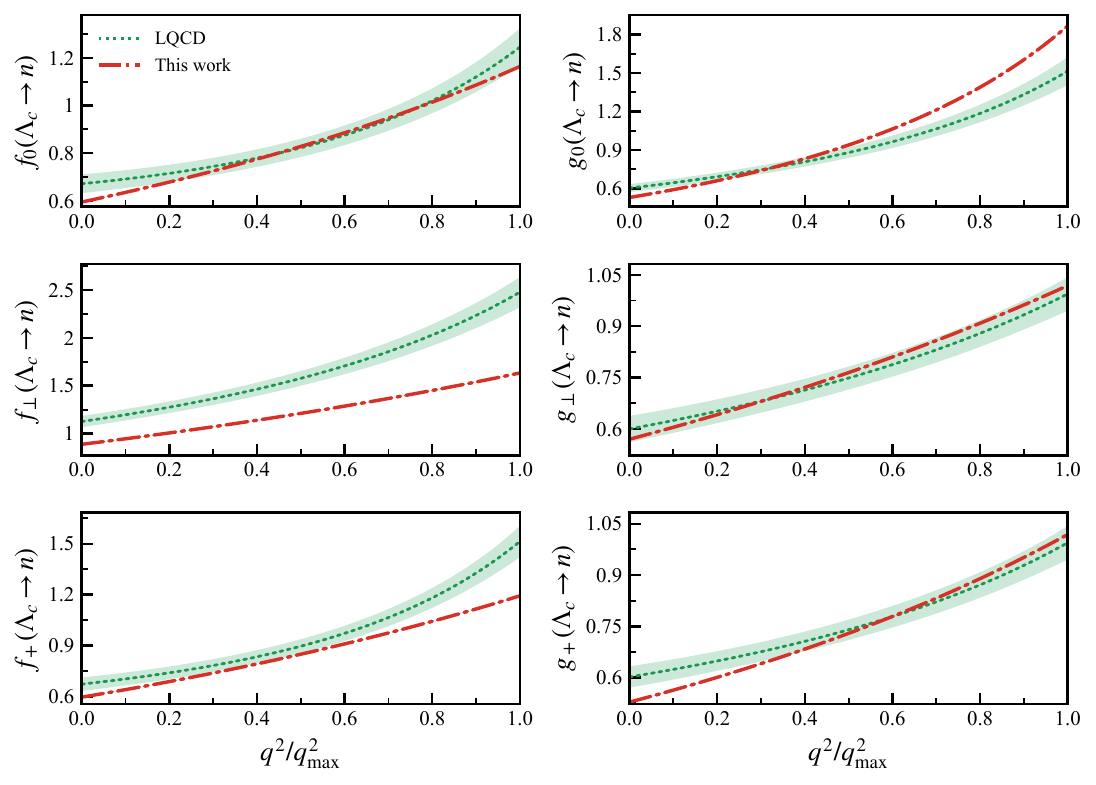}
	\vspace{-0.2 cm}
	\caption{
		Helicity basis form factors for $\Lambda_c^+\to n\ell^+\nu_\ell$ as functions of $q^2/q^2_{\rm max}$.
		Green dotted curves with bands are the LQCD result of Ref.~\cite{Meinel:2017ggx}, and red dash-dotted curves are our R3QM results.
	}\label{fig:Lambdac_n_form_factor}
\end{figure*}

\begin{figure*}[t]
\centering
\includegraphics[width=15.0cm]{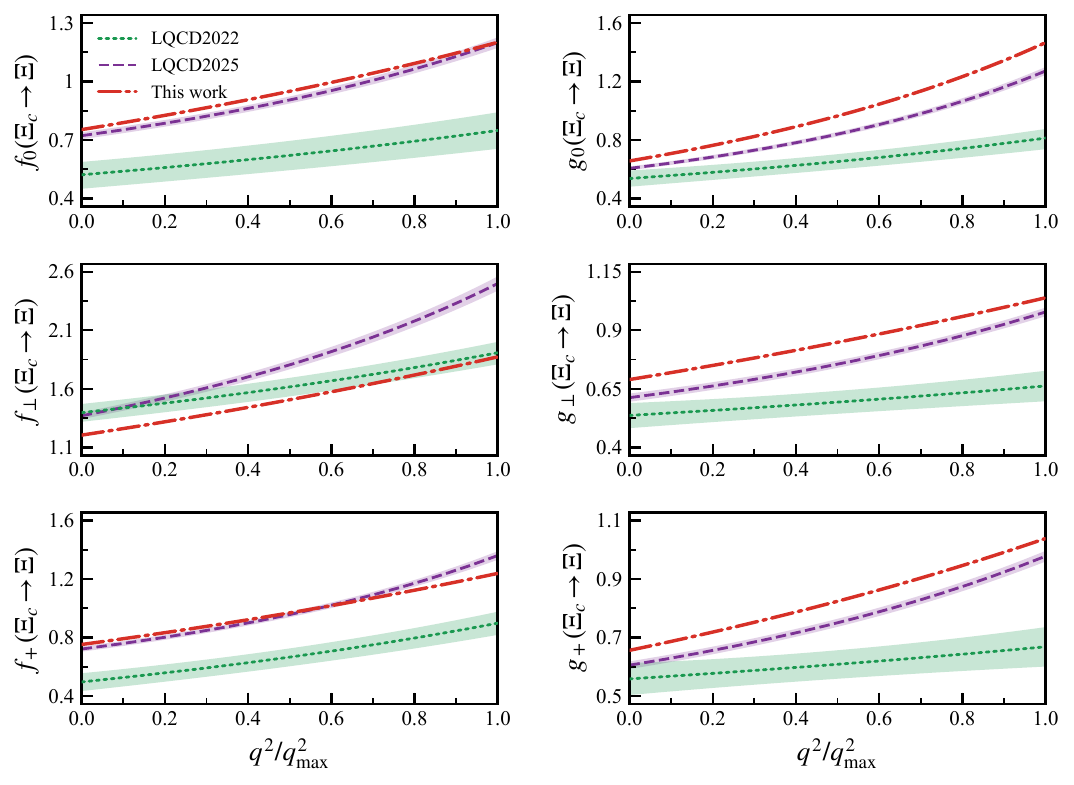}\vspace{-0.2 cm}
\caption{
Helicity basis form factors for the $\Xi_c\to\Xi$ semileptonic transitions as functions of $q^2/q^2_{\max}$.
The green dotted curves with bands give the continuum-extrapolated LQCD result of Ref.~\cite{Zhang:2021oja}, and the purple dashed curves with bands give the domain-wall LQCD result of Ref.~\cite{Farrell:2025gis}.
The red dash-dotted curve gives our R3QM result for $\Xi_c^0\to\Xi^-$.
The corresponding result for $\Xi_c^+\to\Xi^0$ is not drawn separately because the form factors of the two charge channels in the present calculation are nearly coincident on this scale.
}\label{fig:Xic_Xi_form_factor}
\end{figure*}

\begin{figure*}[t]
	\centering
	\includegraphics[width=15cm]{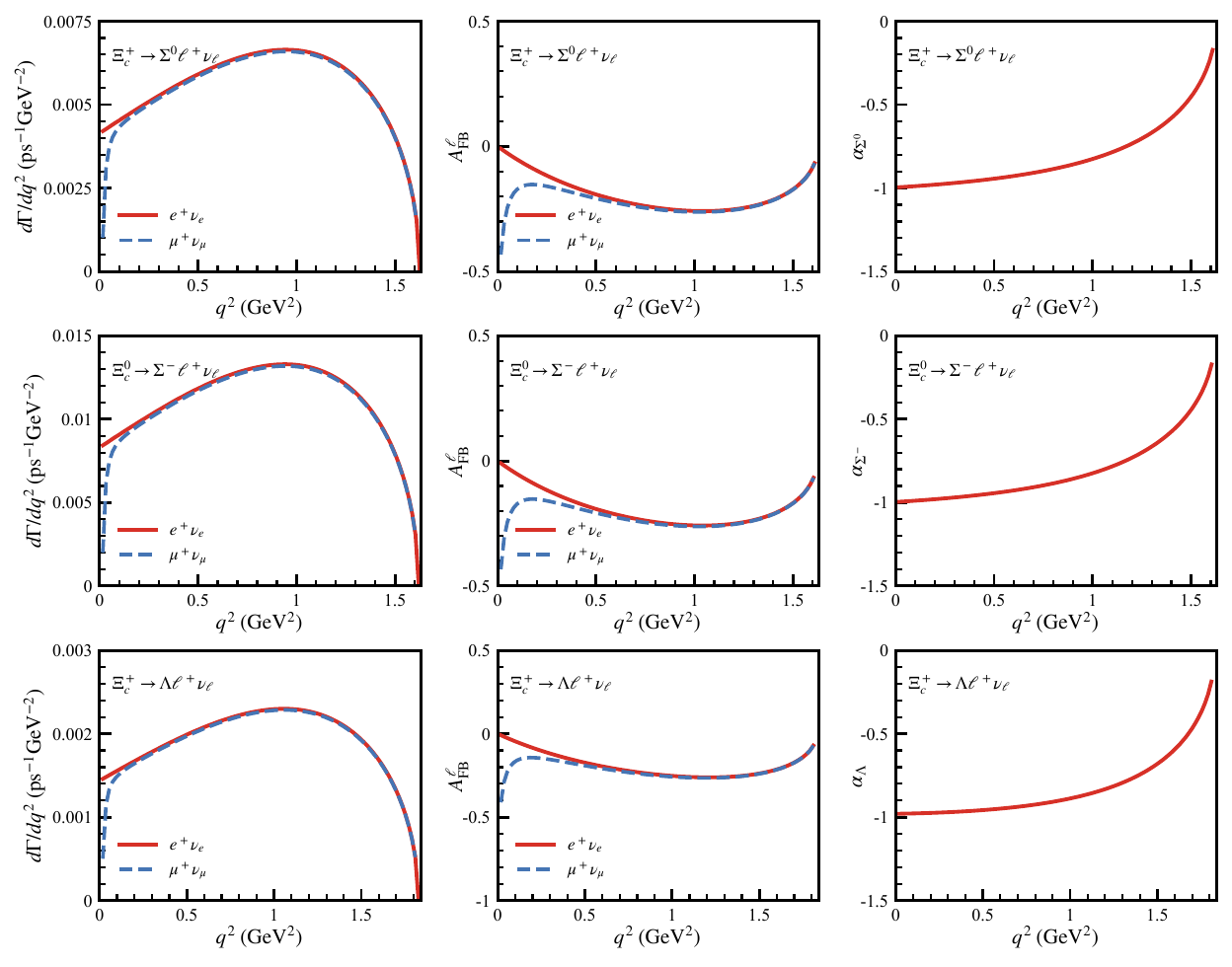}
	\vspace{-0.2 cm}
	\caption{
		Differential decay rates and angular observables for the $\Xi_c$ semileptonic transitions as functions of $q^2$.
		The electron and muon modes are plotted separately for $d\Gamma/dq^2$ and $A_{\rm FB}^{\ell}$, while the polarization $\alpha_{B_f}$ is given as the rate-weighted average over the kinematically allowed lepton modes.
	}\label{fig:Xic_Sigma_Lambda_observables}
\end{figure*}

\begin{figure*}[t]
	\centering
	\includegraphics[width=15cm]{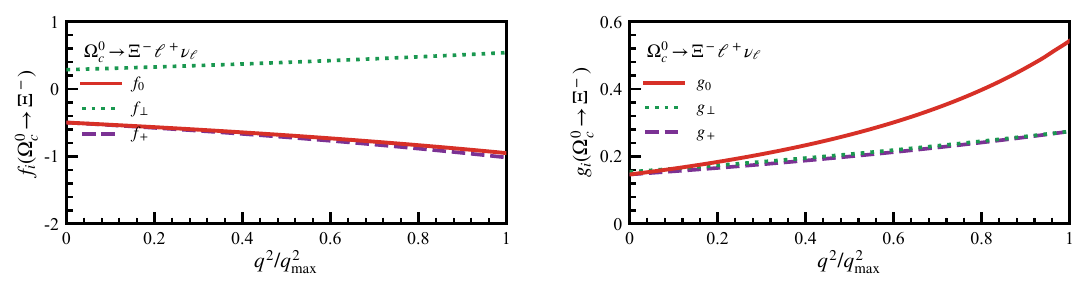}
	\vspace{-0.2 cm}
	\caption{
		Helicity basis form factors for the Cabibbo-suppressed transition $\Omega_c^0\to\Xi^-\ell^+\nu_\ell$ as functions of $q^2/q^2_{\rm max}$.
	}\label{fig:Omegac_Xi_minus_form_factor}
\end{figure*}

In this section, we systematically analyze the singly charmed baryon semileptonic decays and discuss the corresponding numerical results. 
We summarize the branching fractions of all the relevant channels in Fig.~\ref{fig:Charm_Baryon_Ratio}. 
In Tables~\ref{tab:Heavy_Baryon_Decays_Lambdac}--\ref{tab:Heavy_Baryon_Decays_Omegac_Xi}, we list the detailed branching fractions and angular observables alongside the $f_i(0)$ and $g_i(0)$ form factors. 
To illustrate the $q^2$-dependent dynamics, we plot the differential decay rates, the angular observables, and the helicity form factors. 
Specifically, these kinematic distributions are displayed in Figs.~\ref{Lambda_c_decay}, \ref{fig:Lambdac_Lambda_form_factor}, \ref{fig:Lambdac_n_observables}, and \ref{fig:Lambdac_n_form_factor} for the $\Lambda_c^+\to\Lambda\ell^+\nu_\ell$ and $\Lambda_c^+\to n\ell^+\nu_\ell$ transitions; in Fig.~\ref{fig:Xic_Xi_form_factor} for the Cabibbo-favored $\Xi_c^0\to\Xi^-\ell^+\nu_\ell$ and $\Xi_c^+\to\Xi^0\ell^+\nu_\ell$ transitions; and in Figs.~\ref{fig:Xic_Sigma_Lambda_form_factor}, \ref{fig:Xic_Sigma_Lambda_observables}, \ref{fig:Omegac_Xi_minus_form_factor}, and \ref{fig:Omegac_Xi_minus_observables} for the Cabibbo-suppressed $\Xi_c^+\to\Lambda\ell^+\nu_\ell$, $\Xi_c^+\to\Sigma^0\ell^+\nu_\ell$, $\Xi_c^0\to\Sigma^-\ell^+\nu_\ell$, and $\Omega_c^0\to\Xi^-\ell^+\nu_\ell$ transitions.

\begin{table*}[t]
	\centering
	\setlength{\tabcolsep}{0pt}
	\renewcommand{\arraystretch}{1.3}
	\caption{Branching fractions, $R_{\mu e}$, averaged angular observables, and the $f_i(0)$ and $g_i(0)$ form factors for $\Lambda_c^+\to\Lambda\ell^+\nu_\ell$. The model abbreviations are CCQM (covariant confined quark model), RQM (relativistic quark model), LFQM (light-front quark model), HBM (homogeneous bag model), MBM (MIT bag model), and LFCQM (light-front constituent quark model).}
	\label{tab:Heavy_Baryon_Decays_Lambdac}
	\begin{tabular*}{\textwidth}{@{\extracolsep{\fill}}cccccccc}
		\toprule \toprule
		\multirow{2}{*}{Method} & \multicolumn{2}{c}{$\mathcal{B}\;(10^{-2})$} & \multirow{2}{*}{$R_{\mu e}$} & \multicolumn{2}{c}{$\langle A_{FB} \rangle$} & \multicolumn{2}{c}{$\langle \alpha_{\Lambda} \rangle$} \\
		\cmidrule(lr){2-3} \cmidrule(lr){5-6} \cmidrule(lr){7-8}
		& $\ell = e$ & $\ell = \mu$ & & $\ell = e$ & $\ell = \mu$ & $\ell = e$ & $\ell = \mu$ \\
		\midrule
		CCQM~\cite{Gutsche:2015rrt}   & $2.78$          & $2.69$          & $0.968$ & $-0.21$          & $-0.24$          & $-0.87$          & $-0.87$           \\
		RQM~\cite{Faustov:2016yza}    & $3.25$          & $3.14$          & $0.966$ & $-0.209$         & $-0.242$         & $-0.86$          & $-0.86$           \\
		LFQM~\cite{Li:2021qod}        & $4.04(75)$      & $3.90(73)$      & $0.965$ & $-0.20(5)$       & $-0.16(4)$       & $-0.87(9)$       & $-0.87(9)$          \\
		HBM~\cite{Geng:2022fsr}       & $3.78(25)$      & $3.67(23)$      & $0.971$ & $-0.176(5)$      & $-0.143(6)$      & $-0.826$         & $-0.823$           \\
		QCDSR~\cite{Zhang:2023nxl}    & $3.49(65)$      & $3.37(54)$      & $0.966$ & $-0.20(1)$       & $-0.24(1)$       & $-0.90(3)$       & $-0.90(2)$ \\
		LQCD~\cite{Meinel:2016dqj}    & $3.80(22)$      & $3.69(22)$      & $0.969$ & ---              & ---              & ---              & --- \\
		RPP~\cite{ParticleDataGroup:2024cfk}
		                                  & $3.56(13)$      & $3.48(17)$      & $0.98(6)$ & ---              & ---              & \multicolumn{2}{c}{$-0.875(33)$} \\
		\textbf{This work}                 & $\mathbf{3.56}$ & $\mathbf{3.45}$ & $\mathbf{0.969}$ & $\mathbf{-0.175}$ & $\mathbf{-0.204}$ & $\mathbf{-0.822}$ & $\mathbf{-0.819}$ \\
	\end{tabular*}
	\vspace{1.0mm}
	\renewcommand{\arraystretch}{1.22}
	\begin{tabular*}{\textwidth}{@{\extracolsep{\fill}}ccccccc}
		\toprule
		Method & $f_1(0)$ & $f_2(0)$ & $f_3(0)$ & $g_1(0)$ & $g_2(0)$ & $g_3(0)$ \\
		\midrule
		LCSR~\cite{Liu:2009sn}
		& $0.665$ & $0.285$ & --- & $0.665$ & $-0.285$ & --- \\
		CCQM~\cite{Gutsche:2015rrt}
		& $0.511$ & $0.289$ & $-0.014$ & $0.466$ & $-0.025$ & $-0.400$ \\
		RQM~\cite{Faustov:2016yza}
		& $0.700$ & $0.295$ & $0.222$ & $0.448$ & $-0.135$ & $-0.832$ \\
		LFQM~\cite{Zhao:2018zcb}
		& $0.468$ & $0.222$ & --- & $0.407$ & $0.035$ & --- \\
		MBM~\cite{Geng:2020fng}
		& $0.54$ & $-0.22$ & --- & $0.52$ & $0.06$ & --- \\
		LFCQM~\cite{Geng:2020gjh}
		& $0.67(1)$ & $0.76(2)$ & --- & $0.59(1)$ & $3.8(12)\times10^{-3}$ & --- \\
		LFQM~\cite{Li:2021qod}
		& $0.71(7)$ & $0.36(4)$ & $-0.29(3)$
		& $0.62(6)$ & $0.11(1)$ & $-0.60(6)$ \\
		HBM~\cite{Geng:2022fsr}
		& $0.604(39)$ & $0.209(5)$ & --- & $0.566(34)$ & $0.012(2)$ & --- \\
		LQCD~\cite{Meinel:2016dqj}
		& $0.643(23)$ & $0.308(36)$ & ---
		& $0.572(15)$ & $-0.001(45)$ & --- \\
		LQCD~\cite{Bahtiyar:2021voz}
		& $0.687(138)$ & $0.327(79)$ & $0.110(54)$
		& $0.539(101)$ & $-0.261(67)$ & $-0.241(190)$ \\
		\textbf{This work}
		& $\mathbf{0.600}$ & $\mathbf{0.206}$ & $\mathbf{0.019}$ & $\mathbf{0.537}$ & $\mathbf{-0.058}$ & $\mathbf{-1.079}$ \\
		\bottomrule\bottomrule
	\end{tabular*}
\end{table*}

\begin{table*}[t]
	\centering
	\setlength{\tabcolsep}{0pt}
	\renewcommand{\arraystretch}{1.3}
	\caption{Branching fractions, angular observables, and the $f_i(0)$ and $g_i(0)$ form factors for $\Lambda_c^+\to n\ell^+\nu_\ell$.
		For the constituent quark model (CQM) row, the decay rate of Ref.~\cite{Pervin:2005ve} is converted to a branching fraction using the value $\tau_{\Lambda_c^+}=0.2026~{\rm ps}$ from RPP~\cite{ParticleDataGroup:2024cfk}.}
	\label{tab:Heavy_Baryon_Decays_Lambdac_n}
	\begin{tabular*}{\textwidth}{@{\extracolsep{\fill}}cccccccc}
		\toprule \toprule
		\multirow{2}{*}{Method} & \multicolumn{2}{c}{$\mathcal{B}\;(10^{-3})$} & \multirow{2}{*}{$R_{\mu e}$} & \multicolumn{2}{c}{$\langle A_{FB} \rangle$} & \multicolumn{2}{c}{$\langle \alpha_n \rangle$} \\
		\cmidrule(lr){2-3} \cmidrule(lr){5-6} \cmidrule(lr){7-8}
		& $\ell = e$ & $\ell = \mu$ & & $\ell = e$ & $\ell = \mu$ & $\ell = e$ & $\ell = \mu$ \\
		\midrule
		CQM~\cite{Pervin:2005ve}      & $2.74$          & ---             & ---     & ---              & ---              & ---              & --- \\
		CCQM~\cite{Gutsche:2014zna}   & $2.07$          & $2.02$          & $0.976$ & $-0.236$         & $-0.260$         & ---              & --- \\
		SU(3)~\cite{Lu:2016ogy}       & $2.93(34)$      & ---             & ---     & ---              & ---              & ---              & --- \\
		RQM~\cite{Faustov:2016yza}    & $2.68$          & $2.62$          & $0.978$ & $-0.251$         & $-0.276$         & $-0.91$          & $-0.90$ \\
		LFQM~\cite{Zhao:2018zcb}      & $2.01$          & ---             & ---     & ---              & ---              & ---              & --- \\
		SU(3)~\cite{Geng:2019bfz}     & $5.1(4)$        & ---             & ---     & ---              & ---              & $-0.89(4)$       & --- \\
		MBM~\cite{Geng:2020fng}       & $2.79$          & $2.73$          & $0.978$ & ---              & ---              & $-0.87$          & $-0.87$ \\
		LFCQM~\cite{Geng:2020fng}     & $3.6(15)$       & $3.4(14)$       & $0.944$ & ---              & ---              & $-0.96(4)$       & $-0.96(4)$ \\
		QCDSR~\cite{Zhang:2023nxl}    & $2.81(56)$      & $2.75(55)$      & $0.979$ & $-0.23(1)$       & $-0.25(2)$       & $-0.93(3)$       & $-0.93(3)$ \\
		LQCD~\cite{Meinel:2017ggx}    & $4.10(26)$      & $4.00(26)$      & $0.977$ & ---              & ---              & ---              & --- \\
		BESIII~\cite{BESIII:2024mgg}  & $3.57(34)(14)$  & ---             & ---     & ---              & ---              & ---              & --- \\
		\textbf{This work}                 & $\mathbf{3.79}$ & $\mathbf{3.71}$ & $\mathbf{0.979}$ & $\mathbf{-0.211}$ & $\mathbf{-0.232}$ & $\mathbf{-0.847}$ & $\mathbf{-0.845}$ \\
	\end{tabular*}
	\vspace{1.0mm}
	\renewcommand{\arraystretch}{1.22}
	\begin{tabular*}{\textwidth}{@{\extracolsep{\fill}}ccccccc}
		\toprule
		Method & $f_1(0)$ & $f_2(0)$ & $f_3(0)$ & $g_1(0)$ & $g_2(0)$ & $g_3(0)$ \\
		\midrule
		CCQM~\cite{Gutsche:2014zna}
		& $0.470$ & $0.246$ & $0.039$ & $0.414$ & $-0.073$ & $-0.328$ \\
		RQM~\cite{Faustov:2016yza}
		& $0.627$ & $0.259$ & $0.179$ & $0.433$ & $-0.118$ & $-0.744$ \\
		LFQM~\cite{Zhao:2018zcb}
		& $0.513$ & $0.266$ & --- & $0.443$ & $0.034$ & --- \\
		MBM~\cite{Geng:2020fng}
		& $0.40$ & $-0.22$ & --- & $0.43$ & $0.07$ & --- \\
		LFCQM~\cite{Geng:2020gjh}
		& $0.83(1)$ & $1.05(2)$ & --- & $0.71(1)$ & $0.27(1)$ & --- \\
		HBM~\cite{Geng:2022fsr}
		& $0.570(56)$ & $0.210(1)$ & --- & $0.526(50)$ & $0.015(2)$ & --- \\
		QCDSR~\cite{Zhang:2023nxl}
		& $0.53(4)$ & $0.25(3)$ & --- & $0.53(4)$ & $0.25(3)$ & --- \\
		LQCD~\cite{Meinel:2017ggx}
		& $0.672(39)$ & $0.321(38)$ & --- & $0.602(31)$ & $0.003(52)$ & --- \\
		\textbf{This work}
		& $\mathbf{0.595}$ & $\mathbf{0.207}$ & $\mathbf{0.021}$ & $\mathbf{0.529}$ & $\mathbf{-0.070}$ & $\mathbf{-1.115}$ \\
		\bottomrule\bottomrule
	\end{tabular*}
\end{table*}

\subsection{$\Lambda_c^+$ baryon}

Among charmed baryon semileptonic decays, $\Lambda_c^+\to\Lambda\ell^+\nu_\ell$ has the most complete experimental information. 
Experimentally, the form factor ratio~\cite{CLEO:2004txf}, the absolute branching fractions, the differential distributions~\cite{BESIII:2022ysa}, the lepton flavor universality ratio, and the angular asymmetries~\cite{BESIII:2023jxv} have been sequentially measured in both the electron and muon modes. 
Theoretically, the $c\to s$ form factors have been calculated in LQCD~\cite{Meinel:2016dqj,Bahtiyar:2021voz}. 
For the Cabibbo-suppressed partner $\Lambda_c^+\to n\ell^+\nu_\ell$, the electron mode branching fraction was recently measured by the BESIII Collaboration~\cite{BESIII:2024mgg}. 
The corresponding differential rates and form factors have also been investigated in LQCD~\cite{Meinel:2017ggx}.

For $\Lambda_c^+\to\Lambda\ell^+\nu_\ell$, we obtain the branch ratios $\mathcal B_e=3.56\times10^{-2}$ and $\mathcal B_\mu=3.45\times10^{-2}$, consistent with the RPP averages~\cite{ParticleDataGroup:2024cfk} and the LQCD result of Ref.~\cite{Meinel:2016dqj}. 
For $\Lambda_c^+\to n\ell^+\nu_\ell$, we obtain $\mathcal B_e=3.79\times10^{-3}$ and $\mathcal B_\mu=3.71\times10^{-3}$; the electron mode is compatible with the measurement by the BESIII Collaboration~\cite{BESIII:2024mgg}, and both lepton modes agree with the LQCD predictions of Ref.~\cite{Meinel:2017ggx}. 
These neutron branching fractions are roughly one tenth of the Cabibbo-favored $\Lambda_c^+\to\Lambda\ell^+\nu_\ell$ branching fractions. 
This relative rate is approximately twice the pure CKM penalty $|V_{cd}/V_{cs}|^2 \simeq 0.05$, reflecting a substantial phase space compensation driven by the lighter neutron mass. 
This kinematic compensation is best visualized in the differential decay rates $d\Gamma/dq^2$. 
As shown in Fig.~\ref{Lambda_c_decay}, the differential width shapes of $\Lambda_c^+\to\Lambda\ell^+\nu_\ell$ calculated from our model is rather close to both the measured data and the LQCD results. 
The neutron differential rate similarly matches the LQCD results as shown in Fig.~\ref{fig:Lambdac_n_observables}, although its larger recoil manifests as a slight suppression relative to the lattice central value in the lower-$q^2$ region.

For the $(e,\mu)$ modes, we obtain the average asymmetries $\langle A_{FB}^{\ell}\rangle=(-0.175,-0.204)$ in $\Lambda_c^+\to\Lambda\ell^+\nu_\ell$, and notably larger magnitudes $(-0.211,-0.232)$ in $\Lambda_c^+\to n\ell^+\nu_\ell$. 
As shown in Figs.~\ref{Lambda_c_decay} and \ref{fig:Lambdac_n_observables}, the longitudinal-transverse helicity interference vanishes at zero recoil and grows only at finite recoil momenta. 
The extended kinematic range of the neutron mode therefore provides a wider active integration region, driving the magnitude enhancement. 
This kinematic enhancement is also found in the longitudinal polarization $\alpha_B$ of the final baryon, which shifts from $\langle\alpha_{\Lambda}\rangle_e=-0.822$ to $\langle\alpha_n\rangle_e=-0.847$. 
In the Cabibbo-favored $\Lambda_c^+\to\Lambda$ channel, the self-analyzing $\Lambda\to p\pi^-$ decay directly accesses this spin information. 
The neutron channel does not offer the same experimental handle, since direct measurements of the final baryon polarization are experimentally difficult. 
Therefore, the lepton distribution $A_{FB}^{\ell}$ serves as the primary practical probe of the underlying helicity interference. 
In this context, our calculated asymmetry provides a theoretical reference for future experimental tests.

Since the weak current acts on the charm quark, the scalar $ud$ spectator pair keeps its spin-$0$ configuration. 
As detailed in Appendix~\ref{app:static_form_factors}, this spin structure gives the static SU(6) relation $g_1^{\text{stat.}} = f_1^{\text{stat.}}$. 
At maximum recoil ($q^2=0$), we obtain $f_1(0)=0.600~(0.595)$ and $g_1(0)=0.537~(0.529)$ for the $\Lambda_c^+\to\Lambda$ and $\Lambda_c^+\to n$ transitions, respectively. 
The ratio $g_1(0)/f_1(0) \simeq 0.9$ therefore stays close to the static SU(6) pattern. 
The remaining difference between $g_1$ and $f_1$ is mainly generated by the active $c\to s$ or $c\to d$ quark line. 
At large recoil, the lower components of the active quark spinor and its Wigner rotation suppress the axial-vector contribution.
The similar values obtained in the two channels further show that the leading form factors are not very sensitive to the replacement of the final $\Lambda$ by the neutron. 
With these comparable hadronic matrix elements, the smaller $\Lambda_c^+\to n\ell^+\nu_\ell$ rate mainly comes from the CKM suppression and the phase space difference.

At maximum recoil, the calculated $f_1(0)$ is close to the LQCD results for both $\Lambda_c^+\to\Lambda$~\cite{Meinel:2016dqj} and $\Lambda_c^+\to n$~\cite{Meinel:2017ggx}, as listed in Tables~\ref{tab:Heavy_Baryon_Decays_Lambdac} and \ref{tab:Heavy_Baryon_Decays_Lambdac_n}.
In the helicity basis, $f_0$ and $f_+$ are also in good agreement with the LQCD results, whereas $f_\perp$ lies systematically below the corresponding LQCD values, as shown in Figs.~\ref{fig:Lambdac_Lambda_form_factor} and \ref{fig:Lambdac_n_form_factor}.
According to Eq.~\eqref{eq:helicity_invariant_map_charm}, $f_2$ contributes to $f_\perp$ with the coefficient $(M_i+M_f)/M_i$, while its contribution to $f_+$ is proportional to $q^2/[M_i(M_i+M_f)]$ and $f_0$ does not contain $f_2$.
Since $f_2$ is positive in both channels, its smaller value in the present calculation mainly lowers $f_\perp$, with a much weaker effect on $f_+$.
For this reason, we further compare $f_2(0)/f_1(0)$ with the values inferred from the experimental fits.

Experimentally, the determination of $f_2(0)/f_1(0)$ relies on a simultaneous fit of the form factors to the full four-dimensional decay distribution in $q^2$ and the three decay angles.
At $q^2=0$, Eq.~\eqref{eq:helicity_invariant_map_charm} gives $f_+(0)=f_1(0)$ and $f_\perp(0)=f_1(0)+(M_{\Lambda_c}+M_\Lambda)f_2(0)/M_{\Lambda_c}$, and hence $f_2(0)/f_1(0)=M_{\Lambda_c}[f_\perp(0)/f_+(0)-1]/(M_{\Lambda_c}+M_\Lambda)$.
The CLEO Collaboration~\cite{CLEO:2004txf} obtained $R_{\rm HQET}\equiv f_2^{\rm HQET}/f_1^{\rm HQET}=-0.31(5)(4)$ from a fit in the heavy quark effective theory (HQET) basis.
Using the relations obtained from the Gordon identity, $f_1=f_1^{\rm HQET}+(M_\Lambda/M_{\Lambda_c})f_2^{\rm HQET}$ and $f_2=-f_2^{\rm HQET}$~\cite{Korner:1991ph,Korner:1994nh}, the CLEO fit corresponds to $[f_2(0)/f_1(0)]_{\rm CLEO}=0.365(89)$ in the present convention.

The BESIII Collaboration~\cite{BESIII:2022ysa,BESIII:2023jxv} fitted the helicity form factors with a linear $z$ expansion.
Since $f_\perp$ and $f_+$ were assigned the same $q^2$ dependence, their fitted ratio gives $[f_2(0)/f_1(0)]_{\rm BESIII}\simeq0.37(14)$ through the relation above.
For $\Lambda_c^+\to\Lambda$, the present R3QM calculation gives $f_2(0)/f_1(0)=0.343$, while LQCD gives $0.479$~\cite{Meinel:2016dqj}.
The CLEO and BESIII central values are both close to $0.37$ and lie between these two results.
The uncertainties of the two fits remain sizable and do not yet allow a precise determination of $f_2(0)/f_1(0)$.

For the axial-vector form factors, the leading term $g_1$ agrees well with the LQCD results~\cite{Meinel:2016dqj,Meinel:2017ggx} in both channels. 
The calculated $g_2$ is slightly different from the lattice central values, but its contribution to $g_\perp$ and $g_+$ is suppressed by the kinematic coefficients in Eq.~\eqref{eq:helicity_invariant_map_charm}. 
Since $g_1$ gives the main contribution, the resulting $g_\perp$ and $g_+$ are close to the lattice curves in Figs.~\ref{fig:Lambdac_Lambda_form_factor} and \ref{fig:Lambdac_n_form_factor}. 
The small $f_3$ term also has little effect on the longitudinal vector form factor $f_0$. 
The induced pseudoscalar form factor $g_3$ has a different origin. 
In the present calculation, it is reconstructed from the pseudoscalar pole prescription, which gives $g_0(q^2) = g_1(q^2)m_P^2/(m_P^2-q^2)$. 
For $\Lambda_c^+\to n$, the pole mass is smaller ($m_D < m_{D_s}$), while the allowed $q^2$ range is larger. 
The denominator $m_D^2-q^2$ therefore decreases faster near the endpoint, leading to a steeper rise of $g_0^n$ than that of $g_0^\Lambda$. 
This pole effect has little impact on the available electron and muon branching fractions, because the $g_0$ contribution is helicity-suppressed by $m_\ell^2/q^2$ in the decay rate.

From these results, the $\Lambda_c^+\to\Lambda$ and $\Lambda_c^+\to n$ modes have the same scalar $ud$ spectator but different active weak currents. 
The close values of $f_1$ and $g_1$ show that the leading overlap is little changed when the final $\Lambda$ is replaced by the neutron. 
Thus, the leading form factors and their $q^2$ dependence show similar behavior in the two channels, while the smaller $\Lambda_c^+\to n\ell^+\nu_\ell$ rate mainly reflects the CKM suppression and the phase space difference. 
Future precise measurements of the $\Lambda_c^+\to n\ell^+\nu_\ell$ recoil spectrum and lepton angular distribution would test whether the same transverse pattern also holds in the Cabibbo-suppressed channel.

\begin{table*}[t]
	\centering
	\setlength{\tabcolsep}{0pt}
	\renewcommand{\arraystretch}{1.3}
	\caption{Branching fractions, angular observables, and the $f_i(0)$ and $g_i(0)$ form factors for $\Xi_c^0\to\Xi^-\ell^+\nu_\ell$.}
	\label{tab:Heavy_Baryon_Decays_Xic_Xim}
	\begin{tabular*}{\textwidth}{@{\extracolsep{\fill}}cccccccc}
		\toprule \toprule
		\multirow{2}{*}{Method} & \multicolumn{2}{c}{$\mathcal{B}\;(10^{-2})$} & \multirow{2}{*}{$R_{\mu e}$} & \multicolumn{2}{c}{$\langle A_{FB} \rangle$} & \multicolumn{2}{c}{$\langle \alpha_{\Xi^-} \rangle$} \\
		\cmidrule(lr){2-3} \cmidrule(lr){5-6} \cmidrule(lr){7-8}
		& $\ell = e$ & $\ell = \mu$ & & $\ell = e$ & $\ell = \mu$ & $\ell = e$ & $\ell = \mu$ \\
		\midrule
		RQM~\cite{Faustov:2019ddj}    & $2.38$          & $2.31$          & $0.969$ & $-0.208$         & $-0.235$         & $-0.795$         & $-0.791$ \\
		SU(3)~\cite{Geng:2019bfz}     & $3.0(3)$        & $2.9(4)$        & $0.967$ & --- & --- & \multicolumn{2}{c}{$-0.86(4)$} \\
		                                  & $2.4(3)$        & $2.4(3)$        & $1.000$ & --- & --- & \multicolumn{2}{c}{$-0.86(4)$} \\
		                                  & $2.7(2)$        & $2.7(2)$        & $1.000$ & --- & --- & \multicolumn{2}{c}{$-0.83(4)$} \\
		LFQM~\cite{Zhao:2018zcb}      & $1.35$          & ---             & ---     & ---              & ---              & ---              & --- \\
		LFCQM~\cite{Geng:2020gjh}     & $3.49(95)$      & $3.34(94)$      & $0.957$ & ---              & ---              & $-0.98(2)$      & $-0.98(2)$ \\
		LCSR~\cite{Aliev:2021wat}     & $1.85(56)$      & $1.79(54)$      & $0.968$ & ---              & ---              & ---              & --- \\
		LCSR~\cite{Duan:2022yia}      & $2.81({}^{+17}_{-15})$ & $2.72({}^{+17}_{-15})$ & $0.968$ & --- & --- & --- & --- \\
		LCSR~\cite{Aliev:2025zbk}     & $3.73(104)$     & $3.59(101)$     & $0.963$ & ---              & ---              & ---              & --- \\
		LQCD~\cite{Zhang:2021oja}     & $2.38(44)$      & $2.29(42)$      & $0.962$ & ---              & ---              & ---              & --- \\
		LQCD~\cite{Farrell:2025gis}   & $3.58(12)$      & $3.47(12)$      & $0.969$ & ---              & ---              & ---              & --- \\
		RPP~\cite{ParticleDataGroup:2024cfk}
		                                  & $1.05(20)$      & $1.01(21)$      & ---     & ---              & ---              & ---              & --- \\
		\textbf{This work}                 & $\mathbf{4.03}$ & $\mathbf{3.91}$ & $\mathbf{0.968}$ & $\mathbf{-0.166}$ & $\mathbf{-0.196}$ & $\mathbf{-0.818}$ & $\mathbf{-0.815}$ \\
	\end{tabular*}
	\vspace{1.0mm}
	\renewcommand{\arraystretch}{1.22}
	\begin{tabular*}{\textwidth}{@{\extracolsep{\fill}}ccccccc}
		\toprule
		Method & $f_1(0)$ & $f_2(0)$ & $f_3(0)$ & $g_1(0)$ & $g_2(0)$ & $g_3(0)$ \\
		\midrule
		LFQM~\cite{Zhao:2018zcb}
		& $0.567$ & $0.305$ & --- & $0.491$ & $0.046$ & --- \\
		RQM~\cite{Faustov:2019ddj}
		& $0.590$ & $0.441$ & $0.388$ & $0.582$ & $-0.184$ & $-1.144$ \\
		LFCQM~\cite{Geng:2020gjh}
		& $0.74(2)$ & $0.96(2)$ & ---
		& $0.69(1)$ & $6.8(3)\times10^{-3}$ & --- \\
		LCSR~\cite{Aliev:2025zbk}
		& $0.84(11)$ & $0.49(6)$ & $-0.35(11)$
		& $0.84(11)$ & $0.49(6)$ & $-0.42(8)$ \\
		LQCD~\cite{Zhang:2021oja}
		& $0.50(9)$ & $0.58(11)$ & ---
		& $0.56(8)$ & $0.05(35)$ & --- \\
		LQCD~\cite{Farrell:2025gis}
		& $0.722(18)$ & $0.422(17)$ & ---
		& $0.606(13)$ & $-0.014(26)$ & --- \\
		\textbf{This work}
		& $\mathbf{0.753}$ & $\mathbf{0.294}$ & $\mathbf{0.009}$
		& $\mathbf{0.657}$ & $\mathbf{-0.071}$ & $\mathbf{-1.588}$ \\
		\bottomrule\bottomrule
	\end{tabular*}
\end{table*}

\begin{table*}[t]
	\centering
	\setlength{\tabcolsep}{0pt}
	\renewcommand{\arraystretch}{1.3}
	\caption{Branching fractions, angular observables, and the $f_i(0)$ and $g_i(0)$ form factors for $\Xi_c^+\to\Xi^0\ell^+\nu_\ell$.}
	\label{tab:Heavy_Baryon_Decays_Xic_Xi0}
	\begin{tabular*}{\textwidth}{@{\extracolsep{\fill}}cccccccc}
		\toprule \toprule
		\multirow{2}{*}{Method} & \multicolumn{2}{c}{$\mathcal{B}\;(10^{-2})$} & \multirow{2}{*}{$R_{\mu e}$} & \multicolumn{2}{c}{$\langle A_{FB} \rangle$} & \multicolumn{2}{c}{$\langle \alpha_{\Xi^0} \rangle$} \\
		\cmidrule(lr){2-3} \cmidrule(lr){5-6} \cmidrule(lr){7-8}
		& $\ell = e$ & $\ell = \mu$ & & $\ell = e$ & $\ell = \mu$ & $\ell = e$ & $\ell = \mu$ \\
		\midrule
		RQM~\cite{Faustov:2019ddj}    & $9.40$          & $9.11$          & $0.969$ & $-0.208$         & $-0.235$         & $-0.795$         & $-0.791$ \\
		SU(3)~\cite{Geng:2019bfz}     & $11.9(13)$      & $11.6(17)$      & $0.975$ & --- & --- & \multicolumn{2}{c}{$-0.86(4)$} \\
		                                  & $9.8(11)$       & $9.8(11)$       & $1.000$ & --- & --- & \multicolumn{2}{c}{$-0.86(4)$} \\
		                                  & $10.7(9)$       & $10.8(9)$       & $1.009$ & --- & --- & \multicolumn{2}{c}{$-0.83(4)$} \\
		LFQM~\cite{Zhao:2018zcb}      & $5.39$          & ---             & ---     & ---              & ---              & ---              & --- \\
		LFCQM~\cite{Geng:2020gjh}     & $11.3(34)$      & $10.8(33)$      & $0.956$ & ---              & ---              & $-0.97(3)$       & $-0.97(3)$ \\
		LCSR~\cite{Aliev:2021wat}     & $5.51(165)$     & $5.34(161)$     & $0.969$ & ---              & ---              & ---              & --- \\
		LCSR~\cite{Duan:2022yia}      & $8.43({}^{+52}_{-45})$ & $8.16({}^{+50}_{-43})$ & $0.968$ & --- & --- & --- & --- \\
		LCSR~\cite{Aliev:2025zbk}     & $11.20(325)$    & $10.8(31)$      & $0.964$ & ---              & ---              & ---              & --- \\
		LQCD~\cite{Zhang:2021oja}     & $7.18(133)$     & $6.91(127)$     & $0.962$ & ---              & ---              & ---              & --- \\
		LQCD~\cite{Farrell:2025gis}   & $10.94(34)$     & $10.61(33)$     & $0.969$ & ---              & ---              & ---              & --- \\
		RPP~\cite{ParticleDataGroup:2024cfk}
		                                  & $7(4)$          & ---             & ---     & ---              & ---              & ---              & --- \\
		\textbf{This work}                 & $\mathbf{12.3}$ & $\mathbf{11.9}$ & $\mathbf{0.969}$ & $\mathbf{-0.166}$ & $\mathbf{-0.197}$ & $\mathbf{-0.818}$ & $\mathbf{-0.815}$ \\
	\end{tabular*}
	\vspace{1.0mm}
	\renewcommand{\arraystretch}{1.22}
	\begin{tabular*}{\textwidth}{@{\extracolsep{\fill}}ccccccc}
		\toprule
		Method & $f_1(0)$ & $f_2(0)$ & $f_3(0)$ & $g_1(0)$ & $g_2(0)$ & $g_3(0)$ \\
		\midrule
		LFQM~\cite{Zhao:2018zcb}
		& $0.567$ & $0.305$ & --- & $0.491$ & $0.046$ & --- \\
		RQM~\cite{Faustov:2019ddj}
		& $0.590$ & $0.441$ & $0.388$ & $0.582$ & $-0.184$ & $-1.144$ \\
		LFCQM~\cite{Geng:2020gjh}
		& $0.77(2)$ & $0.96(2)$ & ---
		& $0.69(1)$ & $6.8(3)\times10^{-3}$ & --- \\
		LQCD~\cite{Zhang:2021oja}
		& $0.50(9)$ & $0.58(11)$ & ---
		& $0.56(8)$ & $0.05(35)$ & --- \\
		LQCD~\cite{Farrell:2025gis}
		& $0.721(18)$ & $0.422(17)$ & ---
		& $0.605(13)$ & $-0.014(26)$ & --- \\
		\textbf{This work}
		& $\mathbf{0.751}$ & $\mathbf{0.292}$ & $\mathbf{0.009}$
		& $\mathbf{0.654}$ & $\mathbf{-0.071}$ & $\mathbf{-1.577}$ \\
		\bottomrule\bottomrule
	\end{tabular*}
\end{table*}

\subsection{$\Xi_c^{0,+}$ baryon}

\subsubsection{$\Xi_c^0 \to \Xi^- \ell^+ \nu_\ell$ and $\Xi_c^+ \to \Xi^0 \ell^+ \nu_\ell$}

The flavor antitriplet state $\Xi_c$ and flavor sextet state $\Xi_c'$ both have $J^P=1/2^+$, where the light quark pair is in the spin $S=0$ and $S=1$ states, respectively. 
Because of the mass difference between the strange and light quarks ($m_s > m_{u,d}$), the color magnetic spin--spin interaction can mix the flavor antitriplet and flavor sextet configurations. 
In our model, this mixing angle is determined to be $\theta_c \simeq 2.03^\circ$ from the baryon mass spectrum~\cite{Ni:2026arb}. 
Therefore, in calculating the $\Xi_c$ semileptonic decay amplitudes, both the dominant flavor antitriplet and the small flavor sextet components are coherently taken into account.

In the exact SU(2) isospin limit, the Cabibbo-favored transitions $\Xi_c^0\to\Xi^-\ell^+\nu_\ell$ and $\Xi_c^+\to\Xi^0\ell^+\nu_\ell$ have the same weak transition matrix element, since the active $c\to s$ weak current is an isospin singlet ($\Delta I = 0$). 
If the minor difference in phase space is neglected, the decay widths of the two channels are approximately equal. 
However, the corresponding branching fractions differ by a factor of about three due to the different lifetimes of the two \(\Xi_c\) baryons ($\tau_{\Xi_c^+}/\tau_{\Xi_c^0} \simeq 3.02$), which are given by, $\mathcal{B}_{e}[\Xi_c^0\to\Xi^-]\simeq 4.03\%$ and $\mathcal{B}_{e}[\Xi_c^+\to\Xi^0]\simeq 12.3\%$, and the ratio is $\mathcal{B}_{e}[\Xi_c^+\to\Xi^0]/\mathcal{B}_{e}[\Xi_c^0\to\Xi^-]\simeq 3.05$. 
Both the absolute branching fractions and this ratio agree well with the recent LQCD calculation with domain-wall fermions~\cite{Farrell:2025gis}, which gives $\mathcal{B}_{e}[\Xi_c^0\to\Xi^-]=3.58(12)\%$ and $\mathcal{B}_{e}[\Xi_c^+\to\Xi^0]=10.94(34)\%$. 
An earlier LQCD study~\cite{Zhang:2021oja} gives smaller branching fractions, $\mathcal{B}_{e}[\Xi_c^0\to\Xi^-]=2.38(44)\%$ and $\mathcal{B}_{e}[\Xi_c^+\to\Xi^0]=7.18(133)\%$, while keeping the charged-to-neutral ratio close to $3.0$. 
By comparison, the current RPP averages~\cite{ParticleDataGroup:2024cfk}, $\mathcal{B}_{e}[\Xi_c^0\to\Xi^-]=1.05(20)\%$ and $\mathcal{B}_{e}[\Xi_c^+\to\Xi^0]=7(4)\%$, are much smaller than the theory values. 
This difference is especially clear in the neutral channel, where the experimental uncertainty is already relatively small.

Since the dominant flavor antitriplet $\Xi_c$ component has the same spin-$0$ spectator as the $\Lambda_c$ state, the leading form factors in the static $\mathrm{SU}(6)$ limit satisfy $f_1^{\text{stat.}}[\Xi_c\to\Xi] / f_1^{\text{stat.}}[\Lambda_c\to\Lambda] = g_1^{\text{stat.}}[\Xi_c\to\Xi] / g_1^{\text{stat.}}[\Lambda_c\to\Lambda] = \sqrt{3/2} \simeq 1.225$.
At $q^2=0$, the full calculation (which includes the coherent flavor sextet contribution) gives $1.255$ for $f_1(0)$ and $1.223$ for $g_1(0)$. 
Both ratios remain close to the static value. 
Thus, the recoil effects on the active quark change the $\Xi_c\to\Xi$ and $\Lambda_c\to\Lambda$ matrix elements in nearly the same way. 
They make $g_1$ smaller than $f_1$ within each channel, but this effect largely cancels in the ratio between the two transitions. 
The leading form factors therefore still follow the SU(3) flavor scaling rather well.

In the present model, the two transitions of $\Xi_c\to\Xi$ and $\Lambda_c^+\to\Lambda$ are rather similar and have similar leading form factor and phase spaces, thus, the $\Xi_c\to\Xi$ rate is not expected to be several times smaller.
However, the neutral mode RPP average~\cite{ParticleDataGroup:2024cfk} in Table~\ref{tab:Heavy_Baryon_Decays_Xic_Xim} is nearly a factor of four below both our quark model result and LQCD predictions.
In the literature, $\Xi_c$--$\Xi_c'$ mixing has often been proposed as a possible mechanism to suppress the $\Xi_c\to\Xi\ell^+\nu_\ell$ decay rate.
Our spectrum calculation gives an effective mixing angle of $\theta_c \simeq 2.03^\circ$, which is consistent with the small mixing of a few degrees obtained in quark models~\cite{Franklin:1996ve}, heavy hadron chiral perturbation theory~\cite{Boyd:1996cd}, QCD sum rules~\cite{Aliev:2010ra,Sun:2023noo}, HQET~\cite{Matsui:2020wcc}, and LQCD~\cite{Liu:2023feb,Liu:2023pwr}.
In fact, even when a much larger value $|\theta_c|\simeq25^\circ$ was invoked in Ref.~\cite{Geng:2022yxb}, the $\Xi_c\to\Xi e^+\nu_e$ rate was reduced by only about $20\%$.
In the present model, the few degree mixing ($\theta_c \simeq 2.03^\circ$) determined from the mass spectrum gives a negligible suppression, and $\Xi_c$--$\Xi_c'$ mixing alone is insufficient to account for the factor of four difference between our prediction and the current RPP average.

It is also useful to recall how the present experimental branching fraction is obtained.
In the ALICE Collaboration~\cite{ALICE:2021bli} and Belle Collaboration~\cite{Belle:2021crz} measurements, the absolute branching fraction of $\Xi_c^0\to\Xi^-\ell^+\nu_\ell$ is not measured directly.
It is inferred from ratios involving the hadronic reference mode $\Xi_c^0\to\Xi^-\pi^+$.
The extracted semileptonic branching fraction therefore depends on the absolute branching fraction of this reference mode.
Experimentally, a direct absolute measurement of $\Xi_c^0\to\Xi^-\pi^+$ would provide a useful check of the normalization used to extract the $\Xi_c$ semileptonic branching fraction.

The form factor comparison gives a cleaner check of the $\Xi_c\to\Xi$ weak transition matrix element, since it is not tied to the experimental branching fraction.
As shown in Fig.~\ref{fig:Xic_Xi_form_factor}, the calculated helicity form factors follow a pattern similar to that in the $\Lambda_c^+$ channels when compared with LQCD.
This agreement is most visible in the vector form factors $f_0$ and $f_+$.
The main difference in the vector sector appears in the transverse form factor $f_\perp$.
In our calculation, the smaller weak magnetism form factor $f_2$ suppresses this transverse component.
In the axial-vector sector, $g_+$ and $g_\perp$ have a similar $q^2$ dependence but are somewhat larger than the recent LQCD central values~\cite{Farrell:2025gis}.
These larger axial-vector amplitudes also make the total decay width remain above the LQCD central value.

Clarifying this rate puzzle therefore requires not only a more precise absolute branching fraction, but also the $q^2$ spectra and angular observables of the $\Xi_c\to\Xi\ell^+\nu_\ell$ decays.
A change in the absolute branching fraction would mainly shift the overall scale of $d\Gamma/dq^2$.
It would not change the relative $q^2$ dependence or the dimensionless angular observables.
As summarized in Tables~\ref{tab:Heavy_Baryon_Decays_Xic_Xim} and \ref{tab:Heavy_Baryon_Decays_Xic_Xi0}, we predict negative forward-backward asymmetries, $\langle A_{FB} \rangle \simeq -0.17$, and large negative final baryon polarizations, $\langle\alpha_{\Xi}\rangle \simeq -0.82$.
These quantities are determined by the relative strengths and $q^2$ dependence of the helicity amplitudes.
Future measurements of the $q^2$ spectra and angular observables, together with a more precise absolute branching fraction, would test the $\mathrm{SU}(3)$ flavor relations and the dominant scalar spectator dynamics in the $\Xi_c\to\Xi$ transitions.

\begin{table*}[t]
	\centering
	\setlength{\tabcolsep}{0pt}
	\renewcommand{\arraystretch}{1.3}
	\caption{Branching fractions, angular observables, and the $f_i(0)$ and $g_i(0)$ form factors for $\Xi_c^+\to\Lambda\ell^+\nu_\ell$. The SU(3) form factors of Ref.~\cite{Geng:2026bnk} are reconstructed from helicity form factors. NRQM denotes the nonrelativistic quark model.}
	\label{tab:Heavy_Baryon_Decays_Xic_Lambda}
	\begin{tabular*}{\textwidth}{@{\extracolsep{\fill}}cccccccc}
		\toprule \toprule
		\multirow{2}{*}{Method} & \multicolumn{2}{c}{$\mathcal{B}\;(10^{-3})$} & \multirow{2}{*}{$R_{\mu e}$} & \multicolumn{2}{c}{$\langle A_{FB} \rangle$} & \multicolumn{2}{c}{$\langle \alpha_{\Lambda} \rangle$} \\
		\cmidrule(lr){2-3} \cmidrule(lr){5-6} \cmidrule(lr){7-8}
		& $\ell = e$ & $\ell = \mu$ & & $\ell = e$ & $\ell = \mu$ & $\ell = e$ & $\ell = \mu$ \\
		\midrule
		NRQM~\cite{Perez-Marcial:1989sch} & $0.884$ & ---             & ---     & ---              & ---              & ---              & --- \\
		LFQM~\cite{Zhao:2018zcb}      & $0.82$          & ---             & ---     & ---              & ---              & ---              & --- \\
		SU(3)~\cite{Geng:2019bfz}     & $1.03(15)$      & ---             & ---     & ---              & ---              & $-0.86(4)$       & --- \\
		\quad                         & $1.66(18)$      & ---             & ---     & ---              & ---              & $-0.86(4)$       & --- \\
		\quad                         & $2.18(18)$      & ---             & ---     & ---              & ---              & $-0.86(4)$       & --- \\
		RQM~\cite{Faustov:2019ddj}    & $1.27$          & $1.24$          & $0.980$ & $-0.266$         & $-0.283$         & $-0.842$         & $-0.841$ \\
		LFCQM~\cite{Geng:2020gjh}     & $1.2(4)$        & $1.1(5)$        & $0.92$  & ---              & ---              & $-0.98(2)$       & $-0.98(2)$ \\
		LCSR~\cite{Aliev:2021wat}     & $0.92(28)$      & $0.89(27)$      & $0.967$ & ---              & ---              & ---              & --- \\
		SU(3)~\cite{He:2021qnc}       & $0.67(13)$      & $0.69(21)$      & $1.03$  & ---              & ---              & ---              & --- \\
		SU(3)~\cite{Geng:2026bnk}     & $1.17(14)$      & $1.15(13)$      & $0.983$ & ---              & ---              & $-0.908(102)$    & $-0.907(101)$ \\
		\textbf{This work}            & $\mathbf{1.58}$ & $\mathbf{1.55}$ & $\mathbf{0.980}$ & $\mathbf{-0.199}$ & $\mathbf{-0.221}$ & $\mathbf{-0.850}$ & $\mathbf{-0.848}$ \\
	\end{tabular*}
	\vspace{1.0mm}
	\renewcommand{\arraystretch}{1.22}
	\begin{tabular*}{\textwidth}{@{\extracolsep{\fill}}ccccccc}
		\toprule
		Method & $f_1(0)$ & $f_2(0)$ & $f_3(0)$ & $g_1(0)$ & $g_2(0)$ & $g_3(0)$ \\
		\midrule
		LFQM~\cite{Zhao:2018zcb}
		& $0.207$ & $0.122$ & --- & $0.177$ & $0.016$ & --- \\
		RQM~\cite{Faustov:2019ddj}
		& $0.203$ & $0.165$ & $0.120$ & $0.201$ & $-0.054$ & $-0.427$ \\
		LFCQM~\cite{Geng:2020gjh}
		& $0.28(1)$ & $0.38(1)$ & ---
		& $0.25(1)$ & $0.04(1)$ & --- \\
		SU(3)~\cite{Geng:2026bnk}
		& $0.216(155)$ & $0.111(189)$ & ---
		& $0.145(85)$ & $-0.089(241)$ & --- \\
		\textbf{This work}
		& $\mathbf{0.255}$ & $\mathbf{0.085}$ & $\mathbf{-0.002}$
		& $\mathbf{0.208}$ & $\mathbf{-0.029}$ & $\mathbf{-0.526}$ \\
		\bottomrule\bottomrule
	\end{tabular*}
\end{table*}

\begin{table*}[t]
\centering
\setlength{\tabcolsep}{0pt}
	\renewcommand{\arraystretch}{1.15}
	\caption{Branching fractions, angular observables, and the $f_i(0)$ and $g_i(0)$ form factors for the $\Xi_c^+\to\Sigma^0\ell^+\nu_\ell$ decay. The SU(3) row of Ref.~\cite{Geng:2026bnk} is reconstructed from helicity form factors.}
	\label{tab:Heavy_Baryon_Decays_Xic_Sigma}
	\begin{tabular*}{\textwidth}{@{\extracolsep{\fill}}cccccccc}
		\toprule \toprule
		\multirow{2}{*}{Method} & \multicolumn{2}{c}{$\mathcal{B}\;(10^{-3})$} & \multirow{2}{*}{$R_{\mu e}$} & \multicolumn{2}{c}{$\langle A_{FB} \rangle$} & \multicolumn{2}{c}{$\langle \alpha_{\Sigma^0} \rangle$} \\
		\cmidrule(lr){2-3} \cmidrule(lr){5-6} \cmidrule(lr){7-8}
		& $\ell = e$ & $\ell = \mu$ & & $\ell = e$ & $\ell = \mu$ & $\ell = e$ & $\ell = \mu$ \\
		\midrule
		NRQM~\cite{Perez-Marcial:1989sch} & $2.21$ & ---             & ---     & ---              & ---              & ---              & --- \\
		LFQM~\cite{Zhao:2018zcb}      & $1.87$          & ---             & ---     & ---              & ---              & ---              & --- \\
		SU(3)~\cite{Geng:2019bfz}     & $3.1(4)$        & ---             & ---     & ---              & ---              & $-0.86(4)$       & --- \\
		\quad                         & $3.8(4)$        & ---             & ---     & ---              & ---              & $-0.86(4)$       & --- \\
		\quad                         & $4.6(4)$        & ---             & ---     & ---              & ---              & $-0.85(4)$       & --- \\
		LFCQM~\cite{Geng:2020gjh}     & $3.3(9)$        & $3.1(9)$        & $0.94$  & ---              & ---              & $-0.98(1)$       & $-0.98(2)$ \\
		SU(3)~\cite{He:2021qnc}       & $4.96(46)$      & $4.81(44)$      & $0.970$ & ---              & ---              & ---              & --- \\
		SU(3)~\cite{Geng:2026bnk}     & $2.82(30)$      & $2.75(29)$      & $0.975$ & ---              & ---              & $-0.899(93)$     & $-0.897(92)$ \\
		\textbf{This work}            & $\mathbf{4.08}$ & $\mathbf{3.98}$ & $\mathbf{0.976}$ & $\mathbf{-0.199}$ & $\mathbf{-0.222}$ & $\mathbf{-0.829}$ & $\mathbf{-0.827}$ \\
		\bottomrule
	\end{tabular*}
	\vspace{0.6mm}
	\renewcommand{\arraystretch}{1.10}
	\begin{tabular*}{\textwidth}{@{\extracolsep{\fill}}ccccccc}
		Method & $f_1(0)$ & $f_2(0)$ & $f_3(0)$ & $g_1(0)$ & $g_2(0)$ & $g_3(0)$ \\
		\midrule
		LFQM~\cite{Zhao:2018zcb}
		& $0.359$ & $0.211$ & --- & $0.307$ & $0.027$ & --- \\
		LFCQM~\cite{Geng:2020gjh}
		& $0.52(1)$ & $0.70(2)$ & ---
		& $0.45(1)$ & $0.08(1)$ & --- \\
		SU(3)~\cite{Geng:2026bnk}
		& $0.385(202)$ & $0.214(247)$ & ---
		& $0.289(111)$ & $-0.116(335)$ & --- \\
		\textbf{This work}
		& $\mathbf{0.435}$ & $\mathbf{0.205}$ & $\mathbf{0.011}$
		& $\mathbf{0.398}$ & $\mathbf{-0.049}$ & $\mathbf{-1.029}$ \\
		\bottomrule\bottomrule
	\end{tabular*}
\end{table*}

\begin{table*}[t]
\centering
\setlength{\tabcolsep}{0pt}
	\renewcommand{\arraystretch}{1.15}
	\caption{Branching fractions, angular observables, and the $f_i(0)$ and $g_i(0)$ form factors for the Cabibbo-suppressed $\Xi_c^0\to\Sigma^-\ell^+\nu_\ell$ decay. The SU(3) row of Ref.~\cite{Geng:2026bnk} is reconstructed from helicity form factors.}
	\label{tab:Heavy_Baryon_Decays_Xic_Sigmam}
	\begin{tabular*}{\textwidth}{@{\extracolsep{\fill}}cccccccc}
		\toprule\toprule
		\multirow{2}{*}{Method} & \multicolumn{2}{c}{$\mathcal{B}\;(10^{-3})$} & \multirow{2}{*}{$R_{\mu e}$} & \multicolumn{2}{c}{$\langle A_{FB} \rangle$} & \multicolumn{2}{c}{$\langle \alpha_{\Sigma^-} \rangle$} \\
		\cmidrule(lr){2-3} \cmidrule(lr){5-6} \cmidrule(lr){7-8}
		& $\ell = e$ & $\ell = \mu$ & & $\ell = e$ & $\ell = \mu$ & $\ell = e$ & $\ell = \mu$ \\
		\midrule
		NRQM~\cite{Perez-Marcial:1989sch} & $1.12$         & ---             & ---     & ---              & ---              & ---              & --- \\
		LFQM~\cite{Zhao:2018zcb}      & $0.95$          & ---             & ---     & ---              & ---              & ---              & --- \\
		SU(3)~\cite{Geng:2019bfz}     & $1.57(22)$      & ---             & ---     & ---              & ---              & $-0.86(4)$       & --- \\
		\quad                         & $1.92(21)$      & ---             & ---     & ---              & ---              & $-0.86(4)$       & --- \\
		\quad                         & $2.32(19)$      & ---             & ---     & ---              & ---              & $-0.85(4)$       & --- \\
		LFCQM~\cite{Geng:2020gjh}     & $2.2(6)$        & $2.1(6)$        & $0.95$  & ---              & ---              & $-0.98(2)$       & $-0.98(2)$ \\
		SU(3)~\cite{He:2021qnc}       & $3.33(31)$      & $3.23(29)$      & $0.970$ & ---              & ---              & ---              & --- \\
		SU(3)~\cite{Geng:2026bnk}     & $1.86(20)$      & $1.82(19)$      & $0.978$ & ---              & ---              & $-0.898(93)$     & $-0.897(92)$ \\
		\textbf{This work}            & $\mathbf{2.69}$ & $\mathbf{2.63}$ & $\mathbf{0.976}$ & $\mathbf{-0.198}$ & $\mathbf{-0.222}$ & $\mathbf{-0.829}$ & $\mathbf{-0.827}$ \\
		\bottomrule
	\end{tabular*}
	\vspace{0.6mm}
	\renewcommand{\arraystretch}{1.10}
	\begin{tabular*}{\textwidth}{@{\extracolsep{\fill}}ccccccc}
		Method & $f_1(0)$ & $f_2(0)$ & $f_3(0)$ & $g_1(0)$ & $g_2(0)$ & $g_3(0)$ \\
		\midrule
		LFQM~\cite{Zhao:2018zcb}
		& $0.507$ & $0.298$ & --- & $0.434$ & $0.038$ & --- \\
		LFCQM~\cite{Geng:2020gjh}
		& $0.73(1)$ & $0.99(2)$ & ---
		& $0.63(1)$ & $0.11(1)$ & --- \\
		SU(3)~\cite{Geng:2026bnk}
		& $0.544(286)$ & $0.303(349)$ & ---
		& $0.408(157)$ & $-0.163(474)$ & --- \\
		\textbf{This work}
		& $\mathbf{0.617}$ & $\mathbf{0.292}$ & $\mathbf{0.015}$
		& $\mathbf{0.564}$ & $\mathbf{-0.069}$ & $\mathbf{-1.463}$ \\
		\bottomrule\bottomrule
	\end{tabular*}
\end{table*}

\begin{figure*}[t]
	\centering
	\includegraphics[width=15cm]{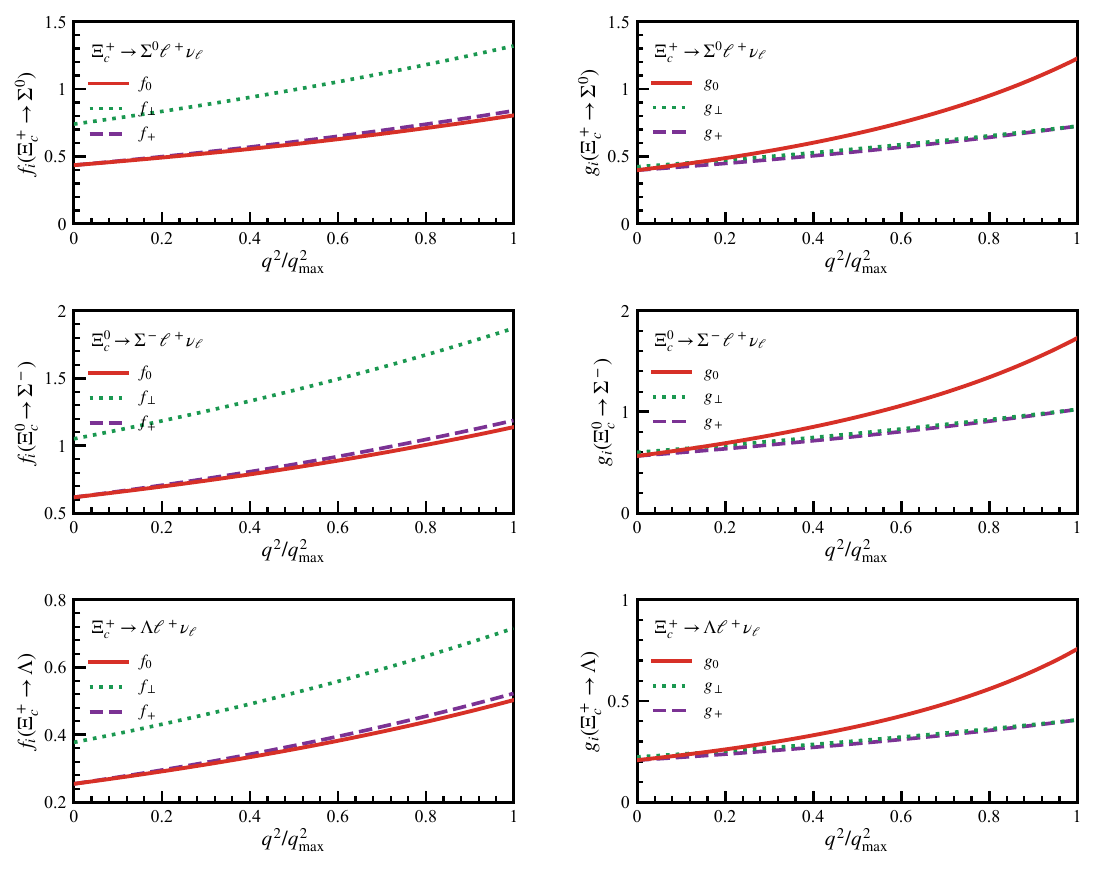}
	\vspace{-0.2 cm}
	\caption{
		R3QM predictions for the helicity basis form factors of the Cabibbo-suppressed $\Xi_c\to\Sigma,\Lambda$ semileptonic transitions as functions of $q^2/q^2_{\rm max}$.
	}\label{fig:Xic_Sigma_Lambda_form_factor}
\end{figure*}

\subsubsection{$\Xi_c^+ \to \Lambda \ell^+ \nu_\ell$, $\Xi_c^+ \to \Sigma^0 \ell^+ \nu_\ell$, and $\Xi_c^0 \to \Sigma^- \ell^+ \nu_\ell$}

The three Cabibbo-suppressed channels $\Xi_c^+\to\Lambda\ell^+\nu_\ell$, $\Xi_c^+\to\Sigma^0\ell^+\nu_\ell$, and $\Xi_c^0\to\Sigma^-\ell^+\nu_\ell$ contain the same $c\to d$ weak current, but the produced $d$ quark is coupled to different light quark configurations in the final baryon.
In the physical $\Xi_c$ baryon, the dominant flavor antitriplet component contains a spin-$0$ $us$ or $ds$ spectator pair, together with the small flavor sextet admixture described above.
After the weak transition, the dominant scalar component combines with the produced $d$ quark to form the final $\Lambda$ or $\Sigma$ baryon.
The calculated branching fractions for the electron modes, listed in Tables~\ref{tab:Heavy_Baryon_Decays_Xic_Lambda}--\ref{tab:Heavy_Baryon_Decays_Xic_Sigmam}, are $\mathcal{B}_{e}[\Xi_c^+\to\Lambda] = 1.58\times 10^{-3}$, $\mathcal{B}_{e}[\Xi_c^+\to\Sigma^0] = 4.08\times 10^{-3}$, and $\mathcal{B}_{e}[\Xi_c^0\to\Sigma^-] = 2.69\times 10^{-3}$.
After the lifetime difference of the initial baryons is removed, we obtain $\Gamma[\Xi_c^0\to\Sigma^-\ell^+\nu_\ell]/\Gamma[\Xi_c^+\to\Sigma^0\ell^+\nu_\ell]=1.99$.
Since the two initial states are related by the isospin exchange of the spectator quark, $d\leftrightarrow u$, this value is close to the isospin expectation of $2$.
The small departure from $2$ comes from the isospin-breaking mass differences in the phase space.

The comparison between $\Xi_c^+\to\Sigma^0$ and $\Xi_c^+\to\Lambda$ then isolates the role of the final light baryon.
These two transitions have the same initial baryon, but the final state is projected onto the $I=1$ $\Sigma$ state or the $I=0$ $\Lambda$ state.
We obtain $\Gamma[\Xi_c^+\to\Sigma^0\ell^+\nu_\ell]/\Gamma[\Xi_c^+\to\Lambda\ell^+\nu_\ell]=2.58$, below the static SU(3) spin-flavor value of $3$ given by the squared ratio of the corresponding coefficients in Table~\ref{tab:slot_singlecharm_sf}.
The smaller value mainly comes from the phase space: the heavier $\Sigma^0$ baryon lowers the endpoint to $q^2_{\max}\simeq 1.63~{\rm GeV}^2$, compared with $q^2_{\max}\simeq 1.83~{\rm GeV}^2$ for $\Xi_c^+\to\Lambda$.
Together with the SU(3) breaking in the spatial wave functions, this phase space effect suppresses the $\Xi_c^+\to\Sigma^0$ rate relative to the $\Xi_c^+\to\Lambda$ rate.

The leading form factors have the same spin-flavor origin as the partial width ratios discussed above.
For the dominant scalar spectator component, the $c\to d$ current acts on the charm quark, and the difference among the $\Lambda$, $\Sigma^0$, and $\Sigma^-$ channels comes from the SU(6) spin-flavor projection of the final baryon.
In the static limit this projection gives
$f_1^{\text{stat.}}[\Xi_c^+\to\Lambda] : f_1^{\text{stat.}}[\Xi_c^+\to\Sigma^0] : f_1^{\text{stat.}}[\Xi_c^0\to\Sigma^-] = 1 : \sqrt{3} : \sqrt{6}$,
with $g_1^{\text{stat.}} = f_1^{\text{stat.}}$ (see Table~\ref{tab:slot_singlecharm_sf}).
This is a relation among the leading form factor amplitudes.
If the phase spaces and spatial wave functions are further taken to be the same, the partial widths are obtained by squaring these coefficients, giving $1:3:6$ for the $\Lambda$, $\Sigma^0$, and $\Sigma^-$ final states.
Equivalently, one obtains $\Gamma[\Xi_c^+\to\Sigma^0]/\Gamma[\Xi_c^+\to\Lambda]=3$ and $\Gamma[\Xi_c^0\to\Sigma^-]/\Gamma[\Xi_c^+\to\Sigma^0]=2$.
The full calculation at $q^2=0$ tests how much this spin-flavor relation is changed by the baryon masses, spatial wave functions, and coherent $\Xi_c$--$\Xi_c'$ admixture.
Using the $\Xi_c^0\to\Sigma^-$ coefficient as the reference, the three leading vector form factors become $\sqrt{6} f_1(0)[\Xi_c^+\to\Lambda] \simeq 0.625$, $\sqrt{2} f_1(0)[\Xi_c^+\to\Sigma^0] \simeq 0.615$, and $f_1(0)[\Xi_c^0\to\Sigma^-] = 0.617$.
In the exact symmetry limit these three numbers would be equal.
Their remaining $1$--$2\%$ spread shows that the leading form factors still follow the static spin-flavor relation, with moderate SU(3) breaking from the final baryon masses and spatial wave functions.

The relation between $f_2$, $g_2$ and the helicity form factors follows from Eq.~\eqref{eq:helicity_invariant_map_charm}.
For the vector form factors, $f_0$ is governed by $f_1$ and the small $f_3$ term, while the $f_2$ term in $f_+$ is multiplied by $q^2/[M_i(M_i+M_f)]$.
The transverse form factor $f_\perp$, instead, contains $f_2$ with the coefficient $(M_i+M_f)/M_i$.
Since the coefficient in $f_\perp$ is much larger over the physical region, the $f_2$ term is more visible in $f_\perp$ than in $f_+$, leading to $f_\perp>f_+\gtrsim f_0$ in Fig.~\ref{fig:Xic_Sigma_Lambda_form_factor}.
For the axial-vector form factors, $g_\perp$ and $g_+$ differ from $g_1$ only through the induced tensor form factor $g_2$.
Since $g_2$ is small in these scalar spectator channels, the two axial-vector helicity form factors stay close to each other, $g_\perp\simeq g_+$.
After the relative spin-flavor factors are divided out, the three channels show very similar $q^2$ behavior.
The main exception is $g_0$.
The $\Xi_c^+\to\Lambda\ell^+\nu_\ell$ channel has the largest endpoint, so its physical region lies closer to the $D$ meson pole and the $g_0$ curve has a larger curvature near zero recoil.

Besides the decay rates, the angular observables test the same dominant scalar spectator structure, including the small mixed state correction, through the relative helicity amplitudes.
Both $A_{FB}$ and $\alpha$ are ratios of partial angular or helicity rates, so the common spin-flavor coefficients cancel to a large extent.
The lepton forward-backward asymmetry is determined by the part of the angular distribution that changes sign under $\cos\theta_\ell\to-\cos\theta_\ell$.
For the $V-A$ current, this odd angular term is sensitive to the cross term between the vector and axial-vector helicity amplitudes.
Schematically, for a fixed helicity amplitude, $|H_V-H_A|^2=|H_V|^2+|H_A|^2-2\mathrm{Re}(H_VH_A^*)$.
The last term is the vector--axial-vector interference.
The longitudinal polarization of the final baryon, in contrast, is fixed by the normalized difference between the $\lambda_f=+1/2$ and $\lambda_f=-1/2$ rates.
With the same dominant scalar spectator structure and similar rescaled helicity amplitudes, the three channels give nearly the same large negative polarization, $\langle\alpha\rangle \simeq -0.84$.
As listed in Tables~\ref{tab:Heavy_Baryon_Decays_Xic_Lambda}--\ref{tab:Heavy_Baryon_Decays_Xic_Sigmam}, the small channel dependence is mainly seen in $\langle A_{FB}^{e}\rangle$.
The larger kinematic range in the $\Lambda$ channel changes the $q^2$ average of the vector--axial-vector interference slightly, leading to $\langle A_{FB}^{e} \rangle \simeq -0.20$, compared with $\langle A_{FB}^{e} \rangle \simeq -0.20$ for the $\Xi_c^+\to\Sigma^0\ell^+\nu_\ell$ and $\Xi_c^0\to\Sigma^-\ell^+\nu_\ell$ transitions.

At present, the absolute branching fractions of these Cabibbo-suppressed channels have not been measured.
Once measured, their partial widths can be compared with the static SU(3) spin-flavor ratios discussed above.
For the two $\Xi_c^+$ modes, this comparison can be made directly with the branching fraction ratio.
Ratios involving $\Xi_c^0\to\Sigma^-\ell^+\nu_\ell$ should instead use $\Gamma=\mathcal B/\tau$, because $\Xi_c^0$ and $\Xi_c^+$ have different lifetimes.
The $q^2$ dependence of $d\Gamma/dq^2$, together with $A_{FB}$ and $\alpha$, would test another part of the calculation.
These observables depend on how the helicity amplitudes vary with $q^2$ and on their relative signs, while the partial widths mainly test the total strength of the transition.
Future measurements of these branching fractions, $q^2$ spectra, and angular observables will provide crucial tests for the spin-flavor recoupling and recoil dynamics in the Cabibbo suppressed $\Xi_c$ decays, clarifying whether the same theoretical framework consistently describes both the $c\to s$ and $c\to d$ transition sectors.

\subsection{$\Omega_c^0$ baryon}

\begin{figure*}[t]
	\centering
	\includegraphics[width=15cm]{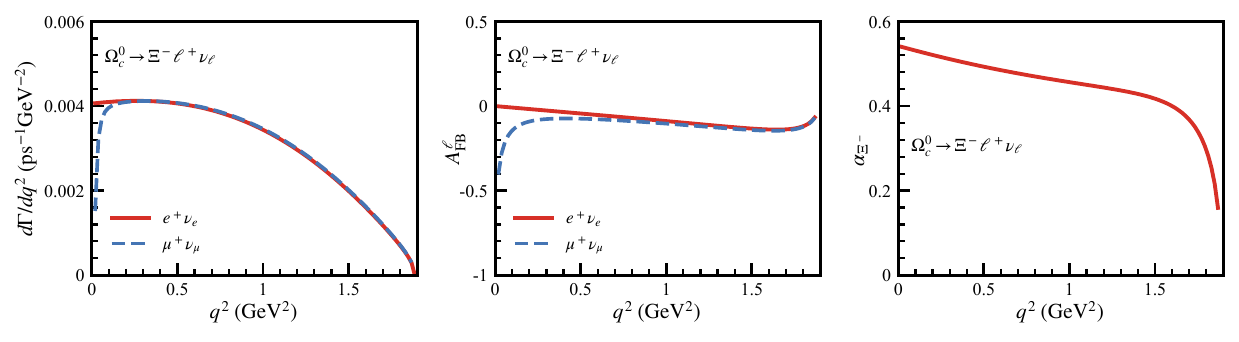}
	\vspace{-0.2 cm}
	\caption{
		Differential decay rates and angular observables for $\Omega_c^0\to\Xi^-\ell^+\nu_\ell$ as functions of $q^2$.
		The electron and muon modes are plotted separately for $d\Gamma/dq^2$ and $A_{\rm FB}^{\ell}$, while the polarization $\alpha_{\Xi^-}$ is given as the rate-weighted average over the kinematically allowed lepton modes.
	}\label{fig:Omegac_Xi_minus_observables}
\end{figure*}

In the Cabibbo-suppressed $\Omega_c^0(css)\to\Xi^-(dss)\ell^+\nu_\ell$ decay, the weak current changes the charm quark into a $d$ quark, while the two $s$ quarks remain as the spectator pair.
For the two identical $s$ quarks, the flavor wave function is symmetric.
In the ground state, the spatial wave function is also symmetric with angular momentum $L=0$.
Together with the antisymmetric color wave function, Fermi statistics then requires the spin part to be symmetric, giving an total spin $S=1$ spectator.
Since this pair is in an $S$ wave, it has positive parity and corresponds to an axial-vector spectator with $J^P=1^+$.
This spin assignment separates the $\Omega_c^0\to\Xi^-$ transition from the $\Lambda_c$ and $\Xi_c$ antitriplet decays, where the light spectator pair is scalar with $J^P=0^+$.
For this channel, Table~\ref{tab:Heavy_Baryon_Decays_Omegac_Xi} gives $\mathcal{B}_{e}[\Omega_c^0\to\Xi^-]=1.59\times10^{-3}$ and $\mathcal{B}_{\mu}[\Omega_c^0\to\Xi^-]=1.56\times10^{-3}$.
The available phase space is large, and the muon mass changes the rate only slightly.
The electron and muon modes then have nearly the same branching fractions, $R_{\mu e}=0.983$, and very similar $q^2$ distributions, except near the muon threshold in Fig.~\ref{fig:Omegac_Xi_minus_observables}.
At present, no experimental measurement is available for this decay.
The predicted branching fractions and spectra can be tested in future measurements and compared with other model calculations~\cite{Pervin:2006ie,Zhao:2018zcb,Duan:2020xcc}.

For the spin-$1$ $ss$ spectator, Table~\ref{tab:slot_singlecharm_sf} gives the static SU(6) values $f_1^{\text{stat.}} = -1$ and $g_1^{\text{stat.}} = 1/3$.
Although the signs of the individual $f_i(q^2)$ and $g_i(q^2)$ form factors depend on the phase convention of the final baryon, the static ratio $g_1^{\text{stat.}} / f_1^{\text{stat.}} = -1/3$ is convention independent.
At maximum recoil, we obtain $g_1(0)/f_1(0) = -0.296$, close to this value.
At $q^2=0$, the negative $f_1(0)=-0.497$ and the positive weak magnetism form factor $f_2(0)=0.526$ partly cancel in $f_\perp$, which has the opposite sign to $f_0$ and $f_+$ at low $q^2$.
No analogous cancellation occurs for the axial-vector form factors because the induced tensor form factor is very small, $g_2(0)=-0.015$.

The cancellation in $f_\perp$ suppresses the transverse vector amplitude and its interference with the axial-vector amplitude in the $\cos\theta_\ell$-odd part of the lepton angular distribution.
This angular term also contains longitudinal--timelike contributions proportional to $m_\ell^2/q^2$, which are strongly suppressed for electrons but remain non-negligible for muons.
We obtain $\langle A_{FB}^{e}\rangle=-0.066$ and $\langle A_{FB}^{\mu}\rangle=-0.106$, as listed in Table~\ref{tab:Heavy_Baryon_Decays_Omegac_Xi} and shown in Fig.~\ref{fig:Omegac_Xi_minus_observables}.
The spectator spin recoupling also affects the longitudinal polarization of the final baryon.
In scalar spectator channels, the light pair carries no spin, so the final baryon spin follows the active light quark in the static limit.
The corresponding ratio $g_1^{\text{stat.}}/f_1^{\text{stat.}}=+1$ makes the leading vector--axial-vector interference favor the $\lambda_f=-1/2$ rate.

For $\Omega_c^0\to\Xi^-$, the active $d$ quark is instead coupled to the spin-$1$ $ss$ pair.
In the static limit, with the spin quantization axis chosen along the recoil direction, the $J_z=\pm1/2$ states are given by the Clebsch--Gordan expansions $|1/2, +1/2\rangle = \sqrt{2/3}\,|1, +1\rangle_{ss} |d, \downarrow\rangle - \sqrt{1/3}\,|1, 0\rangle_{ss} |d, \uparrow\rangle$ and $|1/2, -1/2\rangle = \sqrt{1/3}\,|1, 0\rangle_{ss} |d, \downarrow\rangle - \sqrt{2/3}\,|1, -1\rangle_{ss} |d, \uparrow\rangle$, where $\uparrow$ and $\downarrow$ denote the active quark spin projections $s_z=+1/2$ and $-1/2$, respectively.
Because the initial and final baryons share the same $1\otimes1/2\to1/2$ spin recoupling and the one-body current leaves the $ss$ spectator state unchanged, the spin part of the axial-vector matrix element is determined by the expectation value of the active quark Pauli matrix $\sigma_z^{(d)}$.
For the $J_z=+1/2$ state, the squared Clebsch--Gordan coefficients give probabilities of $2/3$ and $1/3$ for the $|d, \downarrow\rangle$ and $|d, \uparrow\rangle$ components, yielding $\langle\sigma_z^{(d)}\rangle_{J_z=+1/2}=(2/3)(-1)+(1/3)(+1)=-1/3$.
The vector charge gives a unit spin overlap, and the common $c\to d$ flavor overlap cancels exactly between the two currents.
These spin factors therefore give $g_1^{\text{stat.}}/f_1^{\text{stat.}}=(-1/3)/1=-1/3$.

Relative to the scalar spectator channels, this negative ratio reverses the sign of the leading $f_1$--$g_1$ interference, thereby favoring the $\lambda_{\Xi^-}=+1/2$ state in the static limit.
In the full calculation, this $\lambda_{\Xi^-}=+1/2$ dominance persists over the entire physical region, yielding a positive integrated polarization,
$\langle\alpha_{\Xi^-}\rangle = [\Gamma(+1/2) - \Gamma(-1/2)]/[\Gamma(+1/2) + \Gamma(-1/2)] \simeq +0.47$.
Experimentally, the $\Xi^-$ polarization can be extracted from the daughter $\Lambda$ angular distribution in $\Xi^-\to\Lambda\pi^-$ using the established decay asymmetry parameter, allowing a direct test of the positive $\alpha_{\Xi^-}$ predicted for the spin-$1$ spectator configuration.

\begin{table*}[t]
	\centering
	\setlength{\tabcolsep}{0pt}
	\renewcommand{\arraystretch}{1.3}
	\caption{Branching fractions, angular observables, and the $f_i(0)$ and $g_i(0)$ form factors for $\Omega_c^0\to\Xi^-\ell^+\nu_\ell$. The CQM range of Ref.~\cite{Pervin:2006ie} is converted with the RPP lifetime~\cite{ParticleDataGroup:2024cfk}.}
	\label{tab:Heavy_Baryon_Decays_Omegac_Xi}
	\begin{tabular*}{\textwidth}{@{\extracolsep{\fill}}cccccccc}
		\toprule \toprule
		\multirow{2}{*}{Method} & \multicolumn{2}{c}{$\mathcal{B}\;(10^{-3})$} & \multirow{2}{*}{$R_{\mu e}$} & \multicolumn{2}{c}{$\langle A_{FB} \rangle$} & \multicolumn{2}{c}{$\langle \alpha_{\Xi^-} \rangle$} \\
		\cmidrule(lr){2-3} \cmidrule(lr){5-6} \cmidrule(lr){7-8}
		& $\ell = e$ & $\ell = \mu$ & & $\ell = e$ & $\ell = \mu$ & $\ell = e$ & $\ell = \mu$ \\
		\midrule
		CQM~\cite{Pervin:2006ie}      & $0.93\text{--}1.78$ & ---         & ---     & ---              & ---              & ---              & --- \\
		LFQM~\cite{Zhao:2018zcb}      & $0.22$          & ---             & ---     & ---              & ---              & ---              & --- \\
		LCSR~\cite{Duan:2020xcc}      & $3.06(15)$      & ---             & ---     & ---              & ---              & ---              & --- \\
		\textbf{This work}                 & $\mathbf{1.59}$ & $\mathbf{1.56}$ & $\mathbf{0.983}$ & $\mathbf{-0.066}$ & $\mathbf{-0.106}$ & $\mathbf{0.474}$ & $\mathbf{0.471}$ \\
	\end{tabular*}
	\vspace{0.5mm}
	\renewcommand{\arraystretch}{1.22}
	\begin{tabular*}{\textwidth}{@{\extracolsep{\fill}}ccccccc}
		\toprule
		Method & $f_1(0)$ & $f_2(0)$ & $f_3(0)$ & $g_1(0)$ & $g_2(0)$ & $g_3(0)$ \\
		\midrule
		LFQM~\cite{Zhao:2018zcb}
		& $-0.377$ & $0.358$ & --- & $0.105$ & $0.001$ & --- \\
		\textbf{This work}
		& $\mathbf{-0.497}$ & $\mathbf{0.526}$ & $\mathbf{0.273}$ & $\mathbf{0.147}$ & $\mathbf{-0.015}$ & $\mathbf{-0.454}$ \\
		\bottomrule\bottomrule
	\end{tabular*}
\end{table*}

\section{Summary}
\label{sec:summary}

We have studied the spin-$1/2\to1/2$ semileptonic decays of the singly charmed baryons $\Lambda_c^+$, $\Xi_c^{0,+}$, and $\Omega_c^0$ into the light baryon octet within the relativistic three-quark model. 
In our framework, all constituent quark masses and rest frame wave functions are determined by the baryon mass spectrum, where the coherent $\Xi_c$--$\Xi_c'$ mixing is fixed by the mass eigenstates. 
Consequently, no adjustable parameters are introduced in calculating the weak decay amplitudes.
From the calculated helicity amplitudes, we obtain the branching fractions, $q^2$ distributions, longitudinal polarizations, and form factors.
Such a unified description avoids distorting the underlying parameters of the theoretical model through fits to experimental data, and instead allows one to examine the intrinsic shortcomings of the theory itself by comparing its predictions with experimental and LQCD results, thereby deepening our understanding of hadronic properties and structure.

The $\Lambda_c^+\to\Lambda\ell^+\nu_\ell$ and $\Lambda_c^+\to n\ell^+\nu_\ell$ modes first test the scalar spectator part of the calculation.
They have the same initial baryon and the same scalar $ud$ pair, while the active weak current changes from $c\to s$ to $c\to d$.
Our calculated decay rates are consistent with the available experimental data and LQCD results. 
For the form factors, the leading $f_1$ and $g_1$ components agree well with LQCD, while the transverse weak magnetism form factor $f_2$ is smaller. 
A similar suppression of $f_2$ was also found in our previous study of hyperon semileptonic decays~\cite{Ni:2026arb}, indicating that it is closely related to the scalar spectator structure in the present model.

The Cabibbo favored $\Xi_c\to\Xi\ell^+\nu_\ell$ modes retain a predominantly scalar spectator containing a strange quark, with a small spectrum determined flavor sextet admixture.
In the isospin limit, the two $\Xi_c\to\Xi$ modes have the same weak transition matrix element.
Their branching fractions differ mainly because the $\Xi_c^+$ lifetime is much longer than the $\Xi_c^0$ lifetime, with only a small correction from the phase space.
The two modes still have nearly the same $q^2$ dependence and angular distributions.
Our predicted decay rates agree with recent LQCD calculations, but the theoretical values remain well above the present experimental average for $\Xi_c^0\to\Xi^- e^+\nu_e$.
Further theoretical and experimental studies are needed to clarify this discrepancy and the possible role of dynamical suppression beyond the $\Xi_c$--$\Xi_c'$ mixing.

For the Cabibbo-suppressed $\Xi_c^+\to\Lambda\ell^+\nu_\ell$, $\Xi_c^+\to\Sigma^0\ell^+\nu_\ell$, and $\Xi_c^0\to\Sigma^-\ell^+\nu_\ell$ modes, the weak current changes to $c\to d$, whereas the light spectator pair remains predominantly scalar.
The leading $f_1$ and $g_1$ form factors are then governed mainly by the same static spin-flavor overlaps, which give the static SU(6) coefficients in the ratio $1:\sqrt{3}:\sqrt{6}$.
Departures from this limit appear in the recoil-sensitive $f_2$ and $g_2$ terms and in the different phase spaces of the $\Lambda$ and $\Sigma$ final states.
The $\Omega_c^0\to\Xi^-\ell^+\nu_\ell$ mode gives a complementary test, because the spectator spin is no longer scalar.
In this channel, the scalar light pair is replaced by a spin-$1$ $ss$ spectator.
The corresponding spin recoupling changes the relative vector and axial-vector helicity amplitudes, leading to the static $g_1/f_1\simeq -1/3$ behavior and to a positive longitudinal polarization of the final $\Xi^-$.

The lepton flavor ratios mainly reflect the available phase space, while the $q^2$ spectra, lepton forward-backward asymmetries, and final baryon polarizations are sensitive to the relative strengths and $q^2$ dependence of the vector and axial-vector helicity amplitudes.
For $\Xi_c^0\to\Xi^-e^+\nu_e$, most theoretical predictions for the branching fraction lie above the current experimental average.
A direct measurement of $\mathcal B(\Xi_c^0\to\Xi^-\pi^+)$ would provide a useful check of the normalization used to obtain the absolute semileptonic branching fraction from the measured branching fraction ratio.
Beyond the total rate, the $q^2$ spectra and angular observables in $\Xi_c\to\Xi\ell^+\nu_\ell$ will test the predicted helicity structure and scalar spectator dynamics.
For the $\Lambda_c^+\to\Lambda\ell^+\nu_\ell$ decay, a more precise measurement of the four-dimensional distribution would reduce the uncertainty in the fitted $f_2(0)/f_1(0)$ ratio.
For the unmeasured or poorly measured channels, we have systematically predicted the decay rates, $q^2$ distributions, and polarization observables within the R3QM.
We hope that these predictions will provide useful references for future measurements at Belle II, BESIII, and LHCb, especially for examining the dependence of these observables on the spectator spin configuration.

\begin{acknowledgments}
This work is supported by the National Natural Science Foundation of China (Grant Nos. 12547111 and 12221005), the Chinese Academy of Sciences under Grant No. YSBR-101, and the National Key Research and Development Program of China under Contract No. 2025YFA1613900.
\end{acknowledgments}

\section*{Data Availability}
All data supporting the findings of this article are available within the article.

\appendix

\section{Wave functions and transition matrix elements}
\label{app:slot_flavor_spin_wfs}

In this appendix, we summarize the spatial, spin, and flavor wave functions for the initial singly charmed and final light octet baryons, and give the explicit transition matrix elements $[\mathcal G_{\kappa_f\kappa_i}^{\tau}(j)]_{\alpha\beta}$ in the decay amplitude in Eq.~\eqref{eq:sc_full_amplitude_charm}.

\subsection{Spatial and spin-flavor wave functions}
\label{app:charm_wave_functions}

Because the color wave function of a baryon is totally antisymmetric, the combined spatial, spin, and flavor wave function must be totally symmetric under the permutation of any two quarks.
In the constituent quark model, the internal relative motions of the three quarks are described by the Jacobi momenta $\bm k_\rho$ and $\bm k_\lambda$.
For each mode $\zeta\in\{\rho,\lambda\}$, the single mode harmonic oscillator wave function in momentum space takes the form $\psi_{nlm}(\bm k_\zeta)=\mathcal R_{nl}(k_\zeta;\alpha_\zeta)Y_{lm}(\hat{\bm k}_\zeta)$, with the radial wave function
\begin{align}
&\mathcal R_{nl}(k_\zeta;\alpha_\zeta) \nonumber \\
&=
\frac{(-1)^n (-i)^l}{\alpha_\zeta^{3/2}} \sqrt{\frac{2n!}{\Gamma(n+l+3/2)}} \left(\frac{k_\zeta}{\alpha_\zeta}\right)^l e^{-\frac{k_\zeta^2}{2\alpha_\zeta^2}} L_n^{l+1/2}\left(\frac{k_\zeta^2}{\alpha_\zeta^2}\right),
\label{eq:radial_ho_wfs}
\end{align}
where $\alpha_\zeta$ is the oscillator width parameter and $L_n^{l+1/2}$ is the associated Laguerre polynomial.
Coupling the two Jacobi modes to total orbital angular momentum $L$ with projection $M_L$, the spatial basis states are given by
\begin{align}
&\Psi_{N_{\mathrm{osc}}LM_L}^{(n_\rho l_\rho,n_\lambda l_\lambda)}
(\bm k_\rho,\bm k_\lambda)
\nonumber \\
&= \sum_{m_\rho,m_\lambda}
\langle l_\rho m_\rho\,l_\lambda m_\lambda|LM_L\rangle
\psi_{n_\rho l_\rho m_\rho}(\bm k_\rho)
\psi_{n_\lambda l_\lambda m_\lambda}(\bm k_\lambda),
\label{eq:coupled_spatial_basis}
\end{align}
where $N_{\mathrm{osc}}=2n_\rho+l_\rho+2n_\lambda+l_\lambda$ denotes the total harmonic oscillator shell number.
For the ground state ($J^P=1/2^+$) baryons studied in this work, we truncate the basis up to $N_{\mathrm{osc}}\leq2$ with $L=0$~\cite{Ni:2026arb}. This truncation includes four spatial configurations $(n_\rho l_\rho,n_\lambda l_\lambda)$: the ground state $(00,00)$, the two radially excited states $(10,00)$ and $(00,10)$, and the internal orbital excitation $(01,01)$ with $l_\rho=l_\lambda=1$.

For a light octet baryon, these spatial states are combined into wave functions with definite permutation symmetries,
\begin{align}
\Psi_{000}^{s}
&=\Psi_{000}^{(00,00)},
\nonumber\\
\Psi_{200}^{s}
&=\frac{1}{\sqrt{2}}
\left(\Psi_{200}^{(10,00)}+\Psi_{200}^{(00,10)}\right),
\nonumber\\
\Psi_{200}^{\rho}
&=-\Psi_{200}^{(01,01)},
\nonumber\\
\Psi_{200}^{\lambda}
&=\frac{1}{\sqrt{2}}
\left(-\Psi_{200}^{(10,00)}+\Psi_{200}^{(00,10)}\right),
\label{eq:spatial_symmetry_basis}
\end{align}
where the superscript $s$ denotes the totally symmetric state, while $\rho$ and $\lambda$ denote the mixed symmetry states that are antisymmetric and symmetric under the permutation of the first two quarks ($1\leftrightarrow2$), respectively.

In the $\mathrm{SU}(6)$ spin-flavor symmetry, the spin-flavor wave functions of the light octet baryons are classified into the symmetric $\mathbf{56}$-plet and mixed symmetric $\mathbf{70}$-plet representations,
\begin{align}
\Phi_{\mathbf{56},f}^{s}
&=
\frac{1}{\sqrt2}
\left(
\phi_f^\rho\chi_{1/2}^\rho
+
\phi_f^\lambda\chi_{1/2}^\lambda
\right),
\nonumber\\
\Phi_{\mathbf{70},f}^{\rho}
&=
\frac{1}{\sqrt2}
\left(
\phi_f^\rho\chi_{1/2}^\lambda
+
\phi_f^\lambda\chi_{1/2}^\rho
\right),
\nonumber\\
\Phi_{\mathbf{70},f}^{\lambda}
&=
\frac{1}{\sqrt2}
\left(
\phi_f^\rho\chi_{1/2}^\rho
-
\phi_f^\lambda\chi_{1/2}^\lambda
\right),
\label{eq:app_charm_octet_spin_flavor_basis}
\end{align}
where $\phi_f^{\rho,\lambda}$ and $\chi_{1/2}^{\rho,\lambda}$ are the flavor and spin wave functions with antisymmetric ($\rho$) and symmetric ($\lambda$) properties under $1\leftrightarrow2$~\cite{Isgur:1978xj}.

Combining the spatial states in Eq.~\eqref{eq:spatial_symmetry_basis} with the spin, flavor functions in Eq.~\eqref{eq:app_charm_octet_spin_flavor_basis} to form totally symmetric wave functions yields three configuration basis states for the final light octet baryon,
\begin{align}
\left|B_{f,1}^{\mathbf 8},\frac12^+\right\rangle
&=
\Psi_{000}^{s,(f)}
\Phi_{\mathbf{56},f}^{s},
\nonumber\\
\left|B_{f,2}^{\mathbf 8},\frac12^+\right\rangle
&=
\Psi_{200}^{s,(f)}
\Phi_{\mathbf{56},f}^{s},
\nonumber\\
\left|B_{f,3}^{\mathbf 8},\frac12^+\right\rangle
&=
\frac{1}{\sqrt2}
\left(
\Psi_{200}^{\rho,(f)}
\Phi_{\mathbf{70},f}^{\rho}
+
\Psi_{200}^{\lambda,(f)}
\Phi_{\mathbf{70},f}^{\lambda}
\right).
\label{eq:app_charm_octet_configuration_basis}
\end{align}
Here, the first two states ($\kappa_f=1,2$) belong to the symmetric $\mathbf{56}$-plet, while the third state ($\kappa_f=3$) arises from the mixed symmetric $\mathbf{70}$-plet.

For a singly charmed baryon, we take the state with the charm quark in slot 3 as the reference configuration $[3]$, where the two light quarks occupy slots $(12)$ with flavor symmetry $\tau\in\{A,S\}$ ($\phi_{i,A}^{[3]}$ for the antisymmetric antitriplet and $\phi_{i,S}^{[3]}$ for the symmetric sextet).
Because the color wave function of the light quark pair is antisymmetric, Fermi statistics requires its spatial, spin, and flavor wave function to be symmetric under $1\leftrightarrow2$.
Since the spatial states with $l_\rho=0$ ($\kappa_i=1,2,3$) are even under $1\leftrightarrow2$ while the state with $l_\rho=1$ ($\kappa_i=4$) is odd, the spin of the light quark pair $S_{12}$ is uniquely fixed for each configuration.

For the flavor antitriplet ($\tau=A$), the four configuration basis states are given by
\begin{align}
\left|B_{i,1}^{A,[3]},\frac12^+\right\rangle
&=
\Psi_{000}^{(00,00),(i)[3]}
\phi_{i,A}^{[3]}\chi_{1/2}^{\rho,[3]},
\nonumber\\
\left|B_{i,2}^{A,[3]},\frac12^+\right\rangle
&=
\Psi_{200}^{(10,00),(i)[3]}
\phi_{i,A}^{[3]}\chi_{1/2}^{\rho,[3]},
\nonumber\\
\left|B_{i,3}^{A,[3]},\frac12^+\right\rangle
&=
\Psi_{200}^{(00,10),(i)[3]}
\phi_{i,A}^{[3]}\chi_{1/2}^{\rho,[3]},
\nonumber\\
\left|B_{i,4}^{A,[3]},\frac12^+\right\rangle
&=
\Psi_{200}^{(01,01),(i)[3]}
\phi_{i,A}^{[3]}\chi_{1/2}^{\lambda,[3]},
\label{eq:app_charm_A_configuration_basis}
\end{align}
where the light pair has $S_{12}=0$ ($\chi_{1/2}^{\rho,[3]}$) for $\kappa_i=1,2,3$ and $S_{12}=1$ ($\chi_{1/2}^{\lambda,[3]}$) for $\kappa_i=4$.
For the flavor sextet ($\tau=S$), the spin coupling is reversed ($S_{12}=1$ for $\kappa_i=1,2,3$ and $S_{12}=0$ for $\kappa_i=4$), yielding
\begin{align}
\left|B_{i,1}^{S,[3]},\frac12^+\right\rangle
&=
\Psi_{000}^{(00,00),(i)[3]}
\phi_{i,S}^{[3]}\chi_{1/2}^{\lambda,[3]},
\nonumber\\
\left|B_{i,2}^{S,[3]},\frac12^+\right\rangle
&=
\Psi_{200}^{(10,00),(i)[3]}
\phi_{i,S}^{[3]}\chi_{1/2}^{\lambda,[3]},
\nonumber\\
\left|B_{i,3}^{S,[3]},\frac12^+\right\rangle
&=
\Psi_{200}^{(00,10),(i)[3]}
\phi_{i,S}^{[3]}\chi_{1/2}^{\lambda,[3]},
\nonumber\\
\left|B_{i,4}^{S,[3]},\frac12^+\right\rangle
&=
\Psi_{200}^{(01,01),(i)[3]}
\phi_{i,S}^{[3]}\chi_{1/2}^{\rho,[3]}.
\label{eq:app_charm_S_configuration_basis}
\end{align}
The fully symmetrized physical states $|B_{i,\kappa_i}^{\tau}\rangle$ are then obtained by summing over the three charm slot components according to Eq.~\eqref{eq:single_charm_slot_state}.

In the reference slot $[3]$, the mixed symmetry spin wave functions $\chi_{1/2}^{\rho,[3]}$ and $\chi_{1/2}^{\lambda,[3]}$ for a spin-$1/2$ baryon with spin projection $\lambda_z=\pm1/2$ are explicitly given by
\begin{align}
\chi_{1/2}^{\rho,[3]}\!\left(\frac12\right)
&=\frac{1}{\sqrt2}\bigl(\ket{\uparrow\downarrow\uparrow}-\ket{\downarrow\uparrow\uparrow}\bigr),
\nonumber\\
\chi_{1/2}^{\rho,[3]}\!\left(-\frac12\right)
&=\frac{1}{\sqrt2}\bigl(\ket{\uparrow\downarrow\downarrow}-\ket{\downarrow\uparrow\downarrow}\bigr),
\nonumber\\
\chi_{1/2}^{\lambda,[3]}\!\left(\frac12\right)
&=\frac{1}{\sqrt6}\bigl(2\ket{\uparrow\uparrow\downarrow}-\ket{\uparrow\downarrow\uparrow}-\ket{\downarrow\uparrow\uparrow}\bigr),
\nonumber\\
\chi_{1/2}^{\lambda,[3]}\!\left(-\frac12\right)
&=\frac{1}{\sqrt6}\bigl(\ket{\uparrow\downarrow\downarrow}+\ket{\downarrow\uparrow\downarrow}-2\ket{\downarrow\downarrow\uparrow}\bigr).
\label{eq:app_slot_spin_basis_charm}
\end{align}
The spin and flavor wave functions in slots $[1]$ and $[2]$ are obtained through permutations: $\chi_{1/2}^{\sigma,[1]}=\widehat P_{13}\chi_{1/2}^{\sigma,[3]}$, $\chi_{1/2}^{\sigma,[2]}=\widehat P_{23}\chi_{1/2}^{\sigma,[3]}$ and $\phi_f^{\zeta,[1]}=\widehat P_{13}\phi_f^\zeta$, $\phi_f^{\zeta,[2]}=\widehat P_{23}\phi_f^\zeta$ ($\sigma,\zeta=\rho,\lambda$).

In terms of the three slot basis $(abc)\chi_{1/2}^{\sigma,[j]}(\lambda_i)$, where $abc$ label the quark flavors in slots $(1,2,3)$ and $\chi_{1/2}^{\sigma,[j]}$ denotes the spectator spin wave function, the physical ground configurations ($\kappa_i=1$) for the flavor antitriplet ($\Lambda_c^+, \Xi_c^{+,0}$) and flavor sextet ($\Xi_c^{\prime+,0}, \Omega_c^0$) states are expanded over the three charm slots as~\cite{Perez-Marcial:1989sch}
\begin{widetext}
\begin{align}
|B_{\Lambda_c,1}^{A},\lambda_i\rangle
&=\frac{1}{\sqrt3}\Bigl[
\frac{1}{\sqrt2}|(cdu-cud)\chi_{1/2}^{\rho,[1]}(\lambda_i)\rangle
+\frac{1}{\sqrt2}|(ucd-dcu)\chi_{1/2}^{\rho,[2]}(\lambda_i)\rangle
+\frac{1}{\sqrt2}|(udc-duc)\chi_{1/2}^{\rho,[3]}(\lambda_i)\rangle
\Bigr],
\nonumber\\
|B_{\Xi_c^+,1}^{A},\lambda_i\rangle
&=\frac{1}{\sqrt3}\Bigl[
\frac{1}{\sqrt2}|(csu-cus)\chi_{1/2}^{\rho,[1]}(\lambda_i)\rangle
+\frac{1}{\sqrt2}|(ucs-scu)\chi_{1/2}^{\rho,[2]}(\lambda_i)\rangle
+\frac{1}{\sqrt2}|(usc-suc)\chi_{1/2}^{\rho,[3]}(\lambda_i)\rangle
\Bigr],
\nonumber\\
|B_{\Xi_c^0,1}^{A},\lambda_i\rangle
&=\frac{1}{\sqrt3}\Bigl[
\frac{1}{\sqrt2}|(csd-cds)\chi_{1/2}^{\rho,[1]}(\lambda_i)\rangle
+\frac{1}{\sqrt2}|(dcs-scd)\chi_{1/2}^{\rho,[2]}(\lambda_i)\rangle
+\frac{1}{\sqrt2}|(dsc-sdc)\chi_{1/2}^{\rho,[3]}(\lambda_i)\rangle
\Bigr],
\nonumber\\
|B_{\Xi_c^{\prime+},1}^{S},\lambda_i\rangle
&=\frac{1}{\sqrt3}\Bigl[
\frac{1}{\sqrt2}|(csu+cus)\chi_{1/2}^{\lambda,[1]}(\lambda_i)\rangle
+\frac{1}{\sqrt2}|(ucs+scu)\chi_{1/2}^{\lambda,[2]}(\lambda_i)\rangle
+\frac{1}{\sqrt2}|(usc+suc)\chi_{1/2}^{\lambda,[3]}(\lambda_i)\rangle
\Bigr],
\nonumber\\
|B_{\Xi_c^{\prime 0},1}^{S},\lambda_i\rangle
&=\frac{1}{\sqrt3}\Bigl[
\frac{1}{\sqrt2}|(csd+cds)\chi_{1/2}^{\lambda,[1]}(\lambda_i)\rangle
+\frac{1}{\sqrt2}|(dcs+scd)\chi_{1/2}^{\lambda,[2]}(\lambda_i)\rangle
+\frac{1}{\sqrt2}|(dsc+sdc)\chi_{1/2}^{\lambda,[3]}(\lambda_i)\rangle
\Bigr],
\nonumber\\
|B_{\Omega_c^0,1}^{S},\lambda_i\rangle
&=\frac{1}{\sqrt3}\Bigl[
|css\chi_{1/2}^{\lambda,[1]}(\lambda_i)\rangle
+|scs\chi_{1/2}^{\lambda,[2]}(\lambda_i)\rangle
+|ssc\chi_{1/2}^{\lambda,[3]}(\lambda_i)\rangle
\Bigr].
\label{eq:app_slot_singlecharm_sextet_wfs}
\end{align}
\end{widetext}
For the radially excited configurations ($\kappa_i=2,3$), the spin-flavor wave functions are identical to those of $\kappa_i=1$, with the spatial profiles given by the radial excitations $\Psi_{200}^{(10,00)}$ and $\Psi_{200}^{(00,10)}$.
For the orbitally excited configuration ($\kappa_i=4$), the internal spatial wave function $\Psi_{200}^{(01,01)}$ is antisymmetric under spectator exchange ($l_\rho=1$), which by Fermi statistics requires the spectator spin state to flip: $\chi_{1/2}^{\rho} \leftrightarrow \chi_{1/2}^{\lambda}$. 
The fully symmetrized initial states for $\kappa_i=4$ are then expressed as
\begin{align}
|B_{i,4}^{A},\lambda_i\rangle
&=
\frac{1}{\sqrt3} \sum_{a=1}^{3}
\Psi_{200}^{(01,01),(i)[a]}
\phi_{i,A}^{[a]}
\chi_{1/2}^{\lambda,[a]}(\lambda_i),
\nonumber\\
|B_{i,4}^{S},\lambda_i\rangle
&=
\frac{1}{\sqrt3} \sum_{a=1}^{3}
\Psi_{200}^{(01,01),(i)[a]}
\phi_{i,S}^{[a]}
\chi_{1/2}^{\rho,[a]}(\lambda_i).
\label{eq:app_initial_kappa4_slot_decomp}
\end{align}
Here, the single slot flavor components $\phi_{i,\tau}^{[a]}$ match those in Eq.~\eqref{eq:app_slot_singlecharm_sextet_wfs}, with $\phi_{i,A}^{[3]}=\frac{1}{\sqrt2}(ud-du)c$ for $\Lambda_c^+$, $\frac{1}{\sqrt2}(us-su)c$ for $\Xi_c^+$, $\frac{1}{\sqrt2}(ds-sd)c$ for $\Xi_c^0$, and $\phi_{i,S}^{[3]}=\frac{1}{\sqrt2}(us+su)c$, $\frac{1}{\sqrt2}(ds+sd)c$, $ssc$ for $(\Xi_c^{\prime+}, \Xi_c^{\prime 0}, \Omega_c^0)$, respectively.

For the light octet baryons in the symmetric $\mathbf{56}$-plet ($\kappa_f=1,2$), the spectator quark pair in each slot component $[j]$ possesses definite flavor exchange symmetry.
Consequently, the spin-flavor state for each physical baryon reduces to a single spectator spin mode: the antisymmetric isospin singlet $ud$ pair in $\Lambda$ isolates $\chi_{1/2}^{\rho,[j]}$, whereas the symmetric pairs in $(n, \Sigma^0, \Sigma^-, \Xi^0, \Xi^-)$ select $\chi_{1/2}^{\lambda,[j]}$.
Suppressing the common spatial ground state $\Psi_{000}^s$, the physical ground configurations ($\kappa_f=1$) are expanded over the three slot components as~\cite{Perez-Marcial:1989sch}
\begin{widetext}
\begin{align}
\langle B_{n,1}^{\mathbf 8},\lambda_f|
&=-\frac{1}{\sqrt3}\Bigl[
\langle (udd)\chi_{1/2}^{\lambda,[1]}(\lambda_f)|
+\langle (dud)\chi_{1/2}^{\lambda,[2]}(\lambda_f)|
+\langle (ddu)\chi_{1/2}^{\lambda,[3]}(\lambda_f)|
\Bigr],
\nonumber\\
\langle B_{\Lambda,1}^{\mathbf 8},\lambda_f|
&=\frac{1}{\sqrt3}\Bigl[
\frac{1}{\sqrt2} \langle (sdu-sud)\chi_{1/2}^{\rho,[1]}(\lambda_f)|
+\frac{1}{\sqrt2}\langle (usd-dsu)\chi_{1/2}^{\rho,[2]}(\lambda_f)|
+\frac{1}{\sqrt2}\langle (uds-dus)\chi_{1/2}^{\rho,[3]}(\lambda_f)|
\Bigr],
\nonumber\\
\langle B_{\Sigma^0,1}^{\mathbf 8},\lambda_f|
&=\frac{1}{\sqrt3}\Bigl[
\frac{1}{\sqrt2}\langle (sdu+sud)\chi_{1/2}^{\lambda,[1]}(\lambda_f)|
+\frac{1}{\sqrt2}\langle (usd+dsu)\chi_{1/2}^{\lambda,[2]}(\lambda_f)|
+\frac{1}{\sqrt2}\langle (uds+dus)\chi_{1/2}^{\lambda,[3]}(\lambda_f)|
\Bigr],
\nonumber\\
\langle B_{\Sigma^-,1}^{\mathbf 8},\lambda_f|
&=\frac{1}{\sqrt3}\Bigl[
\langle (sdd)\chi_{1/2}^{\lambda,[1]}(\lambda_f)|
+\langle (dsd)\chi_{1/2}^{\lambda,[2]}(\lambda_f)|
+\langle (dds)\chi_{1/2}^{\lambda,[3]}(\lambda_f)|
\Bigr],
\nonumber\\
\langle B_{\Xi^0,1}^{\mathbf 8},\lambda_f|
&=-\frac{1}{\sqrt3}\Bigl[
\langle (uss)\chi_{1/2}^{\lambda,[1]}(\lambda_f)|
+\langle (sus)\chi_{1/2}^{\lambda,[2]}(\lambda_f)|
+\langle (ssu)\chi_{1/2}^{\lambda,[3]}(\lambda_f)|
\Bigr],
\nonumber\\
\langle B_{\Xi^-,1}^{\mathbf 8},\lambda_f|
&=-\frac{1}{\sqrt3}\Bigl[
\langle (dss)\chi_{1/2}^{\lambda,[1]}(\lambda_f)|
+\langle (sds)\chi_{1/2}^{\lambda,[2]}(\lambda_f)|
+\langle (ssd)\chi_{1/2}^{\lambda,[3]}(\lambda_f)|
\Bigr].
\label{eq:app_slot_light_xi_wfs_charm}
\end{align}
\end{widetext}
For the radially excited configuration ($\kappa_f=2$), the wave function retains the spin-flavor structure of $\kappa_f=1$ with the radial spatial mode $\Psi_{200}^s$.
For the mixed symmetry $\mathbf{70}$-plet configuration ($\kappa_f=3$), the physical states are obtained directly from Eq.~\eqref{eq:app_slot_light_xi_wfs_charm} by replacing the spectator spin wave functions in each slot component as $\chi_{1/2}^{\rho,[a]} \to \frac{1}{\sqrt2}(\Psi_{200}^{\lambda,(f)[a]} \chi_{1/2}^{\rho,[a]} + \Psi_{200}^{\rho,(f)[a]} \chi_{1/2}^{\lambda,[a]})$ and $\chi_{1/2}^{\lambda,[a]} \to \frac{1}{\sqrt2}(\Psi_{200}^{\rho,(f)[a]} \chi_{1/2}^{\rho,[a]} - \Psi_{200}^{\lambda,(f)[a]} \chi_{1/2}^{\lambda,[a]})$.

\subsection{Transition matrix elements}
\label{app:charm_transition_matrix_elements}

For an active quark line $j$, the transition matrix element $[\mathcal G_{\kappa_f\kappa_i}^{\tau}(j)]_{\alpha\beta}$ in Eq.~\eqref{eq:sc_full_amplitude_charm} is defined as
\begin{align}
\big[\mathcal G_{\kappa_f\kappa_i}^{\tau}(j)]_{\alpha\beta}(\lambda_f,\lambda_i)
\equiv
\left\langle B_{f,\kappa_f}^{\mathbf 8},\lambda_f\right|
\hat F_j^{q_f c}\,\hat S_j^{\alpha\beta}
\left|B_{i,\kappa_i}^{\tau,[j]},\lambda_i\right\rangle,
\label{eq:transition_kernel_def_charm}
\end{align}
where $\hat F_j^{q_fc}\equiv\hat b_{q_f}^{\dagger}(j)\hat b_c(j)$ changes the active quark flavor from $c$ to $q_f$ ($q_f=s,d$), and $\hat S_j^{\alpha\beta}\equiv|\alpha\rangle\langle\beta|$ changes its spin projection from $\beta$ to $\alpha$. The bracket $[j]$ denotes the initial state configuration with the charm quark in slot $j$, where the spectator pairs are $(23)$, $(13)$, and $(12)$ for $j=1,2,3$, respectively.

To calculate the transition matrix elements, we define the single line flavor matrix elements and spin transition tensors as
\begin{align}
I_{\zeta_f\tau}^{fi,j}
&\equiv
\left\langle\phi_f^{\zeta_f,[j]}\right|\hat F_j^{q_fc}\left|\phi_{i,\tau}^{[j]}\right\rangle,
\\
(\mathcal C_{\lambda_f\lambda_i}^{\sigma_f\sigma_i,j})_{\alpha\beta}
&\equiv
\left\langle\chi_{\lambda_f}^{\sigma_f,[j]}\right|\hat S_j^{\alpha\beta}\left|\chi_{\lambda_i}^{\sigma_i,[j]}\right\rangle,
\end{align}
where $\zeta_f,\sigma_f,\sigma_i\in\{\rho,\lambda\}$ and $\tau\in\{A,S\}$.
Because the one body weak current acts only on the active quark line $j$, the spectator pair in slots $r,s\ne j$ is unaffected.
The conservation of spectator flavor symmetry and coupled spin $S_{rs}$ leads to the selection rules
\begin{equation}
I_{\lambda A}^{fi,j}=I_{\rho S}^{fi,j}=0,
\qquad
\mathcal C^{\rho\lambda,j}=\mathcal C^{\lambda\rho,j}=0,
\label{eq:app_spectator_selection_rules}
\end{equation}
leaving $I_{\rho A}^{fi,j}$ and $I_{\lambda S}^{fi,j}$ as the only nonzero single line flavor transition matrix elements for $\tau=A$ and $\tau=S$, respectively.

The surviving spin matrix elements reduce to two same-sector blocks: the scalar spectator block $\mathbb I_j$ ($S_{rs}=0$, $\chi^\rho$) and the axial-vector spectator block $\mathbb S_j$ ($S_{rs}=1$, $\chi^\lambda$), defined by
\begin{align}
[\mathbb I_j(\lambda_f,\lambda_i)]_{\alpha\beta}
&\equiv (\mathcal C_{\lambda_f\lambda_i}^{\rho\rho,j})_{\alpha\beta}
= \left\langle \chi_{\lambda_f}^{\rho,[j]}\right| \hat S_j^{\alpha\beta} \left|\chi_{\lambda_i}^{\rho,[j]}\right\rangle,
\nonumber\\
[\mathbb S_j(\lambda_f,\lambda_i)]_{\alpha\beta}
&\equiv (\mathcal C_{\lambda_f\lambda_i}^{\lambda\lambda,j})_{\alpha\beta}
= \left\langle \chi_{\lambda_f}^{\lambda,[j]}\right| \hat S_j^{\alpha\beta} \left|\chi_{\lambda_i}^{\lambda,[j]}\right\rangle.
\label{eq:app_spin_blocks_def}
\end{align}
For a scalar spectator pair ($S_{12}=0$), the active quark spin carries the entire baryon helicity, yielding $[\mathbb I_3]_{\alpha\beta}=\delta_{\alpha,\lambda_f}\delta_{\beta,\lambda_i}$ in the slot-$3$ basis. 
For an axial-vector spectator pair ($S_{12}=1$), the block $\mathbb S_j$ describes the Clebsch--Gordan recoupling. 
Since the slot-$3$ basis matches the $(12)$ spectator pairing adopted in Eqs.~\eqref{eq:app_charm_A_configuration_basis} and \eqref{eq:app_charm_S_configuration_basis}, we first present the explicit transition kernels on line $j=3$.

For the flavor antitriplet initial state ($\tau=A$), the $3\times4$ transition kernel matrix on line $j=3$ takes the form
\begin{widetext}
\begin{align}
&\big[\mathcal G_{\kappa_f\kappa_i}^{A}(3)\big]_{\alpha\beta} \nonumber \\
&=
\frac{I_{\rho A}^{fi,3}}{\sqrt2}
\begin{pmatrix}
\Psi_{000}^{s,(f)*} \Psi_{000}^{(00,00),(i)[3]} [\mathbb I_3]_{\alpha\beta}
&
\Psi_{000}^{s,(f)*} \Psi_{200}^{(10,00),(i)[3]} [\mathbb I_3]_{\alpha\beta}
&
\Psi_{000}^{s,(f)*} \Psi_{200}^{(00,10),(i)[3]} [\mathbb I_3]_{\alpha\beta}
&
0
\\[6pt]
\Psi_{200}^{s,(f)*} \Psi_{000}^{(00,00),(i)[3]} [\mathbb I_3]_{\alpha\beta}
&
\Psi_{200}^{s,(f)*} \Psi_{200}^{(10,00),(i)[3]} [\mathbb I_3]_{\alpha\beta}
&
\Psi_{200}^{s,(f)*} \Psi_{200}^{(00,10),(i)[3]} [\mathbb I_3]_{\alpha\beta}
&
0
\\[6pt]
\dfrac{\Psi_{200}^{\lambda,(f)*} \Psi_{000}^{(00,00),(i)[3]} [\mathbb I_3]_{\alpha\beta}}{\sqrt2}
&
\dfrac{\Psi_{200}^{\lambda,(f)*} \Psi_{200}^{(10,00),(i)[3]} [\mathbb I_3]_{\alpha\beta}}{\sqrt2}
&
\dfrac{\Psi_{200}^{\lambda,(f)*} \Psi_{200}^{(00,10),(i)[3]} [\mathbb I_3]_{\alpha\beta}}{\sqrt2}
&
\dfrac{\Psi_{200}^{\rho,(f)*} \Psi_{200}^{(01,01),(i)[3]} [\mathbb S_3]_{\alpha\beta}}{\sqrt2}
\end{pmatrix}_{\kappa_f\kappa_i},
\label{eq:app_kernel_matrix_A_slot3}
\end{align}
\end{widetext}
where the rows and columns correspond to the final states $\kappa_f=1,2,3$ and initial states $\kappa_i=1,2,3,4$, respectively.
For $\kappa_i=1,2,3$, the scalar spectator pair ($S_{12}=0$) couples through $[\mathbb I_3]_{\alpha\beta}$ to the symmetric final states $\kappa_f=1,2$ and the $\lambda$ component of $\kappa_f=3$.
For $\kappa_i=4$, the axial-vector spectator pair ($S_{12}=1$) couples via $[\mathbb S_3]_{\alpha\beta}$ exclusively to the $\rho$ component of $\kappa_f=3$, causing the $(1,4)$ and $(2,4)$ matrix elements to vanish identically.
For the flavor sextet initial state ($\tau=S$), the corresponding transition kernel matrix is given by
\begin{widetext}
\begin{align}
&\big[\mathcal G_{\kappa_f\kappa_i}^{S}(3)\big]_{\alpha\beta}\nonumber \\
&=
\frac{I_{\lambda S}^{fi,3}}{\sqrt2}
\begin{pmatrix}
\Psi_{000}^{s,(f)*} \Psi_{000}^{(00,00),(i)[3]} [\mathbb S_3]_{\alpha\beta}
&
\Psi_{000}^{s,(f)*} \Psi_{200}^{(10,00),(i)[3]} [\mathbb S_3]_{\alpha\beta}
&
\Psi_{000}^{s,(f)*} \Psi_{200}^{(00,10),(i)[3]} [\mathbb S_3]_{\alpha\beta}
&
0
\\[6pt]
\Psi_{200}^{s,(f)*} \Psi_{000}^{(00,00),(i)[3]} [\mathbb S_3]_{\alpha\beta}
&
\Psi_{200}^{s,(f)*} \Psi_{200}^{(10,00),(i)[3]} [\mathbb S_3]_{\alpha\beta}
&
\Psi_{200}^{s,(f)*} \Psi_{200}^{(00,10),(i)[3]} [\mathbb S_3]_{\alpha\beta}
&
0
\\[6pt]
-\dfrac{\Psi_{200}^{\lambda,(f)*} \Psi_{000}^{(00,00),(i)[3]} [\mathbb S_3]_{\alpha\beta}}{\sqrt2}
&
-\dfrac{\Psi_{200}^{\lambda,(f)*} \Psi_{200}^{(10,00),(i)[3]} [\mathbb S_3]_{\alpha\beta}}{\sqrt2}
&
-\dfrac{\Psi_{200}^{\lambda,(f)*} \Psi_{200}^{(00,10),(i)[3]} [\mathbb S_3]_{\alpha\beta}}{\sqrt2}
&
\dfrac{\Psi_{200}^{\rho,(f)*} \Psi_{200}^{(01,01),(i)[3]} [\mathbb I_3]_{\alpha\beta}}{\sqrt2}
\end{pmatrix}_{\kappa_f\kappa_i}.
\label{eq:app_kernel_matrix_S_slot3}
\end{align}
\end{widetext}
Compared with the $\tau=A$ case, the roles of the spin blocks are reversed: the first three columns contain the axial-vector block $\mathbb S_3$, while the fourth column contains the scalar block $\mathbb I_3$.
The transition kernel $[\mathcal G_{\kappa_f\kappa_i}^\tau(3)]_{\alpha\beta}$ specifies the spatial wave functions and spin coupling for the transitions between the basis states $|B_{i,\kappa_i}^{\tau,[3]}\rangle$ and $|B_{f,\kappa_f}^{\mathbf 8}\rangle$.
Contracting the spin block with the weak current operator $[\mathcal O^\nu_{W,3}]^{\alpha\beta}$ and integrating over the Jacobi momenta $(\bm k_\rho,\bm k_\lambda)$, Eq.~\eqref{eq:sc_full_amplitude_charm} gives the single line matrix elements $H_{\kappa_f\kappa_i}^{\nu;\tau,3}$.
Summing over the initial and final basis configurations with the mixing coefficients $c_{\kappa_i}^{B_i}$ and $c_{\kappa_f}^{B_f}$, one directly obtains the physical hadronic transition amplitude via Eq.~\eqref{eq:charm_configuration_mixed_hadronic_sum}.

The transition kernels on quark lines $j=1$ and $j=2$ are obtained from the reference line $j=3$ through the slot permutation operator $\widehat P_{j3}$, defined as
\begin{equation}
\widehat P_{j3}
=
\begin{cases}
\widehat P_{13}, & j=1,\\
\widehat P_{23}, & j=2,\\
\mathbf 1, & j=3.
\end{cases}
\label{eq:app_permutation_operator_def}
\end{equation}
Under $\widehat P_{j3}$, the initial spatial, flavor, and spin wave functions transform as
\begin{align}
\Psi_{\kappa_i}^{(i)[j]} &= \widehat P_{j3}\Psi_{\kappa_i}^{(i)[3]}, \nonumber\\
\phi_{i,\tau}^{[j]} &= \widehat P_{j3}\phi_{i,\tau}^{[3]}, \nonumber\\
\chi^{\sigma,[j]} &= \widehat P_{j3}\chi^{\sigma,[3]},
\label{eq:app_charm_slot_permutation_rule}
\end{align}
while the single quark transition operator on line $j$ is given by the similarity transformation
\begin{equation}
\hat F_j^{q_fc}\hat S_j^{\alpha\beta}
=
\widehat P_{j3}
\bigl(\hat F_3^{q_fc}\hat S_3^{\alpha\beta}\bigr)
\widehat P_{j3}^{-1}.
\label{eq:app_operator_similarity_trans}
\end{equation}

Because the final light octet states $|B_{f,\kappa_f}^{\mathbf 8}\rangle$ are totally symmetric under $S_3$, one has $\langle B_{f,\kappa_f}^{\mathbf 8}|\widehat P_{j3} = \langle B_{f,\kappa_f}^{\mathbf 8}|$.
Combining this relation with Eq.~\eqref{eq:app_charm_slot_permutation_rule} and Eq.~\eqref{eq:app_operator_similarity_trans}, the transition kernel on line $j$ reduces to
\begin{align}
\big[\mathcal G_{\kappa_f\kappa_i}^{\tau}(j)\big]_{\alpha\beta}
&=
\left\langle B_{f,\kappa_f}^{\mathbf 8}\right|
\widehat P_{j3}
\bigl(\hat F_3^{q_fc}\hat S_3^{\alpha\beta}\bigr)
\widehat P_{j3}^{-1}
\left|B_{i,\kappa_i}^{\tau,[j]}\right\rangle
\nonumber\\
&=
\left\langle B_{f,\kappa_f}^{\mathbf 8}\right|
\hat F_3^{q_fc}\hat S_3^{\alpha\beta}
\left|B_{i,\kappa_i}^{\tau,[3]}\right\rangle
=
\big[\mathcal G_{\kappa_f\kappa_i}^{\tau}(3)\big]_{\alpha\beta}.
\label{eq:app_kernel_line_identity}
\end{align}
This relation shows that in the basis adapted to the spectator pair of line $j$, the algebraic $3\times4$ transition kernel $[\mathcal G_{\kappa_f\kappa_i}^{\tau}(j)]_{\alpha\beta}$ is identical to the reference matrix on line $j=3$ in Eqs.~\eqref{eq:app_kernel_matrix_A_slot3} and \eqref{eq:app_kernel_matrix_S_slot3}.

In the zero recoil limit, the single line matrix elements for the three active lines are strictly equal due to permutation symmetry.
At finite recoil, the recoil momentum transfer and Wigner rotations in Eq.~\eqref{eq:sc_full_amplitude_charm} are partitioned differently among the internal Jacobi coordinates for different active lines, which introduces slight kinematic differences, leading to $H_{\kappa_f\kappa_i}^{\nu;\tau,j} \simeq H_{\kappa_f\kappa_i}^{\nu;\tau,3}$.
Therefore, the coherent sum over the three active lines is approximately given by $\sum_{j=1}^3 H_{\kappa_f\kappa_i}^{\nu;\tau,j} \simeq \sqrt3 \langle B_{f,\kappa_f}^{\mathbf 8}| J_3^\nu(0) |B_{i,\kappa_i}^{\tau,[3]}\rangle$, where the factor $\sqrt3 = 3 \times (1/\sqrt3)$ arises from summing the three nearly equal quark line contributions together with the $1/\sqrt3$ normalization factor of the initial state in Eq.~\eqref{eq:single_charm_slot_state}.
In the practical numerical calculation, we keep the exact coherent sum over all three lines in Eq.~\eqref{eq:charm_configuration_mixed_hadronic_sum}.
For the $\Xi_c$--$\Xi'_c$ system, the $\tau=A$ and $\tau=S$ components are summed over the three active lines separately, and then coherently superposed with the mixing coefficients $c_{\kappa_i,\tau}^{\Xi_c^{(\prime)}}$ via Eq.~\eqref{eq:charm_xi_configuration_mixed_hadronic_sum}.

\section{Static $\mathrm{SU}(6)$ limit and form factors}
\label{app:static_form_factors}

In the static $\mathrm{SU}(6)$ limit, we take the zero recoil limit ($|\bm P_f|\to0$), set the spatial overlap between the normalized ground configurations to unity, and neglect the relativistic Wigner rotations ($D=\mathbf 1$).
The single line transition matrix element between the ground states ($\kappa_i=\kappa_f=1$) then reduces to
\begin{equation}
\mathcal C_O^{fi,j}(\lambda_f,\lambda_i)
=
\frac{1}{\sqrt6}
I_{\sigma\tau}^{fi,j}
\mathcal S_{O,(j)}^{\sigma\sigma}(\lambda_f,\lambda_i),
\label{eq:static_slot_factor_charm}
\end{equation}
where $(\sigma,\tau)=(\rho,A)$ for the flavor antitriplet ($S_{12}=0$) and $(\lambda,S)$ for the flavor sextet ($S_{12}=1$).
Here, $I_{\sigma\tau}^{fi,j}$ is the single line flavor overlap, and the prefactor $1/\sqrt6 = (1/\sqrt3)\times(1/\sqrt2)$ combines the charm slot and $\mathrm{SU}(6)$ spin-flavor normalizations.

In the spin basis adapted to the spectator pair of line $j$, the spin recoupling blocks $\mathcal S_{O,(j)}^{\sigma\sigma}(\lambda_f,\lambda_i)\equiv\langle\chi_{1/2}^{\sigma,[j]}(\lambda_f)|\hat O^{(j)}|\chi_{1/2}^{\sigma,[j]}(\lambda_i)\rangle$ are given by
\begin{align}
\mathcal S_{O,(j)}^{\rho\rho}(\lambda_f,\lambda_i)
&=
\begin{pmatrix}
\mathcal O_{\uparrow\uparrow} & \mathcal O_{\uparrow\downarrow}
\\
\mathcal O_{\downarrow\uparrow} & \mathcal O_{\downarrow\downarrow}
\end{pmatrix}_{\lambda_f\lambda_i},
\label{eq:sc_slot_rr_charm}
\end{align}
and
\begin{align}
\mathcal S_{O,(j)}^{\lambda\lambda}(\lambda_f,\lambda_i)
&=
\frac{1}{3}
\begin{pmatrix}
\mathcal O_{\uparrow\uparrow}+2\mathcal O_{\downarrow\downarrow}
&
-\mathcal O_{\uparrow\downarrow}
\\[4pt]
-\mathcal O_{\downarrow\uparrow}
&
2\mathcal O_{\uparrow\uparrow}+\mathcal O_{\downarrow\downarrow}
\end{pmatrix}_{\lambda_f\lambda_i},
\label{eq:sc_slot_ll_charm}
\end{align}
where $\mathcal O_{\alpha\beta}\equiv\langle\alpha|\hat O|\beta\rangle$ are the single quark spin matrix elements in the Pauli basis ($\lambda_f,\lambda_i=\pm1/2$).
For the scalar spectator pair ($\chi^\rho$), the active quark carries the entire baryon spin, while for the axial-vector spectator pair ($\chi^\lambda$), the matrix elements incorporate the Clebsch-Gordan recoupling with the $S_{12}=1$ core.

Summing over the three charm slots, the static hadronic matrix element between the ground configuration states is given by
\begin{align}
H_O^{fi,(0)}(\lambda_f,\lambda_i)
=
\sum_{j=1}^3 \mathcal C_O^{fi,j}(\lambda_f,\lambda_i)
=
\sqrt{\frac32}\,I_{\sigma\tau}^{fi,3}\,
\mathcal S_{O,(3)}^{\sigma\sigma}(\lambda_f,\lambda_i).
\label{eq:static_total_matrix_element}
\end{align}
For the forward helicity state ($\lambda_i=\lambda_f=+1/2$), the vector ($\hat O=\mathbf 1$) and axial-vector ($\hat O=\sigma_z$) currents give the spin block values $\mathcal S_{\mathbf 1}^{\rho\rho}=\mathcal S_{\mathbf 1}^{\lambda\lambda}=1$, $\mathcal S_{\sigma_z}^{\rho\rho}=1$, and $\mathcal S_{\sigma_z}^{\lambda\lambda}=-1/3$.
The static vector and axial-vector form factors then reduce to
\begin{align}
f_1^{\text{stat.}}
&=
\sqrt{\frac32}
\begin{cases}
I_{\rho A}^{fi,3}, & \tau=A,\\
I_{\lambda S}^{fi,3}, & \tau=S,
\end{cases}
\nonumber\\
g_1^{\text{stat.}}
&=
\sqrt{\frac32}
\begin{cases}
I_{\rho A}^{fi,3}, & \tau=A,\\
-\dfrac13 I_{\lambda S}^{fi,3}, & \tau=S.
\end{cases}
\label{eq:app_static_f1_g1_master}
\end{align}
Consequently, one obtains the static $\mathrm{SU}(6)$ ratios $g_1^{\text{stat.}}/f_1^{\text{stat.}}=1$ for the flavor antitriplet and $g_1^{\text{stat.}}/f_1^{\text{stat.}}=-1/3$ for the flavor sextet.
In Table~\ref{tab:slot_singlecharm_sf}, we summarize the single line factors $\mathcal C_O^{fi,j}$, the total static hadronic matrix elements $H_O^{fi,(0)}$, and the static form factors $f_1^{\text{stat.}}$ and $g_1^{\text{stat.}}$ for all semileptonic channels.

\begin{table*}[t]
	\centering
	\renewcommand{\arraystretch}{1.90}
	\setlength{\tabcolsep}{14pt}
	\caption{Static spin-flavor factors and $\mathrm{SU}(6)$ coefficients for the transitions $B_{i,1}\to B_{f,1}$ between the ground configuration states ($\kappa_i=\kappa_f=1$) of singly charmed and light octet baryons.
		The single line factors $\mathcal C_O^{fi,j}$ and their coherent sum $H_O^{fi,(0)}\equiv\sum_{j=1}^{3}\mathcal C_O^{fi,j}$ are defined in Eqs.~\eqref{eq:static_slot_factor_charm} and \eqref{eq:static_total_matrix_element}, respectively.
		The spin blocks $\mathcal S_O^{\rho\rho}$ and $\mathcal S_O^{\lambda\lambda}$ are given in Eqs.~\eqref{eq:sc_slot_rr_charm} and \eqref{eq:sc_slot_ll_charm}, while the operators $\hat O=\mathbf{1}$ and $\hat O=\sigma_z$ yield the static vector and axial-vector coefficients $f_1^{\text{stat.}}$ and $g_1^{\text{stat.}}$, respectively.}
	\label{tab:slot_singlecharm_sf}
	\begin{tabular}{c c c c c c c c}
		\toprule\toprule
		Channel & Transition & $\mathcal C_O^{fi,1}$ & $\mathcal C_O^{fi,2}$ & $\mathcal C_O^{fi,3}$ & $H_O^{fi,(0)}$ & $f_1^{\text{stat.}}$ & $g_1^{\text{stat.}}$ \\
		\midrule
		$\Lambda_c^+\to\Lambda\ell^+\nu_\ell$ & $c\to s$ & $\dfrac13\mathcal S_O^{\rho\rho}$ & $\dfrac13\mathcal S_O^{\rho\rho}$ & $\dfrac13\mathcal S_O^{\rho\rho}$ & $\mathcal S_O^{\rho\rho}$ & $1$ & $1$ \\
		$\Lambda_c^+\to n\ell^+\nu_\ell$ & $c\to d$ & $\dfrac{1}{\sqrt6}\mathcal S_O^{\rho\rho}$ & $\dfrac{1}{\sqrt6}\mathcal S_O^{\rho\rho}$ & $\dfrac{1}{\sqrt6}\mathcal S_O^{\rho\rho}$ & $\sqrt{\dfrac32}\mathcal S_O^{\rho\rho}$ & $\sqrt{\dfrac32}$ & $\sqrt{\dfrac32}$ \\
		\midrule
		$\Xi_c^+\to\Xi^0\ell^+\nu_\ell$ & $c\to s$ & $\dfrac{1}{\sqrt6}\mathcal S_O^{\rho\rho}$ & $\dfrac{1}{\sqrt6}\mathcal S_O^{\rho\rho}$ & $\dfrac{1}{\sqrt6}\mathcal S_O^{\rho\rho}$ & $\sqrt{\dfrac32}\mathcal S_O^{\rho\rho}$ & $\sqrt{\dfrac32}$ & $\sqrt{\dfrac32}$ \\
		$\Xi_c^0\to\Xi^-\ell^+\nu_\ell$ & $c\to s$ & $\dfrac{1}{\sqrt6}\mathcal S_O^{\rho\rho}$ & $\dfrac{1}{\sqrt6}\mathcal S_O^{\rho\rho}$ & $\dfrac{1}{\sqrt6}\mathcal S_O^{\rho\rho}$ & $\sqrt{\dfrac32}\mathcal S_O^{\rho\rho}$ & $\sqrt{\dfrac32}$ & $\sqrt{\dfrac32}$ \\
		$\Xi_c^+\to\Lambda\ell^+\nu_\ell$ & $c\to d$ & $\dfrac16\mathcal S_O^{\rho\rho}$ & $\dfrac16\mathcal S_O^{\rho\rho}$ & $\dfrac16\mathcal S_O^{\rho\rho}$ & $\dfrac12\mathcal S_O^{\rho\rho}$ & $\dfrac12$ & $\dfrac12$ \\
		$\Xi_c^+\to\Sigma^0\ell^+\nu_\ell$ & $c\to d$ & $\dfrac{1}{2\sqrt3}\mathcal S_O^{\rho\rho}$ & $\dfrac{1}{2\sqrt3}\mathcal S_O^{\rho\rho}$ & $\dfrac{1}{2\sqrt3}\mathcal S_O^{\rho\rho}$ & $\dfrac{\sqrt3}{2}\mathcal S_O^{\rho\rho}$ & $\dfrac{\sqrt3}{2}$ & $\dfrac{\sqrt3}{2}$ \\
		$\Xi_c^0\to\Sigma^-\ell^+\nu_\ell$ & $c\to d$ & $\dfrac{1}{\sqrt6}\mathcal S_O^{\rho\rho}$ & $\dfrac{1}{\sqrt6}\mathcal S_O^{\rho\rho}$ & $\dfrac{1}{\sqrt6}\mathcal S_O^{\rho\rho}$ & $\sqrt{\dfrac32}\mathcal S_O^{\rho\rho}$ & $\sqrt{\dfrac32}$ & $\sqrt{\dfrac32}$ \\
		\midrule
		$\Omega_c^0\to\Xi^-\ell^+\nu_\ell$ & $c\to d$ & $-\dfrac13\mathcal S_O^{\lambda\lambda}$ & $-\dfrac13\mathcal S_O^{\lambda\lambda}$ & $-\dfrac13\mathcal S_O^{\lambda\lambda}$ & $-\mathcal S_O^{\lambda\lambda}$ & $-1$ & $\dfrac13$ \\
		\bottomrule\bottomrule
	\end{tabular}
\end{table*}

\bibliography{ref}
\end{document}